\documentclass[10pt]{iopart}

\usepackage{iopams}  
\usepackage{graphicx,epsfig}
\usepackage{amssymb}
\newcommand{\ba}{\begin{eqnarray}}
\newcommand{\ea}{\end{eqnarray}}
\newcommand{\tfrac}{\frac}

\begin{document}

\title{Geometrical symmetries of nuclear systems: 
${\cal D}_{3h}$ and ${\cal T}_d$ symmetries in light nuclei}

\author{Roelof Bijker}
\address{Instituto de Ciencias Nucleares, 
Universidad Nacional Aut\'onoma de M\'exico, 
A.P. 70-543, 04510 M\'exico, D.F., M\'exico}
\ead{bijker@nucleares.unam.mx}

\begin{abstract}
The role of discrete (or point-group) symmetries in $\alpha$-cluster nuclei is discussed 
in the framework of the algebraic cluster model which describes the relative motion of the 
$\alpha$-particles. Particular attention is paid to the discrete symmetry of the geometric 
arrangement of the $\alpha$-particles, and the consequences for the structure of the 
corresponding rotational bands. The method is applied to study cluster states in the nuclei 
$^{12}$C and $^{16}$O. The observed level sequences can be understood in a simple way as a 
consequence of the underlying discrete symmetry that characterizes the geometrical configuration 
of the $\alpha$-particles, {\it i.e.} an equilateral triangle with ${\cal D}_{3h}$ symmetry for 
$^{12}$C, and a tetrahedron with ${\cal T}_d$ symmetry for $^{16}$O. The structure of rotational 
bands provides a fingerprint of the underlying geometrical configuration of $\alpha$-particles. 
\end{abstract}


\vspace{2pc}

\ioptwocol

\section{Introduction} 

The concept of symmetries has played an important role in nuclear structure physics, 
both continuous and discrete symmetries. Examples of continuous symmetries are isospin 
symmetry \cite{Heisenberg}, Wigner’s combined spin-isospin symmetry \cite{Wigner}, the 
(generalized) seniority scheme \cite{Racah,Talmi71}, the Elliott model \cite{Elliott} 
and the interacting boson model \cite{ibm}. In addition to providing simple analytic 
solutions which can be used to analyze and interpret the structure of nuclei, symmetries 
play a crucial role in establishing the connection between different models of 
nuclear structure, such as the spherical shell model of Goeppert-Mayer \cite{SM1} and Jensen \cite{SM2},  
the geometric collective model of Bohr and Mottelson \cite{BMVol2} and the interacting 
boson model of Arima and Iachello \cite{ibm}. In particular, the Elliott $SU(3)$ model 
provides a link between the spherical shell model and the quadrupole deformation 
of the geometric collective model, and the dynamical symmetries of the interacting boson 
model correspond to the harmonic vibrator, axial rotor and $\gamma$-unstable rotor limits 
of the geometric collective model. An recent review of the relation between symmetries of 
the spherical shell model and quadrupole and octupole deformations of the collective model 
can be found in Ref.~\cite{octupole}. 

On the other hand, discrete symmetries have been used in the context of collective models 
to characterize the intrinsic shape of the nucleus, such as axial symmetry for quadrupole 
deformations \cite{BMVol2}, and tetrahedral \cite{Dudek} and octahedral \cite{Dudek,Piet} 
symmetries for deformations of higher multipoles. A different application is found in the 
context of $\alpha$-particle clustering in light nuclei to describe the geometric configuration 
of the $\alpha$ particles. Early work on $\alpha$-cluster models goes back to the 1930's with 
studies by Wheeler \cite{wheeler}, and Hafstad and Teller \cite{Teller}, followed by later work 
by Brink \cite{Brink1,Brink2} and Robson \cite{Robson}. Recently, there has been a lot of 
renewed interest in the structure of $\alpha$-cluster nuclei, especially for the nucleus $^{12}$C 
\cite{FreerFynbo}. The measurement of new rotational excitations of the ground state \cite{Fre07,C12} 
and the Hoyle state \cite{Itoh,Freer,Gai,Fre11} has stimulated a large theoretical effort to 
understand the structure of $^{12}$C ranging from studies based on the semi-microscopic algebraic 
cluster model \cite{Cseh}, antisymmetrized molecular dynamics \cite{AMD}, fermionic molecular dynamics 
\cite{FMD}, BEC-like cluster model \cite{BEC}, {\it ab initio} no-core shell model \cite{NCSM}, 
lattice effective field theory \cite{EFT}, no-core symplectic model 
\cite{Draayer}, and the algebraic cluster model (ACM) \cite{C12,ACM,O16}. 
A recent review on the structure of $^{12}$C can be found in Ref.~\cite{FreerFynbo}. 

It is the aim of this contribution to discuss the ACM for two-, three- and four-body clusters, 
and study possible applications in $\alpha$-cluster nuclei, like $^{8}$Be, $^{12}$C and $^{16}$O.  
In these applications, it is important to take into account the permutation symmetry between the 
identical $\alpha$ clusters. The manuscript is organized as follows. In Section~2 the radiation problem 
of various configurations of identical charged particles is discussed at the classical level. 
In Sections~3 and 4 some general properties of the ACM are presented which are relevant for  
two-, three- and four-body identical clusters, such as the structure of the Hamiltonian, 
permutation and geometrical symmetries, the classical limit, electromagnetic couplings, dynamical symmetries
and shape-phase transitions. In the next three sections, the ACM for two-, three- and four-body 
clusters is developed in more detail. Particular attention is paid to the cases of the axial rotor, 
the oblate top and the spherical top, and their respective point group symmetries. 
Even though these special solutions do not correspond to a 
dynamical symmetry of the Hamiltonian, approximate formula for energies, form factors and electromagnetic 
transition rates can be derived in a semi-classical mean-field analysis. Finally, I discuss applications 
of the ACM to the cluster states in the nuclei $^{12}$C and $^{16}$O. 

\section{Classical treatment}

In order to appreciate the effect of the geometric configuration of a number of 
identical charges (without magnetic moment, as in the case for the $\alpha$-particle 
model) on the multipole radiation, first consider the classical radiation 
problem for a charge distribution $\rho(\vec{r})$ with multipole moments
\ba
Q_{\lambda \mu} = \int r^{l} \, Y_{\lambda \mu}(\theta,\phi) \, 
\rho(\vec{r}) \, d\vec{r} ~.
\ea
From these multipole moments, the transition probability per unit time can be calculated 
as \cite{Jackson}
\ba
T(E\lambda) = \frac{8\pi c}{\hbar c} \, \frac{\lambda+1}{\lambda \left[ (2\lambda+1)!! \right]^{2}} 
\, k^{2\lambda+1}\,B(E\lambda) ~,
\ea
with
\ba
B(E\lambda) = \sum_{\mu=-\lambda}^{\lambda} Q_{\lambda \mu}^{\ast} Q_{\lambda \mu} ~.
\ea
For a point-like charge distribution of $k$ identical charges with coordinates 
$\vec{r}_i=(r_i,\theta_i,\phi_i)$ ($i=1,\ldots,k$) the charge distribution 
is given by 
\ba
\rho (\vec{r}) = \frac{Ze}{k} \sum_{i=1}^{k} \delta(\vec{r}-\vec{r}_{i}) ~.
\label{rhor}
\ea
The corresponding $B(E \lambda)$ values are given by
\ba
B(E \lambda) = \left( \frac{Ze}{k} \right)^2 \frac{2\lambda+1}{4\pi}  
\sum_{i,j=1}^{k} (r_i r_j)^{\lambda} P_{\lambda}(\cos \theta_{ij}) ~,
\ea
where $\theta_{ij}$ denotes the relative angle between the vectors $\vec{r}_i$ and $\vec{r}_j$ 
\ba
\cos \theta_{ij} =  
\cos \theta_i \cos \theta_j + \sin \theta_i \sin \theta_j \cos(\phi_i-\phi_j) ~. 
\ea

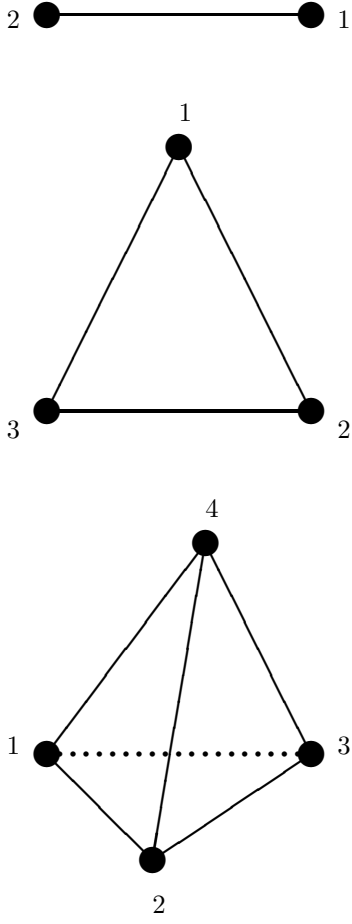
\begin{figure}
\centering
\setlength{\unitlength}{1.0pt}
\begin{picture}(150,375)(0,0)
\thicklines
\put( 65, 30) {\circle*{10}} 
\put( 25, 70) {\circle*{10}}
\put(125, 70) {\circle*{10}}
\put( 85,150) {\circle*{10}}
\multiput( 25, 70)( 5, 0){20}{\circle*{2}}
\put( 65, 30) {\line(-1, 1){40}}
\put( 65, 30) {\line( 3, 2){60}}
\put( 85,150) {\line(-3,-4){60}}
\put( 85,150) {\line( 1,-2){40}}
\put( 85,150) {\line(-1,-6){20}}
\put( 10, 70) {$1$}
\put( 65, 10) {$2$}
\put(135, 70) {$3$}
\put( 85,160) {$4$}
\put ( 25,200) {\circle*{10}}
\put (125,200) {\circle*{10}}
\put ( 75,300) {\circle*{10}}
\put ( 25,200) {\line ( 1,0){100}}
\put ( 25,200) {\line ( 1,2){ 50}}
\put (125,200) {\line (-1,2){ 50}}
\put ( 10,190) {3}
\put (135,190) {2}
\put ( 75,310) {1}
\put ( 25,350) {\circle*{10}}
\put (125,350) {\circle*{10}}
\put ( 25,350) {\line(1,0){100}}
\put ( 10,345) {2}
\put (135,345) {1}
\end{picture}
\caption[]{Geometry of a two-, three- and four-body system}
\label{fewbody}
\end{figure}

For two-body clusters ($k=2$), the coordinates of the two $\alpha$ particles are taken with 
respect to the center of mass: $(r_1,\theta_1,\phi_1)=(\beta,0,-)$ and 
$(r_2,\theta_2,\phi_2)=(\beta,\pi,-)$ (see Fig.~\ref{fewbody}). 
The corresponding $B(E\lambda)$ values are 
\ba
B(E \lambda) = \left( \frac{Ze\beta^{\lambda}}{2} \right)^{2}  
\frac{2\lambda+1}{4\pi} \left[2+2P_{\lambda}(-1) \right] ~,
\label{BEL2}
\ea
which vanish for the odd multipoles, and are equal to 
\ba
B(E\lambda) = \left( Ze\beta^{\lambda} \right)^{2} \, \frac{2\lambda+1}{4\pi} ~,
\ea
for the even multipoles. 

Next consider the case of three identical particles at the vertices of an equilateral triangle 
(point-group symmetry ${\cal D}_{3h}$) (see Fig.~\ref{fewbody}). The origin is placed at the 
center of mass so that the distance from the center is the same for all three particles $r_i=\beta$. 
The spherical coordinates of the three particles $\vec{r}_{i}=(r_{i},\theta_{i},\phi_{i})$ can be 
taken as $(\beta,0,-)$, $(\beta,2\pi/3,0)$ and $(\beta,2\pi/3,\pi)$ for $i=1,2$ and $3$, respectively, 
such that the relative angles $\cos \theta_{ij}=-1/2$ for all $i\neq j$.  
For this configuration, the multipole radiation is given by 
\ba
B(E \lambda) = \left( \frac{Ze\beta^{\lambda}}{3} \right)^{2} 
\frac{2\lambda+1}{4\pi} \left[3+6P_{\lambda}(-\tfrac{1}{2}) \right] ~, 
\label{BEL3} 
\ea
which gives 
\numparts
\ba
B(E1) &=& 0 ~, \\
B(E2) &=& (Ze)^{2} \, \frac{5}{4\pi} \, \frac{1}{4} \, \beta^{4} ~, \\
B(E3) &=& (Ze)^{2} \, \frac{7}{4\pi} \, \frac{5}{8} \, \beta^{6} ~, \\
B(E4) &=& (Ze)^{2} \, \frac{9}{4\pi} \, \frac{9}{64} \, \beta^{8} ~, \\
B(E5) &=& (Ze)^{2} \, \frac{11}{4\pi} \, \frac{35}{128} \, \beta^{10} ~, \\
B(E6) &=& (Ze)^{2} \, \frac{13}{4\pi} \, \frac{281}{512} \, \beta^{12} ~.  
\ea
\endnumparts
The dipole radiation $B(E1)$ vanishes because of the spatial symmetry of the 
charge distribution. 

Finally, consider the case of four identical particles at the vertices of a tetrahedron 
(point-group symmetry ${\cal T}_{d}$) (see Fig.~\ref{fewbody}). Again, the origin is placed at the 
center of mass so that the distance from the center is the same for all four particles $r_i=\beta$. 
The spherical coordinates of the three particles $\vec{r}_{i}=(r_{i},\theta_{i},\phi_{i})$ can be 
taken as $(\beta,0,-)$, $(\beta,\theta,0)$, $(\beta,\theta,2\pi/3)$ and $(\beta,\theta,4\pi/3)$, 
respectively, such that the relative angles satisfies $\cos \theta_{ij}=-1/3$ for all $i\neq j$. 
For this configuration, the multipole radiation is given by 
\ba
B(E\lambda) = \left( \frac{Ze\beta^{\lambda}}{4} \right)^{2}  
\frac{2\lambda +1}{4\pi} \left[ 4+12P_{\lambda}(-\tfrac{1}{3}) \right] ~,
\label{BEL4}
\ea
with 
\numparts
\ba
B(E1) &=& 0 ~, \\
B(E2) &=& 0 ~, \\
B(E3) &=& (Ze)^{2} \frac{7}{4\pi} \frac{5}{9} \beta^{6} ~, \\
B(E4) &=& (Ze)^{2} \frac{9}{4\pi} \frac{7}{27} \beta^{8} ~, \\
B(E5) &=& 0 ~, \\
B(E6) &=& (Ze)^{2} \frac{13}{4\pi} \frac{32}{81}\beta^{12} ~.
\ea
\endnumparts
In this case, the multipoles with $\lambda=1$, $2$ and $5$ vanish as a consequence of the 
tetrahedral symmetry of the configuration of four $\alpha$ particles. 

\section{Algebraic cluster model}

Algebraic models have found useful applications both in many-body and in few-body 
systems. Algebraic methods are based on the general criterion to introduce a 
$U(\nu+1)$ spectrum generating algebra for a bound-state problem with $\nu$ degrees 
of freedom. Well-known examples are the $U(6)$ interacting boson model for the 
$\nu=5$ quadrupole degrees of freedom in collective nuclei \cite{ibm}, and 
the $U(4)$ vibron model for the $\nu=3$ dipole degrees of freedom in diatomic 
molecules \cite{vibron}. 

In this section, I briefly review the basic ingredients of the algebraic cluster 
model (ACM) which was introduced to describe the relative motion of cluster systems 
\cite{ACM}. The relevant degrees of freedom of a system of $k$-body clusters are given 
by the $k-1$ relative Jacobi coordinates 
\numparts
\ba
\vec{\rho}_{1} &=& \frac{1}{\sqrt{2}} \left( \vec{r}_{1} - \vec{r}_{2} \right) ~, \\ 
\vec{\rho}_{2} &=& \frac{1}{\sqrt{6}} \left( \vec{r}_{1} + \vec{r}_{2} - 2\vec{r}_{3} \right) ~, \\
&\vdots& \nonumber\\
\vec{\rho}_{k-1} &=& \frac{1}{\sqrt{(k-1)k}}  
\left( \sum_{j=1}^{k-1} \vec{r}_{j} - (k-1) \vec{r}_{k} \right) ~,
\label{Jacobi}
\ea
\endnumparts
and their conjugate momenta, $\vec{p}_{j}$. Here $\vec{r}_j$ denotes the 
position vector of the $j$-th cluster. The building blocks of the ACM consist 
of a vector boson for each relative coordinate and conjugate momentum  
\numparts
\ba
b^{\dagger}_{j,m} &=& \frac{1}{\sqrt{2}} \left( \rho - ip \right)_{j,m} ~, \\
b_{j,m} &=& \frac{1}{\sqrt{2}} \left( \rho + ip \right)_{j,m} ~,
\ea
\endnumparts
with $j=1,\ldots,k-1$ and $m=-1,0,1$, and a scalar boson, $s^{\dagger}$, $s$. 
In the ACM, cluster states are described in terms of a system of $N$ interacting 
bosons with angular momentum and parity $L^P=1^-$ (dipole or vector bosons) 
and $L^P=0^+$ (monopole or scalar bosons). The $3(k-3)$ components of the vector 
bosons together with the scalar boson span a $(3k-2)$-dimensional space with group 
structure $U(3k-2)$. The many-body states are classified according to the totally 
symmetric irreducible representation $[N]$ of $U(3k-2)$, where $N$ represents the 
total number of bosons $N=n_s+\sum_j n_j$. 

In summary, the ACM is an algebraic treatment of cluster states which is based on 
the spectrum generating algebra of $U(3k-2)$ where $k$ denotes the number of clusters. 
For two-body clusters ($k=2$), it reduces to the $U(4)$ vibron model which was 
introduced originally in molecular physics \cite{vibron}, but which has also found 
applications in nuclear physics \cite{cluster}, and hadronic physics (mesons) 
\cite{mesons}. The ACM for three-body clusters ($k=3$) was introduced to describe 
the relative motion of three-quark configurations in baryons \cite{BIL},  
with applications in molecular physics \cite{BDL} and nuclear physics \cite{C12,ACM}  
as well. More recently, the ACM was extended to four-body clusters ($k=4$) to describe 
the cluster states of the nucleus $^{16}$O in terms of the relative motion of four-alpha 
particles \cite{O16,RB}.  

\subsection{Geometrical symmetries}

In application to $\alpha$-cluster nuclei, like $^{8}$Be, $^{12}$C and $^{16}$O, in which 
the constuent parts are identical, the eigenstates of the Hamiltonian should transform 
according to the symmetric representations of the permutation group ($S_k$ for $k$ identical 
objects). The permutation symmetry of $k$ objects is determined by the transposition $P(12)$ 
and the cyclic permutation $P(12 \cdots k)$. All other permutations can be expressed in 
terms of these elementary ones \cite{KM}. Algebraically, the transposition and cyclic 
permutation can be expressed in terms of the generators 
$b^{\dagger}_{i} b_{j} = \sum_{m} b^{\dagger}_{i,m} b_{j,m}$ that act in 
index space ($i,j=1,\ldots,k-1$), The scalar boson, 
$s^{\dagger}$, transforms as the symmetric representation characterized by Young tableau $[k]$, 
whereas the $k-1$ Jacobi vector bosons, $b^{\dagger}_j$ with $j=1,\ldots,k-1$,   
transform as the components of the mixed symmetry representation $[k-1,1]$. Next, one can 
use the multiplication rules for $S_k$ to to construct physical operators with the 
appropriate symmetry properties. For example, for the bilinear products of the Jacobi bosons, 
one finds
\ba
[k-1,1] \otimes [k-1,1] &=& [k] \oplus [k-1,1] 
\nonumber\\
&&\oplus [k-2,1,1] \oplus [k-2,2] ~,
\ea
for $k \geq 4$. For three clusters ($k=3$) only the first three terms are present, and 
for two clusters ($k=2$) only the first term. 
The isomorphism between the permutation group $S_k$ and the symmetries of a regular simplex 
in $k-1$ dimensions \cite{Coxeter} will be used in the next sections to establish the connection 
with the ${\cal D}_{3h}$ and ${\cal T}_d$ point groups for cluster configurations where the 
clusters are located at the vertices of an equilateral triangle and a regular tetrahedron, respectively. 

\subsection{Hamiltonian}

As a result, for a system of identical clusters the most general one- and two-body Hamiltonian 
that conserves the total number of bosons, angular momentum and parity, and is invariant under 
the permutation group $S_k$, is given by 
\ba
H &=& \epsilon_{0} \, s^{\dagger} \tilde{s}
- \epsilon_{1} \sum_{i} b_{i}^{\dagger} \cdot \tilde{b}_{i}  
\nonumber\\
&& + u_0 \, s^{\dagger} s^{\dagger} \, \tilde{s} \tilde{s}  
- u_1 \sum_{i} s^{\dagger} b_{i}^{\dagger} \cdot \tilde{b}_{i} \tilde{s} 
\nonumber\\
&& + v_0 \left( \sum_i b_{i}^{\dagger} \cdot b_{i}^{\dagger} \, \tilde{s} \tilde{s} + {\rm h.c.} \right)
\nonumber\\
&& + \sum_{L} \sum_{iji'j'} v^{(L)}_{iji'j'} \, [ b_{i}^{\dagger} b_{j}^{\dagger} ]^{(L)} \cdot 
[ \tilde{b}_{i'} \tilde{b}_{j'} ]^{(L)} ~,
\label{HSk}
\ea
with $\tilde{b}_{i,m}=(-1)^{1-m} b_{i,-m}$ and $\tilde{s}=s$. The $\epsilon_0$, $\epsilon_1$, $u_0$, 
$u_1$ and $v_0$ terms in Eq.~(\ref{HSk}) are scalars under $S_k$. The restrictions imposed by the 
permutation symmetry on the coefficients $v^{(L)}_{iji'j'}$ appearing in the last term, will be 
discussed in more detail in the next sections for the cases of two-, three- and four-body clusters. 

In general, the matrix elements of the Hamiltonian of Eq.~(\ref{HSk}) are calculated in a 
set of coupled harmonic oscillator basis states characterized by the total number of bosons 
$N=n_s+\sum_j n_j$, angular momentum $L$ and parity $P$. 
Although for the harmonic oscillator there exists 
a procedure for the explicit construction of states with good permutation symmetry \cite{KM}, in the 
application to the ACM the number of oscillator quanta may be large (up to 10) and moreover the 
oscillator shells are mixed by the $v_0$ term. Therefore, a general procedure was developed in which 
the wave functions with good permutation symmetry $|\psi_t\rangle$ are generated numerically by 
diagonalizing $S_k$ invariant interactions. Subsequently, the permutation 
symmetry $t$ of a given wave function is determined by examining its transformation properties 
under the transposition $P(12)$ and the cyclic permutation $P(12 \cdots k)$.  

\subsection{Geometry}
\label{geometry}

In general, geometric properties of algebraic models such as the interacting boson model 
\cite{ibm}, the vibron model \cite{vibron} and the algebraic cluster model \cite{ACM} 
can be studied with time-dependent mean-field approximations. The mean-field 
equations can be derived by minimizing the action \cite{onno,Levit}
\ba
S = \int_0^T dt \, \langle N;\vec{\alpha} \mid i \frac{\partial}{\partial t} - H 
\mid N;\vec{\alpha} \rangle ~. 
\ea 
Here $\mid N;\vec{\alpha} \rangle$ represents an intrinsic or coherent state 
as a variational wave function in terms of a condensate of $N$ deformed bosons which depend on 
geometric variables $\vec{\alpha}$ \cite{BM,GK,DSI}.  
For the $k$-body ACM, these coherent states are given by 
\ba
\nonumber\\
\mid N;\vec{\alpha} \rangle = \mid N;\vec{\alpha}_{1},\ldots,\vec{\alpha}_{k-1} \rangle 
= \frac{1}{\sqrt{N!}} (b_{c}^{\dagger})^{N}\,|0\rangle ~,
\label{coherent}
\ea
where the condensate boson $b_c^{\dagger}$ is parametrized in terms of $3(k-1)$ complex 
variables 
\ba
b_{c}^{\dagger} = \sqrt{1-\sum_j \vec{\alpha}_{j} \cdot \vec{\alpha}_{j}^{\,\ast}}  
\; s^{\dagger} + \sum_j \vec{\alpha}_{j} \cdot \vec{b}_{j}^{\,\dagger} ~.
\label{bc}
\ea
The variational principle gives Hamilton's equations of motion 
\ba
\dot{\pi}_{j} = -\frac{\partial H_{\rm cl}}{\partial \alpha_{j}} ~,
\hspace{1cm}
\dot{\alpha}_{j} =  \frac{\partial H_{\rm cl}}{\partial \pi_{j}} ~,
\label{eqmot}
\ea
where $\alpha_j$ and $\pi_j=i\alpha^{\ast}_j$ represent canonical coordinates 
and momenta, and $H_{\rm cl}$ denotes the classical limit of the Hamiltonian  
which is defined as the expectation value of the normally ordered Hamiltonian 
in the coherent state of Eq.~(\ref{coherent}) divided by the total number of 
bosons 
\ba
H_{\rm cl} = \frac{1}{N} 
\langle N; \vec{\alpha} \mid \,: H :\, \mid N;\vec{\alpha} \rangle ~.
\label{classical}
\ea
Bound states correspond to periodic classical trajectories $\alpha_j(t)$, 
$\pi_j(t)$ with period $T$ that satisfy a Bohr-Sommerfeld type quantization 
rule \cite{onno}
\ba
N \int_0^T \pi_j \dot{\alpha}_j dt = N \oint \pi_j d\alpha_j = 2\pi n_j ~.
\ea
The energy associated with a periodic classical orbital is independent of 
time and is given by $E/N=H_{\rm cl}(\alpha_j,\pi_j)$.

For the geometric analysis of the ACM Hamiltonian it is convenient to use 
spherical rather than cartesian coordinates and momenta \cite{onno,Levit,OSvR} 
\ba
\alpha _{j,\mu} = \frac{1}{\sqrt{2}} \sum_{\nu} 
{\cal D}_{\mu \nu}^{(1)}(\phi_{j},\theta_{j},0) \, \beta_{j,\nu} ~,
\ea
with 
\ba
\left( \begin{array}{c} \beta_{j,1} \\ \beta_{j,0} \\ \beta_{j,-1} \end{array} \right) 
= \left( \begin{array}{c}
\,[-p_{\phi_{j}}/\sin \theta_{j}-ip_{\theta_{j}}]/q_{j}\sqrt{2} \\ 
\rho_{j}+ip_{j} \\ \,[-p_{\phi_{j}}/\sin \theta_{j}+ip_{\theta_{j}}]/q_{j}\sqrt{2}
\end{array} \right) ~,
\ea
for $j=1,\ldots,k-1$. 

\subsection{Electromagnetic couplings}
\label{ff}

The transition form factor for the excitation of discrete nuclear levels 
is defined as \cite{deForest}  
\ba
F(\gamma LM \rightarrow \gamma' L' M';q) 
\nonumber\\
\;=\; \frac{1}{Ze} \int \mbox{e}^{i \vec{q} \cdot \vec{r}} \, 
\left< \gamma' L' M' \left| \, \hat \rho(\vec{r}) \, \right| \gamma L M \right>  \, d\vec{r} 
\nonumber\\ 
\;=\; \frac{4\pi}{Ze} \sum_{\lambda\mu} i^{\lambda} \, Y_{\lambda\mu}^{\ast}(\hat q) \, 
\left< \gamma' L' M' \left| \, \hat M_{\lambda\mu}(q) \, \right| \gamma L M \right> ~,
\ea
with $q=|\vec{q}|$ and 
\ba
\left< \gamma' L' M' \left| \, \hat M_{\lambda\mu}(q) \, \right| \gamma L M \right> 
\nonumber\\
\;=\; \int j_{\lambda}(qr) \, Y_{\lambda\mu}(\hat r) \, 
\left< \gamma' L' M' \left| \, \hat \rho(\vec{r}) \, \right| \gamma L M \right>  \, d\vec{r} ~. 
\ea
In the long wavelength limit, only one multipole contributes (with $\lambda=|L_i-L_f|$). 
After summing the square of the transition form factor over final and 
averaging over initial magnetic substates, one obtains 
\ba
\frac{1}{2L+1} \sum_{M M'} \left| F(\gamma LM \rightarrow \gamma' L' M';q) \right|^2 
\nonumber\\
\;\rightarrow\; \frac{4\pi}{(Ze)^2} \frac{q^{2\lambda}}{[(2\lambda+1)!!]^2} \, 
B(E\lambda;\gamma L \rightarrow \gamma' L') ~.
\ea
As a consequence, the reduced transition probabilities $B(E\lambda)$ can be extracted 
from the transition form factors in the long wavelength limit 
\ba
B(E\lambda;\gamma L \rightarrow \gamma' L') = \frac{(Ze)^2}{4\pi} \frac{[(2\lambda+1)!!]^{2}}{2L+1} 
\nonumber\\
\hspace{1cm} \lim_{q\rightarrow 0} 
\sum_{M M'} \frac{\left| F(\gamma LM \rightarrow \gamma' L' M';q) \right|^2}{q^{2\lambda}} ~. 
\ea
For the point-like charge distribution of Eq.~(\ref{rhor}) one finds for the transition form factor 
\ba
F(\gamma LM \rightarrow \gamma' L' M';q) 
\nonumber\\
\;=\; \left< \gamma' L'M' \left| \, \frac{1}{k} \sum_{j=1}^k e^{i \vec{q} \cdot \vec{r}_j} \, \right| \gamma LM \right> 
\nonumber\\ 
\;=\; \left< \gamma' L'M' \left| \, e^{i \vec{q} \cdot \vec{r}_k} \, \right| \gamma LM \right> 
\\ 
\;=\; \sum_{M''} {\cal D}^{(L')}_{M'M''}(\hat{q}) {\cal F}(\gamma LM'' \rightarrow \gamma' L'M'';q) \, {\cal D}^{(L)}_{M''M}(-\hat{q}) ~, 
\nonumber
\ea
with 
\ba
{\cal F}(\gamma LM'' \rightarrow \gamma' L'M'';q) 
\nonumber\\
\;=\; \left< \gamma' L'M'' \left| e^{-i q \sqrt{\frac{k-1}{k}} \rho_{k-1,z}} \right| \gamma LM'' \right> ~.
\label{ffme}
\ea
This result was obtained by first using the permutation symmetry of the initial and final wave functions, 
and next carrying out a transformation to center-of-mass and relative Jacobi coordinates, and integrating 
over the center-of-mass coordinate. The reduced transition probabilities $B(E\lambda)$ can be obtained in 
the long wavelength limit by
\ba
B(E\lambda;\gamma L \rightarrow \gamma' L') = \frac{(Ze)^2}{4\pi} \frac{[(2\lambda+1)!!]^{2}}{2L+1} 
\nonumber\\
\hspace{1cm} \lim_{q\rightarrow 0} \sum_{M} 
\frac{\left| {\cal F}(\gamma LM \rightarrow \gamma' L'M;q) \right|^{2}}{q^{2\lambda}} ~. 
\label{belif}
\ea

In order to calculate transition form factors and transition probabilities in the algebraic cluster model 
one has to express the transition operator in terms of an algebraic operator. The matrix elements can be 
obtained algebraically by making the replacement  
\ba
\sqrt{\frac{k-1}{k}} \rho_{k-1,z} \;\rightarrow\; \beta \hat D_{k-1,z}/X_D ~, 
\label{map} 
\ea
where $\beta$ represents the scale of the coordinate and $X_D$ is a normalization 
factor which is related the reduced matrix element of the dipole operator. 
The replacement in Eq.~(\ref{map}) comes from the fact that in the large $N$ 
limit, the dipole operator 
\ba
\hat D_{k-1} = (b^{\dagger}_{k-1} \tilde{s} - s^{\dagger} \tilde{b}_{k-1})^{(1)} ~, 
\label{dipole}
\ea
corresponds to the Jacobi coordinate $\vec{\rho}_{k-1}$ \cite{onno}. 
\ba
\left( D_{k-1,z} \right)_{\rm cl} &=& \frac{1}{N} 
\langle N; \vec{\alpha} \mid \,: \hat D_{k-1,z} :\, \mid N;\vec{\alpha} \rangle 
\nonumber\\
&=& \left( \alpha_{k-1,0} + \alpha_{k-1,0}^{\ast} \right) 
\sqrt{1-\sum_j \vec{\alpha}_{j} \cdot \vec{\alpha}_{j}^{\,\ast}} 
\nonumber\\
&=& \rho_{k-1} \sqrt{2-\sum_j \left( p_{j}^{2}+\rho_{j}^{2}+\frac{L_{j}^{2}}{\rho_{j}^{2}} \right)} ~, 
\label{dcl}
\ea
where $L_{j}^{2}$ is the angular momentum in polar coordinates of the $j$-th oscillator 
\ba
L_{j}^{2} = p_{\theta_{j}}^{2} + \frac{p_{\phi_{j}}^{2}}{\sin^{2}\theta_{j}} ~. 
\ea
The square root factor appearing in Eq.~(\ref{dcl}) is due to the presence of the 
scalar boson in the dipole operator, and is a consequence of the finiteness of the 
model space of the ACM. 

In summary, the transition form factors can be expressed in the ACM in terms of the 
matrix elements 
\ba
{\cal F}(\gamma LM \rightarrow \gamma' L'M;q) = \left< \gamma' L'M \, 
\left| \, \hat T(\epsilon) \, \right| \, \gamma LM \right> ~, 
\ea
with
\ba
\hat T(\epsilon) = e^{i \epsilon \hat{D}_{k-1,z}} = e^{-iq\beta \hat{D}_{k-1,z}/X_{D}} ~. 
\label{trans}
\ea
In general, the transition form factors cannot be obtained in closed analytic form, 
but have to be evaluated numerically. 

The results discussed so far, are for point-like constituent particles with a charge distribution 
given by Eq.~(\ref{rhor}). Next, let's consider the case in which the constituent particles have a 
finite size. With the application to $\alpha$-cluster nuclei in mind, it is reasonable to assume 
a Gaussian form 
\ba
\rho (\vec{r}) = \frac{Ze}{k} \left( \frac{\alpha}{\pi} \right)^{3/2} 
\sum_{i=1}^{k} \exp \left[ -\alpha \left( \vec{r}-\vec{r}_{i} \right)^{2}\right] ~.  
\label{rhor1}
\ea
As a consequence, all form factors are multiplied by an exponential factor $\exp(-q^{2}/4\alpha)$. 
The charge radius can be obtained from the slope of the elastic form 
factor in the origin 
\ba
\langle r^{2} \rangle^{1/2} &=& \left[ -6 \left. 
\frac{d {\cal F}(0_1^{+} \rightarrow 0_1^{+};q)}{dq^{2}} \right|_{q=0} \right]^{1/2} 
\nonumber\\ 
&=& \sqrt{\frac{3}{2\alpha}+\beta^{2}} ~. 
\label{radius}
\ea
Form factors and $B(E\lambda)$ values only depend on the parameters $\alpha$ and $\beta$. 
The coefficient $\beta$ can be determined from the first minimum in the elastic form factor, 
and the charge radius can be used to fix the value of $\alpha$. 

\section{Dynamical symmetries}
\label{limits}

In general, the ACM Hamiltonian has to diagonalized numerically in order to obtain the 
energy eigenvalues and corresponding eigenvectors. It is of general interest to study 
limiting cases of the Hamiltonian of Eq.~(\ref{HSk}), in which the energy spectra, 
electromagnetic transition rates and form factors can be obtained in closed form. 
These special solutions provide benchmarks in which energy spectra and other spectroscopic 
properties can be interpreted in a clear and transparent way. 
These special cases correspond to so-called dynamical symmetries which arise when the 
Hamiltonian has a certain group structure $G$, and it can be expressed in terms of 
Casimir invariants of a chain of subgroups of $G$ only. 
The eigenstates can then be classified uniquely according to the irreducible representations 
of $G$ and its subgroups. The energy eigenvalues are given by the expectation values of the 
Casimir operators. 

The algebraic cluster model has a rich algebraic structure, which includes both continuous 
and discrete symmetries. The ACM Hamiltonian for the $k$-body problem has the group structure 
$G=U(3k-2)$. In this section, I discuss the two dynamical symmetries which are related to the 
group lattice
\ba
U(3k-2) \supset \left\{ \begin{array}{c} U(3k-3) \\ \\ SO(3k-2) \end{array} \right\} 
\supset SO(3k-3) ~.
\label{chain}
\ea
These dynamical symmetries are limiting cases of the ACM, and are called the $U(3k-3)$ and 
$SO(3k-2)$ limit, respectively. In the following, I discuss these special solutions  
for any number of clusters $k$, and will show that, by studying the classical limit, they can be 
interpreted as the harmonic oscillator and the deformed oscillator in $3(k-1)$ dimensions \cite{RB}.  

\subsection{Harmonic oscillator}
\label{hosc}

The first dynamical symmetry corresponds to the group chain 
\ba
\left| \begin{array}{ccccccccc}
U(3k-2) &\supset& U(3k-3) &\supset& SO(3k-3) \\
N &,& n &,& \tau \end{array} \right> \nonumber
\ea
The basis states are classified by the quantum numbers $N$, $n$ and $\tau$, which characterize 
the irreducible representations of $U(3k-2)$, $U(3k-3)$ and $SO(3k-3)$, respectively. Here $N$ 
is the total number of bosons, and $n$ denotes the number of oscillator quanta $n=\sum_j n_j=0,1,\ldots,N$. 
The energy levels are organized into oscillator shells $n$, and are 
further labeled by $\tau$ with $\tau=n,n-2,\ldots,1$ or $0$ for $n$ odd or even. 
The parity of the levels is given by $P=(-1)^{n}$. 

\begin{figure}
\centering
\setlength{\unitlength}{0.7pt} 
\begin{picture}(300,230)(0,0)
\thinlines
\put (  0,  0) {\line(1,0){300}}
\put (  0,230) {\line(1,0){300}}
\put (  0,  0) {\line(0,1){230}}
\put (120,  0) {\line(0,1){230}}
\put (300,  0) {\line(0,1){230}}
\thicklines
\put ( 50, 60) {\line(1,0){20}}
\put ( 50, 90) {\line(1,0){20}}
\put ( 50,120) {\line(1,0){20}}
\put ( 50,149) {\line(1,0){20}}
\put ( 50,151) {\line(1,0){20}}
\put ( 50,178) {\line(1,0){20}}
\put ( 50,180) {\line(1,0){20}}
\put ( 50,182) {\line(1,0){20}}
\thinlines
\put ( 20, 57) {$n$=0}
\put ( 20, 87) {$n$=1}
\put ( 20,117) {$n$=2}
\put ( 20,147) {$n$=3}
\put ( 20,177) {$n$=4}
\put ( 75, 57) {$0$}
\put ( 75, 87) {$1$}
\put ( 75,117) {$2,0$}
\put ( 75,147) {$3,1$}
\put ( 75,177) {$4,2,0$}
\thicklines
\put (150, 60) {\line(1,0){20}}
\put (150, 84) {\line(1,0){20}}
\put (150,114) {\line(1,0){20}}
\put (150,150) {\line(1,0){20}}
\put (150,192) {\line(1,0){20}}
\thinlines
\put (155, 30) {$\sigma$=4}
\put (175, 57) {$0$}
\put (175, 81) {$1$}
\put (175,111) {$2$}
\put (175,147) {$3$}
\put (175,189) {$4$}
\thicklines
\put (195,144) {\line(1,0){20}}
\put (195,168) {\line(1,0){20}}
\put (195,198) {\line(1,0){20}}
\put (240,204) {\line(1,0){20}}
\thinlines
\put (200, 30) {$\sigma$=2}
\put (220,141) {$0$}
\put (220,165) {$1$}
\put (220,195) {$2$}
\put (245, 30) {$\sigma$=0}
\put (265,201) {$0$}
\end{picture}
\caption[]{Comparison of the spectrum of $\tau$ multiplets in the $U(3k-3)$ limit 
(left) and the $SO(3k-2)$ limit (right) for the four-body problem ($k=4$).  
The number of bosons is $N=4$.} 
\label{sphdef}
\end{figure}
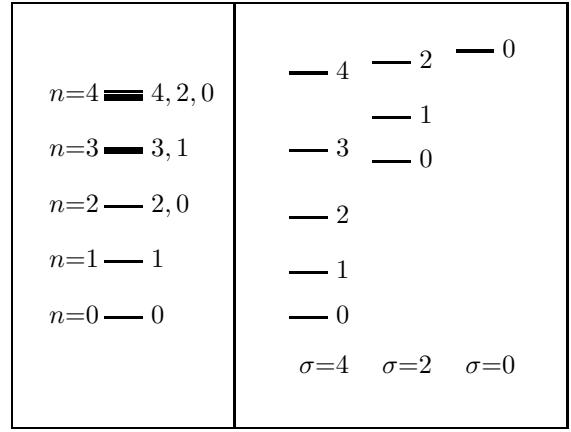

Here, I consider the one-body Hamiltonian 
\ba
H_1 = \epsilon \sum_{j=1}^{k-1} \sum_{m=-1}^{1} b_{j,m}^{\dagger} b_{j,m} = \epsilon \, \hat{C}_{1U(3k-3)} ~, 
\label{ham1}
\ea
with eigenvalues  
\ba
E_1 = \epsilon \, n ~. 
\label{e1}
\ea
The corresponding spectrum of the $\tau$ multiplets is shown in the left panel of Fig.~\ref{sphdef}.  

The classical limit of this Hamiltonian is, according to Eq.~(\ref{classical}),  
given by its coherent state expectation value  
\ba
H_{1,{\rm cl}} = \epsilon \sum_j \vec{\alpha}_{j} \cdot \vec{\alpha}_{j}^{\,\ast} 
= \epsilon \frac{1}{2} \sum_j \left( p_{j}^{2}+\rho_{j}^{2}+\frac{L_{j}^{2}}{\rho_{j}^{2}} \right) ~, 
\ea
where $L_{j}^{2}$ is the angular momentum of the $j$-th oscillator expressed in polar coordinates 
\ba
L_{j}^{2} = p_{\theta_{j}}^{2} + \frac{p_{\phi_{j}}^{2}}{\sin^{2}\theta_{j}} ~. 
\ea
It is convenient to make a change of variables from the coordinates $\rho_j$ to hyperspherical 
coordinates, the hyperradius $\rho$ and the angles $\eta_1,\ldots,\eta_{k-2}$
\numparts
\ba
\rho_{1} &=& \rho \sin \eta_1 \cdots \sin \eta_{k-3} \sin \eta_{k-2} ~, 
\label{hyper1} \\
\rho_{2} &=& \rho \sin \eta_1 \cdots \sin \eta_{k-3} \cos \eta_{k-2} ~, \\
&\vdots& \nonumber\\
\rho_{k-1} &=& \rho \cos \eta_1 ~,
\label{hyper}
\ea
\endnumparts
together with their conjugate momenta, $p$ and $p_{\eta_{j}}$. In the hyperspherical coordinates, 
the classical limit of the $U(3k-3)$ limit is given by 
\ba
H_{1,{\rm cl}} = \epsilon \frac{1}{2}\left( p^{2} + \rho^2 
+ \frac{\Lambda_{k-1}^2}{\rho^2} \right) ~. 
\label{hcl1}
\ea
Here $\Lambda_{k-1}^2$ denotes the generalized angular momentum for rotations in $3(k-1)$ dimensions 
\ba
\Lambda_{k-j}^2 = p_{\eta_j}^{2} + \frac{L_{k-j}^2}{\cos^2 \eta_j} + 
\frac{1}{\sin^2 \eta_j} \Lambda_{k-j-1}^2 ~, 
\label{angmom}
\ea
with $j=1,\ldots,k-2$ and $\Lambda_{1}^2 = L_1^2$. 
It is the classical limit of the $SO(3k-3)$ Casimir operator and is a constant 
of the motion. Therefore, one can first apply the requantization conditions 
to the coordinates and momenta contained in $\Lambda_{k-1}^2$, which yields that  
$\Lambda_{k-1}^2$ be replaced by $\tau^2/N^2$ \cite{onno}. The difference from 
the exact result $\tau(\tau+3k-5)/N^2$ is typical for the semi-classical 
approximation. The quantization condition in the $(p,\rho)$ phase space is given by 
\ba
N \oint p d\rho = 2N \int d\rho \sqrt{\frac{2E}{N\epsilon}-\rho^2-
\frac{\sigma^2}{N^2\rho^2}} = 2\pi n_{\rho} ~.
\ea
The integral can be solved exactly to obtain 
\ba
E = \epsilon \, (2n_{\rho}+\tau) ~, 
\ea
which is identical to the exact result of Eq.~(\ref{e1}) with $n=2n_{\rho}+\tau$. 
This semi-classical analysis shows the correspondence between the $U(3k-3)$ limit 
and the $3(k-1)$-dimensional spherical oscillator. 

In \cite{FormFactors}, the ACM form factors in the $U(3k-3)$ limit were derived 
in closed analytic form. As an example, the elastic form factor is given by 
\ba
{\cal F}(0_1^{+} \rightarrow 0_1^{+};q) = (\cos \epsilon)^{N} 
\rightarrow e^{-q^2\beta^2/6} ~, 
\ea
where $\epsilon$ is given by $\epsilon=-q \beta/X_D$ with $X_D=\sqrt{3N}$. 
The exponential form shown on the right-hand side is obtained in the large $N$ limit 
which is taken such that $q\beta$ remains finite. 

\subsection{Deformed oscillator}
\label{dosc}

For the (an)harmonic oscillator, the number of oscillator quanta $n$ is a good 
quantum number. When $v_{0}\neq 0$ in Eq.~(\ref{HSk}), 
the oscillator shells with $\Delta n=\pm 2$ are mixed, and the
eigenfunctions are spread over many different oscillator shells. A dynamical 
symmetry that involves the mixing between oscillator shells, is provided by
the reduction 
\ba
\left| \begin{array} {ccccccccc} 
U(3k-2) &\supset& SO(3k-2) &\supset& SO(3k-3) \\
N &,& \sigma &,& \tau \end{array} \right> \nonumber
\ea
The basis states are classified by the quantum numbers $N$, $\sigma$ and $\tau$, where $\sigma$ 
characterizes the irreducible representations of $SO(3k-2)$. $N$ and $\tau$ have the same meaning 
as in the $U(3k-3)$ limit. In this case, the energy levels are organized into bands labeled by 
$\sigma$ with $\sigma =N,N-2,\ldots,1$ or $0$ for $N$ odd or even, respectively, The rotational 
excitations are denoted by $\tau$ with $\tau=0,1,\ldots,\sigma$. 

Let's consider a Hamiltonian of the form 
\ba
H_2 &=& \xi \left[ \hat{N}(\hat{N}+3k-4) - \hat{C}_{2SO(3k-2)} 
+ \hat{C}_{2SO(3k-3)} \right] 
\nonumber\\
&=& \xi \left[ ( s^{\dagger} s^{\dagger} 
- \sum_j b_{j}^{\dagger} \cdot b_{j}^{\dagger} ) \, 
( {\rm h.c.} ) + \hat{C}_{2SO(3k-3)} \right] 
\nonumber\\
&=& \xi \left[ P^{\dagger} \tilde{P} + \hat{C}_{2SO(3k-3)} \right] ~,  
\label{ham2}
\ea
where $\hat N$ is the number operator  
\ba
\hat{N} = s^{\dagger}s + \sum_{j=1}^{k-1} \sum_{m=-1}^{1} b_{j,m}^{\dagger} b_{j,m} ~.  
\label{number}
\ea
The difference between the Casimir operators of $SO(3k-2)$ and $SO(3k-3)$ 
corresponds to a dipole-dipole interaction $\sum_j \hat{D}_j \cdot \hat{D}_j$ 
(see Eq.~(\ref{dipole})). The energy spectrum is obtained from 
the expectation values of the Casimir operators as 
\ba
E_2 = \xi \left[ (N-\sigma)(N+\sigma+3k-4) + \tau(\tau+3k-5) \right] ~. 
\nonumber\\ 
\mbox{}
\label{e2}
\ea
The energy spectrum is shown in the right panel of Fig.~\ref{sphdef}.  
Although the size of the model space, and hence the total number of states, 
is the same as for the harmonic oscillator, the ordering and classification 
of the states is different. In the $U(3k-3)$ limit all states are vibrational, 
whereas the $SO(3k-2)$ limit gives rise to a rotation-vibration spectrum, 
where the vibrations are labeled by $\sigma$ and the rotations by $\tau$.  

The classical limit of $H_2$ is given by Eq.~(\ref{classical}) 
\ba
H_{2,{\rm cl}} = \xi (N-1) \left[ \rho^{2}p^{2} + (1-\rho^{2})^{2} + 
\Lambda_{k-1}^2 \right] ~.
\ea
Also in this case, the generalized angular momentum 
$\Lambda_{k-1}^2$ is a constant of the motion, and hence can be requantized first. 
The remaining quantization condition in the $(p,\rho)$ phase space 
\ba
N \oint p d\rho = 2\pi v 
\nonumber\\ 
= 2N \int d\rho \frac{1}{\rho} 
\sqrt{\frac{E-\xi(N-1)\tau^2/N}{\xi N(N-1)}-(1-\rho^2)^2} ~,
\ea
can be solved exactly to obtain 
\ba
E = 4 \xi N \, v \left( 1-\frac{v}{N} \right) 
\left( 1-\frac{1}{N} \right) 
+ \xi \, \tau^2 \left( 1-\frac{1}{N} \right) ~.
\ea
In the large $N$ limit, this expression reduces to the exact one of Eq.~(\ref{e2}), 
if one associates the vibrational quantum number $v$ with $(N-\sigma)/2$ 
\ba
E_2 = \xi \left[ 4Nv \left( 1-\frac{2v-3k+4}{2N} \right) + \tau(\tau+3k-5) \right] ~. 
\nonumber\\
\mbox{} 
\label{ener}
\ea
To leading order in $N$, the vibrational frequency coincides. In conclusion, this 
semi-classical analysis shows that the $SO(3k-2)$ limit corresponds to a deformed 
oscillator in $3(k-1)$ dimensions. 

Also in the $SO(3k-2)$ limit the ACM form factors can be derived in closed analytic form 
\cite{FormFactors}. In this case, the elastic form factor is given in terms of a 
Gegenbauer polynomial which in the large $N$ limit reduces to a spherical (cylindrical) 
Bessel function for $k$ even (odd) 
\ba
{\cal F}(0_1^{+} \rightarrow 0_1^{+};q) = \frac{N!(3k-5)!}{(N+3k-5)!} \, C_{N}^{\frac{3k-4}{2}}(\cos \epsilon)
\nonumber\\
\rightarrow \left\{ \begin{array}{cc} 
(3k-5)!! \frac{j_{\frac{3k-6}{2}}(q\beta \sqrt{k-1})}{(q\beta \sqrt{k-1})^{\frac{3k-6}{2}}} 
& \mbox{  for } k \mbox{ even} \\ \\
(3k-5)!! \frac{J_{\frac{3k-5}{2}}(q\beta \sqrt{k-1})}{(q\beta \sqrt{k-1})^{\frac{3k-5}{2}}} 
& \mbox{  for } k \mbox{ odd} 
\end{array} \right. 
\ea
The coefficient $\epsilon$ is given by $\epsilon=-q \beta/X_D$ with $X_D=\sqrt{N(N+3k-4)/(k-1)}$. 
The large $N$ limit is taken such that $q\beta$ remains finite.  

\subsection{Shape-phase transition}
\label{phases}

\begin{figure}
\vfill 
\begin{minipage}{\linewidth}
\centerline{\epsfig{file=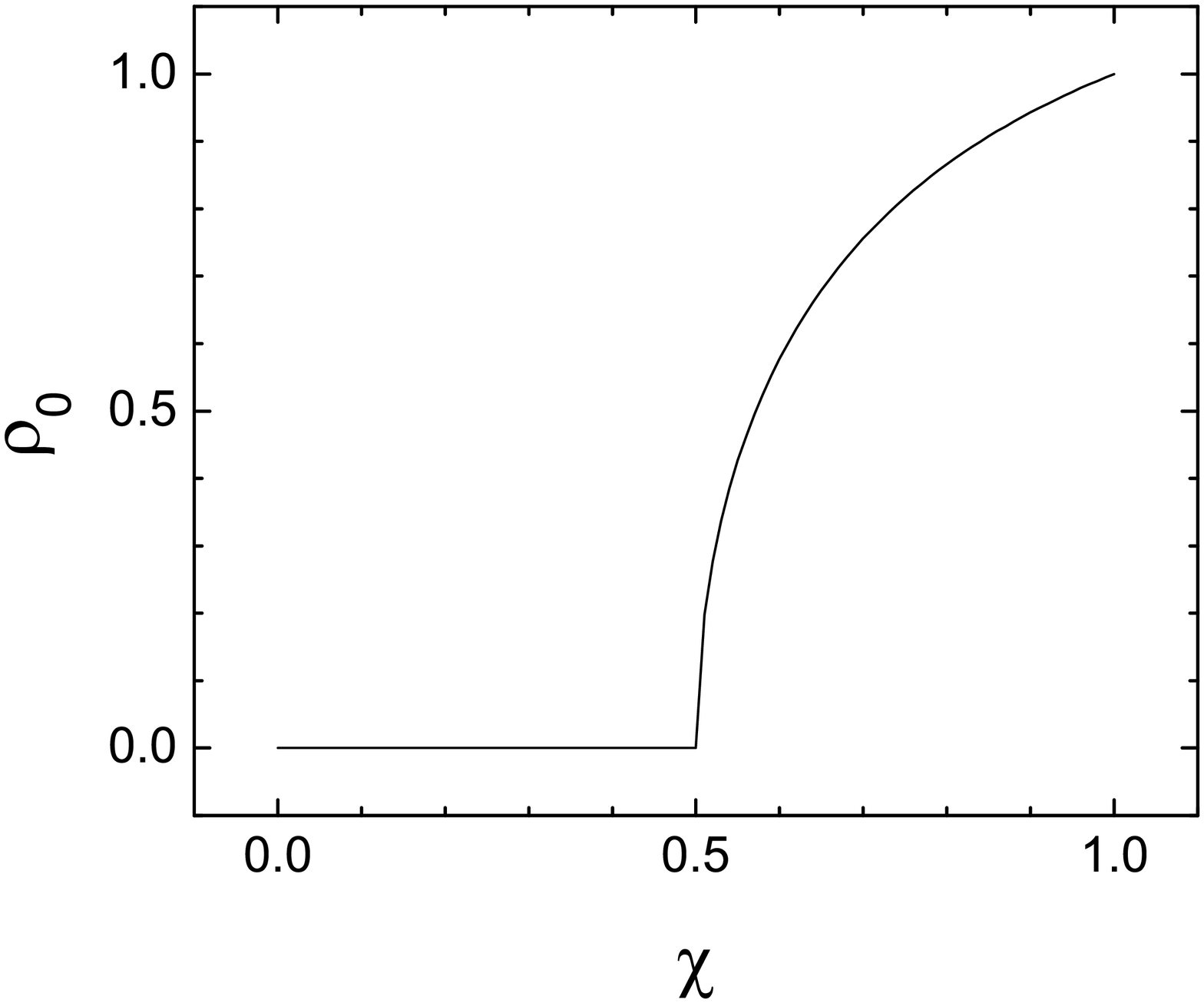,width=\linewidth}}
\end{minipage}\vfill
\begin{minipage}{\linewidth}
\centerline{\epsfig{file=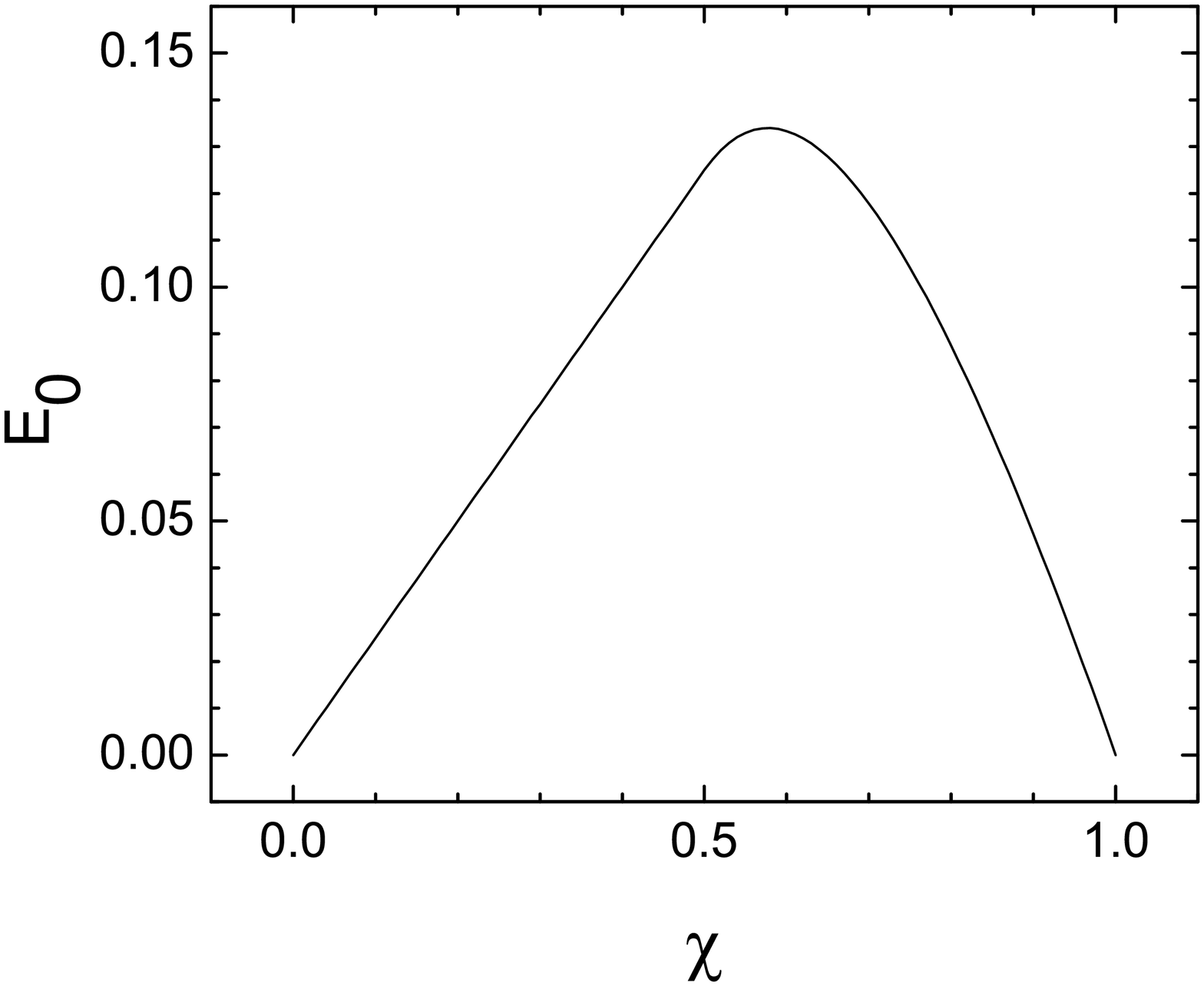,width=\linewidth}}
\end{minipage}
\caption[]{\small The value of the deformation parameter $\rho_{0}$ 
and the ground state energy $E_0$ as a function of $\chi$.}
\label{qmin}
\end{figure}

The method of coherent states is not only useful for a geometric interpretation of 
the dynamical symmetries, but can equally well be applied to the ground state energy of 
a Hamiltonian that describes the transitional region between the harmonic and the 
deformed oscillator. Let's consider the schematic Hamiltonian \cite{OSvR} 
\ba
H = (1-\chi) \sum_{jm} b_{j,m}^{\dagger} b_{j,m} 
+ \frac{\chi}{4(N-1)} \, P^{\dagger} \tilde{P} ~, 
\ea
with $0 \leq \chi \leq 1$. For $\chi=0$ it reduces to the harmonic oscillator and 
for $\chi=1$ to the deformed oscillator. This problem can be solved exactly by a 
numerical diagonalization of $H$. However, a qualitative understanding of the 
transitional region can be obtained by analyzing the classical limit. The potential 
energy surface is obtained from the classical limit by taking all momenta equal to zero
\ba
V_{\rm cl}(\rho) = \frac{1-\chi}{2} \, \rho^2 + \frac{\chi}{4} \, (1-\rho^2)^2 ~. 
\ea
The minimization conditions 
\ba
\frac{d V_{\rm cl}}{d \rho} = 0 ~, \hspace{0.5cm} 
\frac{d^2 V_{\rm cl}}{d \rho^2} > 0 ~,
\ea
give the equilibrium values
\ba
\rho_{0} = \left\{ \begin{array}{lll}
0 & \hspace{1cm} & \chi \leq \frac{1}{2} ~, \\
& & \\
\sqrt{(2\chi-1)/\chi} & & \chi \geq \frac{1}{2} ~. \\
\end{array} \right.
\ea
The structure of the condensate is characterized by $\rho_{0}$ and  
changes from spherical $\rho_{0}=0$ for $\chi \leq 1/2$ to deformed 
$\rho_{0} = \sqrt{(2\chi-1)/\chi}$ for $\chi \geq 1/2$ 
(see top panel of Fig.~\ref{qmin}). 

\begin{figure}
\vfill 
\begin{minipage}{\linewidth}
\centerline{\epsfig{file=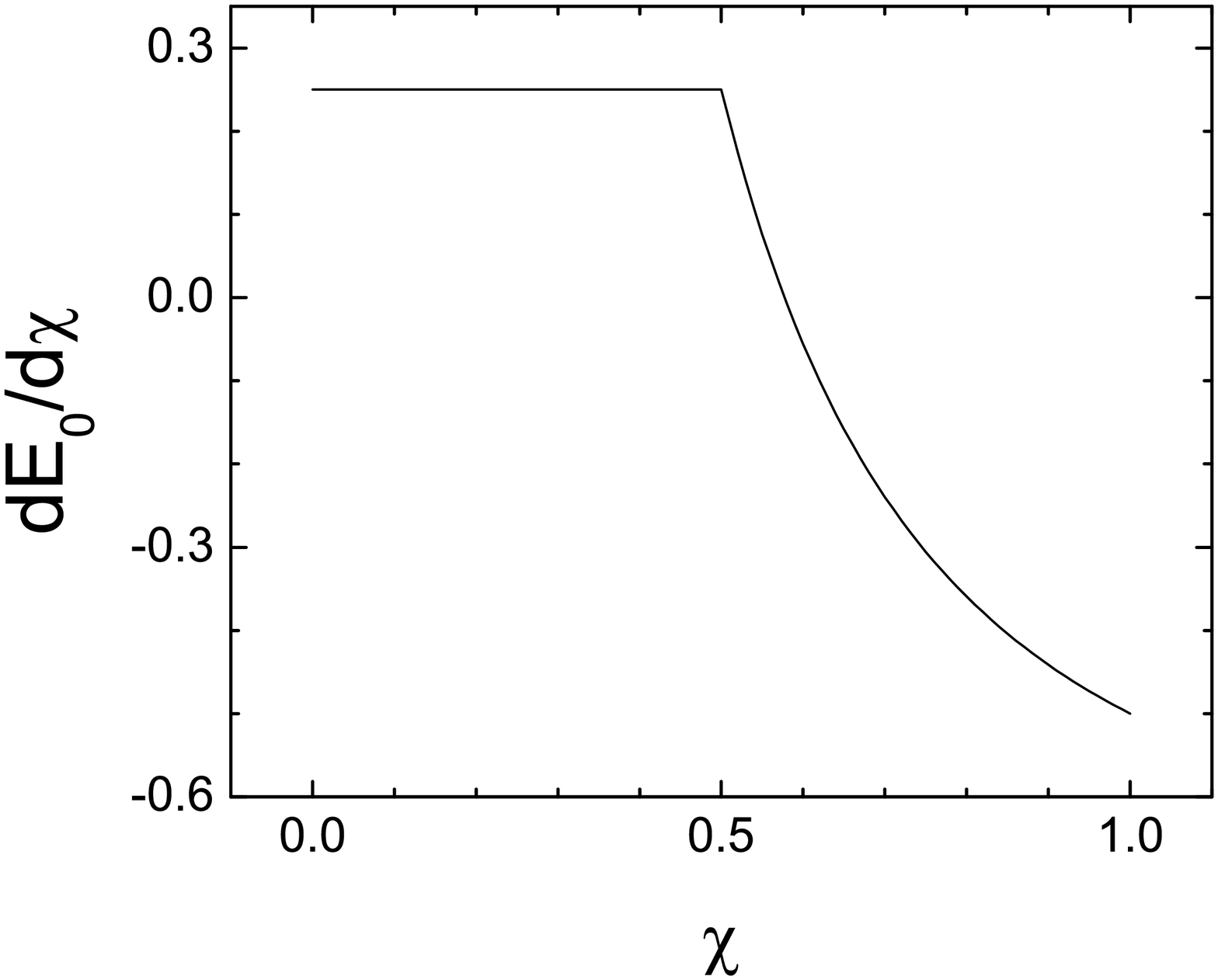,width=\linewidth}}
\end{minipage}\vfill
\begin{minipage}{\linewidth}
\centerline{\epsfig{file=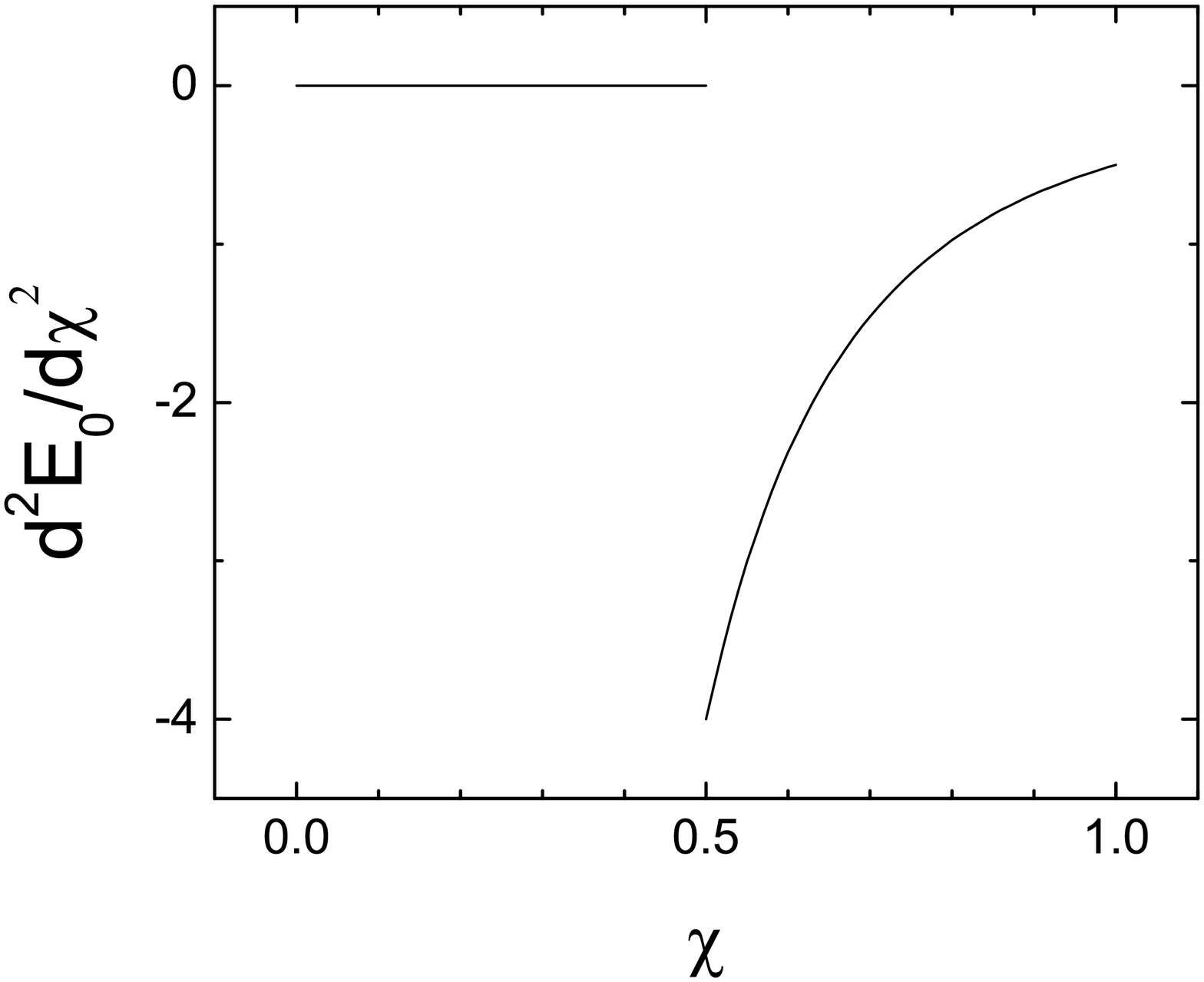,width=\linewidth}}
\end{minipage}
\caption[]{\small The first and second derivatives of the 
ground state energy as a function of $\chi$.}
\label{degs}
\end{figure}

The nature of the phase transition between the spherical region and  
the deformed region can be found by studying the ground state energy  
and its derivatives with respect to the control parameter $\chi$. 
The corresponding ground state energy is (see bottom panel of Fig.~\ref{qmin}) 
\ba
E_0 = \left\{ \begin{array}{lll} 
\chi/4 & \hspace{1cm} & \chi \leq \frac{1}{2} ~, \\
& & \\
(1-\chi)(3\chi-1)/4\chi & & \chi \geq \frac{1}{2} ~. 
\end{array} \right.
\ea
Whereas in the two endpoints ($\chi=0$ and $\chi=1$) the ground state 
energy is zero, it is different from zero in the transitional region. 
The derivatives of the ground state energy with 
respect to the control parameter $\chi$ at the critical point 
$\chi_c=1/2$ determine the nature of the phase transition. 
The first derivative 
\ba
\frac{d E_0}{d \chi} = \left\{ 
\begin{array}{lll}
1/4 & \hspace{1cm} & \chi \leq \frac{1}{2} ~, \\
& & \\
(1-3\chi^2)/4\chi^2 & & \chi \geq \frac{1}{2} ~,
\end{array} \right.
\ea
is continuous for $0 \leq \chi \leq 1$ (see top panel of 
Fig.~\ref{degs}). However, the second derivative 
\ba
\frac{d^2 E_0}{d \chi^2} = \left\{ 
\begin{array}{lll}
0 & \hspace{1cm} & \chi \leq \frac{1}{2} ~, \\
& & \\
-1/2\chi^3 & & \chi \geq \frac{1}{2} ~,
\end{array} \right.
\ea
shows a discontinuity at the critical point $\chi_c=1/2$ (see bottom panel of 
Fig.~\ref{degs}). Hence the transitional region between the harmonic oscillator 
and the deformed oscillator exhibits a second-order phase transition \cite{OSvR}. 

In addition to the dynamical symmetries, there are other examples of $S_k$ 
invariant Hamiltonians for which approximate solutions can be obtained in a 
semi-classical mean-field analysis. In the following sections, I will discuss 
the case of the axial rotor for two-body clusters, the oblate symmetric top for 
three-body clusters and the spherical top with tetrahedral symmetry for four-body clusters.  

\section{Two-body clusters}

The algebraic cluster model for two-body clusters was introduced in the literature 
under the name of the vibron model in applications to diatomic molecules \cite{FI}. 
The vibron model is an algebraic model to describe the relative motion of two clusters. 
The building blocks of the vibron model consist of a vector boson with $L^P=1^-$ 
and a scalar boson with $L^P=0^+$
\ba 
b^{\dagger}_{m} ~, \; s^{\dagger} ~. 
\ea
The group structure of the vibron model is $U(4)$. The model space is characterized 
by the symmetric irreducible representation $[N]$ of $U(4)$ which contains the oscillator 
shells with $n=0,1,2,\ldots, N$. 

\subsection{Geometrical symmetry}

For two identical objects ({\it e.g.} for X$_2$ molecules and $2\alpha$ clusters) 
the Hamiltonian has to be invariant under the permutation group $S_2$. 
The scalar boson,  $s^{\dagger}$, is invariant under the interchange of the two 
coordinates, whereas the vector Jacobi boson, $b^{\dagger}$, changes sign
\ba
P(12) \left( \begin{array}{l} s^{\dagger} \\ b^{\dagger} \end{array} \right) &=& U_{\rm tr} 
\left( \begin{array}{l} s^{\dagger} \\ b^{\dagger} \end{array} \right) U^{-1}_{\rm tr} 
\nonumber\\
&=& \left( \begin{array}{rr} 1 & 0 \\ 0 & -1 \end{array} \right) 
\left( \begin{array}{l} s^{\dagger} \\ b^{\dagger} \end{array} \right) ~, 
\ea
with 
\ba
U_{\rm tr} = \mbox{e}^{i \pi b^{\dagger} b} ~, 
\ea 
where $b^{\dagger} b$, as it appears in the exponent, is a shorthand notation 
for $\sum_m b^{\dagger}_{m} b_{m}$. 
The scalar boson, $s^{\dagger}$, transforms as the symmetric representation $[2]$ of $S_2$, whereas the 
vector Jacobi boson, $b^{\dagger}$ transforms as the antisymmetric representation $[11]$. 

There are two different symmetry classes for the permutation of two objects 
characterized by the Young tableaux
\ba
\begin{array}{ccccc}
\setlength{\unitlength}{1pt}
\begin{picture}(20,20)(0,0)
\put( 0,10) {\line(1,0){20}}
\put( 0, 0) {\line(1,0){20}}
\put( 0, 0) {\line(0,1){10}}
\put(10, 0) {\line(0,1){10}}
\put(20, 0) {\line(0,1){10}}
\end{picture} & : & [2] & \sim & A \\
\setlength{\unitlength}{1pt}
\begin{picture}(20,20)(0,0)
\put( 0, 10) {\line(1,0){10}}
\put( 0,  0) {\line(1,0){10}}
\put( 0,-10) {\line(1,0){10}}
\put( 0,-10) {\line(0,1){20}}
\put(10,-10) {\line(0,1){20}}
\end{picture} & : & [11] & \sim & B \\
& & & & \\ & & & & 
\end{array}
\ea
Since $S_2$ is isomorphic to the point group $S_2 \sim {\cal C}_2$, the irreducible representations 
can also be labeled by $[2] \sim A$ for the symmetric and $[11] \sim B$ for the antisymmetric. 
Next, one can use the multiplication rules for $S_2$ (or ${\cal C}_2$) to construct physical operators 
with the appropriate symmetry properties. For example, for the bilinear products of the  
Jacobi boson, one finds  
\ba
[11] \otimes [11] = [2] ~,
\ea
or, equivalently,
\ba
B \otimes B = A ~. 
\label{bb}
\ea

\subsection{Hamiltonian}

For two identical clusters the ACM Hamiltonian is a scalar under ${\cal C}_2 \sim S_2$, 
and Eq.~(\ref{HSk}) reduces to 
\ba
H &=& \epsilon_{0} \, s^{\dagger} \tilde{s} - \epsilon_{1} \, b^{\dagger} \cdot \tilde{b} 
+ u_0 \, s^{\dagger} s^{\dagger} \tilde{s} \tilde{s} 
\nonumber\\
&& - u_1 \, s^{\dagger} (b^{\dagger} \cdot \tilde{b}) \tilde{s} 
+ v_0 \left[ (b^{\dagger} \cdot b^{\dagger}) \tilde{s} \tilde{s} 
+ s^{\dagger} s^{\dagger} (\tilde{b} \cdot \tilde{b}) \right] 
\nonumber\\
&& + \sum_{L=0,2} a_{L} \, ( b^{\dagger} b^{\dagger} )^{(L)} \cdot 
( \tilde{b} \tilde{b} )^{(L)} ~. 
\label{HS2}
\ea
The first five terms are equivalent to those in Eq.~(\ref{HSk}), whereas the last term  
proportional to $a_L$, corresponds, according to the product of Eq.~(\ref{bb}),  
to interactions involving pairs of vector bosons with $A$ symmetry. 

The eigenfunctions of the Hamiltonian of Eq.~(\ref{HS2}) are labeled by the total number of 
bosons $N=n_s+n$, angular momentum and parity $L^P$, and their transformation property $t$ under 
the point group ${\cal C}_2 \sim S_2$: $t=A$ (even or symmetric) or $t=B$ (odd or antisymmetric). 
Since internal excitations of the clusters are not considered, the two-body wave functions 
arise solely from the relative motion.  Therefore, all states have to be symmetric ($t=A$). 
In this case, the condition of invariance under ${\cal C}_2 \sim S_2$ is equivalent to parity 
conservation. Hence, positive parity state have $A$ symmetry and negative parity states $B$. 

\subsection{Axial rotor}

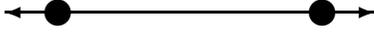
\begin{figure}
\centering
\vspace{15pt}
\setlength{\unitlength}{1pt}
\begin{picture}(200,40)(0,0)
\thicklines
\put( 50,20) {\circle*{10}} 
\put(150,20) {\circle*{10}}
\put( 50,20) {\vector(-1,0){20}} 
\put(150,20) {\vector( 1,0){20}}
\put( 50,20) {\line(1,0){100}}
\end{picture}
\caption[]{Stretching vibration of a linear configuration (point group ${\cal C}_2$).}
\label{fundvib2}
\end{figure}

In this section, I consider a ${\cal C}_2 \sim S_2$ invariant Hamiltonian which is slightly 
more general than that of the $SO(4)$ limit considered in Section~\ref{limits}  
\ba
H_{\rm 2b} = \xi \, P^{\dagger} \tilde{P} + \kappa \, \vec{L} \cdot \vec{L} ~,
\label{H2b}
\ea
where $P^{\dagger}$ denotes a generalized pairing operator 
\ba
P^{\dagger} = R^{2} \, s^{\dagger} s^{\dagger} - b^{\dagger} \cdot b^{\dagger} ~, 
\label{pair2}
\ea
and $\vec{L}$ is the angular momentum in coordinate space 
\ba
\hat{L}_m = \sqrt{2} \, ( b^{\dagger} \tilde{b} )^{(1)}_m ~.
\ea
A comparison with Eq.~(\ref{ham2}) shows that for $R^{2}=1$ the Hamiltonian 
has $U(4) \supset SO(4)$ symmetry and corresponds to a deformed oscillator. 
For $R^{2}=0$ there is no mixing between oscillator shells and the Hamiltonian 
corresponds to an anharmonic vibrator. Here I study the general case with 
$R^{2}\neq 0$ and $\xi>0$. Even though its energy eigenvalues cannot be obtained 
in closed analytic form, it is still possible to derive an approximate energy 
formula using a semi-classical mean-field analysis. 
The potential energy surface can be obtained from the classical limit of the 
Hamiltonian of Eq.~(\ref{H2b}) by setting all momenta equal to zero 
\ba
V_{\rm cl}(\rho) = \xi (N-1) \left[ R^2-\frac{1}{2}\rho^2(1+R^2) \right]^2 ~. 
\ea
Its equilibrium configuration is characterized by 
\ba
\rho^2_{0} = \frac{2R^{2}}{1+R^{2}} ~.
\ea
For small oscillations around the equilibrium value $\rho=\rho_0+\Delta \rho$, 
the classical limit of the Hamiltonian to leading order in $N$ is given by  
\ba
H_{\rm 2b,cl} = \xi N \left[ \frac{2R^{2}}{1+R^{2}} p^{2} 
+ 2R^{2}(1+R^{2}) (\Delta \rho)^{2} \right] ~.  
\ea
Standard quantization gives the vibrational energy 
spectrum one-dimensional harmonic oscillator 
\ba
E_{\rm 2b,vib} = \omega (v+\tfrac{1}{2}) ~, 
\label{e2vib}
\ea
with frequency 
\ba
\omega = 4NR^{2} \xi ~. 
\ea
Here $v$ represents the vibrational quantum number associated with the symmetric 
stretching vibration of Fig.~\ref{fundvib2}.  

\begin{figure}
\centering
\setlength{\unitlength}{1.0pt} 
\begin{picture}(150,265)(0,0)
\thinlines
\put (  0,  0) {\line(1,0){150}}
\put (  0,265) {\line(1,0){150}}
\put (  0,  0) {\line(0,1){265}}
\put ( 75,  0) {\line(0,1){265}}
\put (150,  0) {\line(0,1){265}}
\thicklines
\put ( 20, 30) {\line(1,0){20}}
\put ( 20, 40) {\line(1,0){20}}
\put ( 20, 60) {\line(1,0){20}}
\put ( 20, 90) {\line(1,0){20}}
\put ( 20,130) {\line(1,0){20}}
\put ( 20,180) {\line(1,0){20}}
\put ( 20,240) {\line(1,0){20}}
\put ( 95, 30) {\line(1,0){20}}
\put ( 95, 60) {\line(1,0){20}}
\put ( 95,130) {\line(1,0){20}}
\put ( 95,240) {\line(1,0){20}}
\small
\put ( 45, 23) {$0^+_{A}$}
\put ( 45, 40) {$1^-_{B}$}
\put ( 45, 57) {$2^+_{A}$}
\put ( 45, 87) {$3^-_{B}$}
\put ( 45,127) {$4^+_{A}$}
\put ( 45,177) {$5^-_{B}$}
\put ( 45,237) {$6^+_{A}$}
\put (120, 23) {$0^+_{A}$}
\put (120, 57) {$2^+_{A}$}
\put (120,127) {$4^+_{A}$}
\put (120,237) {$6^+_{A}$}
\end{picture}
\caption{Schematic spectrum of the rotational states of the ground stand vibrational 
band with $v=0$. The states are labeled by $L^P_t$. The right panel shows the symmetric 
states with $t=A$.} 
\label{rotor}
\end{figure}
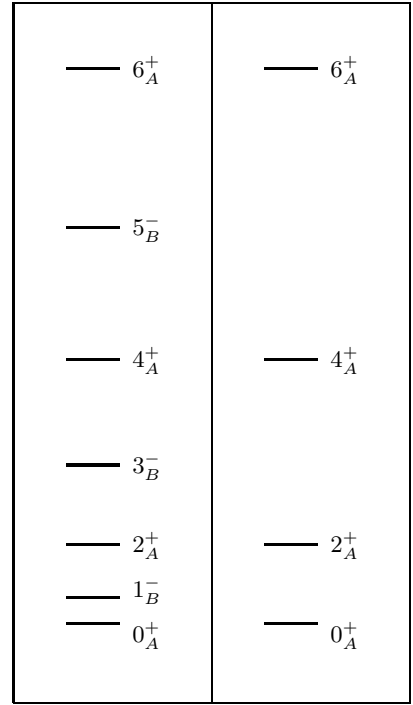

The angular momentum $L$ is an exact symmetry of the $S_{2}$ invariant 
Hamiltonian of Eq.~(\ref{HS2}) and hence also of Eq.~(\ref{H2b}). 
As a consequence, the rotational energies of the 
ground state band are given by 
\ba
E_{\rm 2b,rot} = \kappa L(L+1) ~.
\ea
Fig.~\ref{rotor} shows the structure of the rotational excitations of the ground 
state band $v=0$. For identical bosons, as is the case for a cluster of two $\alpha$-particles, 
the allowed states are symmetric with $t=A$, and therefore the 
states of the ground state band have $L$ even and positive parity,  
$L^P=0^+$, $2^+$, $4^+$, $6^+$, $\ldots$, as shown in the right panel of  
of Fig.~\ref{rotor}. The same holds for the rotational bands built on vibrational 
excitations with $v \neq 0$ (see Fig.~{\ref{axialrotor}). 

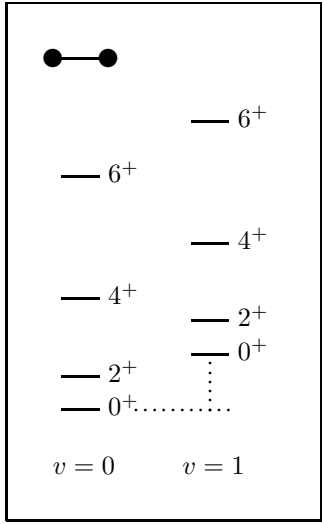
\begin{figure}
\centering
\setlength{\unitlength}{0.7pt} 
\begin{picture}(170,280)(0,0)
\thinlines
\put (  0,  0) {\line(1,0){170}}
\put (  0,280) {\line(1,0){170}}
\put (  0,  0) {\line(0,1){280}}
\put (170,  0) {\line(0,1){280}}
\thicklines
\put ( 30, 60) {\line(1,0){20}}
\put ( 30, 78) {\line(1,0){20}}
\put ( 30,120) {\line(1,0){20}}
\put ( 30,186) {\line(1,0){20}}
\multiput ( 70, 60)(5,0){11}{\circle*{0.1}}
\thinlines
\put ( 25, 25) {$v=0$}
\put ( 55, 57) {$0^+$}
\put ( 55, 75) {$2^+$}
\put ( 55,117) {$4^+$}
\put ( 55,183) {$6^+$}
\thicklines
\put (100, 90) {\line(1,0){20}}
\put (100,108) {\line(1,0){20}}
\put (100,150) {\line(1,0){20}}
\put (100,216) {\line(1,0){20}}
\multiput (110, 60)(0,5){6}{\circle*{0.1}}
\thinlines
\put ( 95, 25) {$v=1$}
\put (125, 87) {$0^+$}
\put (125,105) {$2^+$}
\put (125,147) {$4^+$}
\put (125,213) {$6^+$}
\thicklines
\put(25,250) {\circle*{10}} 
\put(55,250) {\circle*{10}}
\put(25,250) {\line(1,0){30}}
\end{picture}
\caption{Schematic spectrum of an axial rotor. 
All states are symmetric under ${\cal C}_2 \sim S_2$.} 
\label{axialrotor}
\end{figure}

\subsection{Electromagnetic couplings}

In this section I apply the general procedure to study electromagnetic couplings in the ACM 
as discussed in Section~\ref{ff} to the case of two-body clusters. Transition form factors and 
$B(EL)$ values can be obtained from the matrix elements of the transition operator
\numparts
\ba
\hat T(\epsilon) &=& e^{-iq\beta \hat{D}_{z}/X_{D}} ~, \label{trans2} \\
\hat D_{m} &=& (b^{\dagger} \tilde{s} - s^{\dagger} \tilde{b})^{(1)}_m ~. 
\ea
\endnumparts
In general, the transition form factors cannot be obtained in closed analytic form, 
but have to be evaluated numerically. For the axial rotor, the form factors can only be 
obtained in closed form in the large $N$ limit using a technique introduced in \cite{BAS} 
and subsequently applied in \cite{BIL}. The normalization factor $X_D$ which appears in 
the algebraic transition operator of Eq.~(\ref{trans2}), is given by 
\ba
X_D = \frac{2NR}{1+R^2} ~.
\ea 
The elastic form factor can be obtained as  
\ba
{\cal F}(0^+ \rightarrow 0^+;q) &\rightarrow&   
\frac{1}{2} \int d \cos \theta \, e^{-i q \beta \cos \theta} 
\nonumber\\
&=& j_0(q \beta) ~.  
\ea
For transitions along the ground state band $v=0$ the transition 
form factors are given in terms of a spherical Bessel function \cite{BAS}   
\ba
{\cal F}(0^+ \rightarrow L^P;q) \;\rightarrow\; c_L \, j_L(q \beta) ~, 
\ea
with
\ba
c_L^2 &=& \frac{2L+1}{2} \left[ 1+P_{L}(-1) \right] 
\nonumber\\
&=& (2L+1) \frac{1+(-1)^L}{2} ~. 
\ea
For an extended charge distribution according to Eq.~(\ref{rhor1}), the form factors 
are multiplied by an exponential factor $\exp(-q^{2}/4\alpha)$. 

The transition probabilities $B(EL)$ along the ground state band can be extracted 
from the form factors in the long wavelength limit according to Eq.~(\ref{belif})  
\ba
B(EL;0 \rightarrow L) = \frac{(Ze)^2}{4\pi} \, c_L^2 \, \beta^{2L} ~,
\label{BEL}
\ea
in agreement with the results of Eq.~(\ref{BEL2}). 

Form factors and $B(EL)$ values depend only on the parameters $\alpha$ and $\beta$, 
and on the point group symmetry via the coefficients $c_L$. The analytic results given 
in this section provide a set of closed expressions which can be compared with experiment.
The elastic form factor shows an oscillatory behavior as a function of $q\beta$ 
according to $j_0(q\beta)=\sin(q\beta)/q\beta$. Hence, the coefficient $\beta$ can 
be determined from the first minimum in the elastic form factor as $\beta = \pi/q_{\rm min}$. 
Subsequently, the result for the charge radius given in Eq.~(\ref{radius}) can be used to 
determine the value of $\alpha$. 

\subsection{Shape-phase transition}

The ACM describes the relative dynamics of a two-body system and includes  
both the axial rotor and the harmonic oscillator as special limits, as well 
as the region in between these two limiting cases. 
The transitional region can be described by the schematic Hamiltonian
\ba
H = (1-\chi) \sum_{m} b_{m}^{\dagger} b_{m} 
+ \frac{\chi}{4(N-1)} \, P^{\dagger} \tilde{P} ~, 
\label{trans2b}
\ea
with $0 \leq \chi \leq 1$. For $\chi=0$ it reduces to the harmonic 
oscillator and for $\chi=1$ to the axial rotor of the previous section. 
For $\chi=1$ and $R^2=1$ it reduces to the deformed oscillator. 
The transitional region can be studied in the classical limit 
of the Hamiltonian following the methods already discussed in 
Section~\ref{phases}. The only difference is the form of the pairing 
operator which now depends on $R^2$, see Eq.~(\ref{pair2}). 
In this case, the potential energy surface is given by
\ba
V_{\rm cl}(\rho) = \frac{1-\chi}{2} \, \rho^2 
+ \frac{\chi}{4} \left[ R^2-\frac{1}{2}\rho^2(1+R^2) \right]^2 ~. 
\ea
The equilibrium shape is characterized by $\rho_{0}$ which changes from spherical 
$\rho_{0}=0$ for $\chi \leq \chi_c$ to deformed $\rho^2_{0} > 0$ for $\chi \geq \chi_c$  
\ba
\rho^2_{0} = \left\{ \begin{array}{lll}
0 & \hspace{1cm} & \chi \leq \chi_c ~, \\
& & \\
\frac{2R^2}{1+R^2} + \frac{4(\chi-1)}{\chi (1+R^2)^2} 
& & \chi \geq \chi_c ~. \\
\end{array} \right.
\label{deformation}
\ea
Qualitatively, the nature of the phase transition is the same as discussed in 
Section~\ref{phases}. The Hamiltonian of Eq.~(\ref{trans2b}) exhibits a second-order 
phase transition between the spherical and the deformed regions.  
The only difference is that now the critical point depends on $R^2$ 
\ba
\chi_c = \frac{1}{1+R^2(1+R^2)/2} ~.
\label{critical}
\ea
Fig.~\ref{phasetrans} shows the equilibrium shape and the ground state energy 
as a function of $\chi$ and $R^2$. For $R^2=1$ it reduces to the results shown 
in Fig.~\ref{qmin} for the transitional region between the harmonic and the 
deformed oscillator. 

\begin{figure}
\vfill 
\begin{minipage}{\linewidth}
\centerline{\epsfig{file=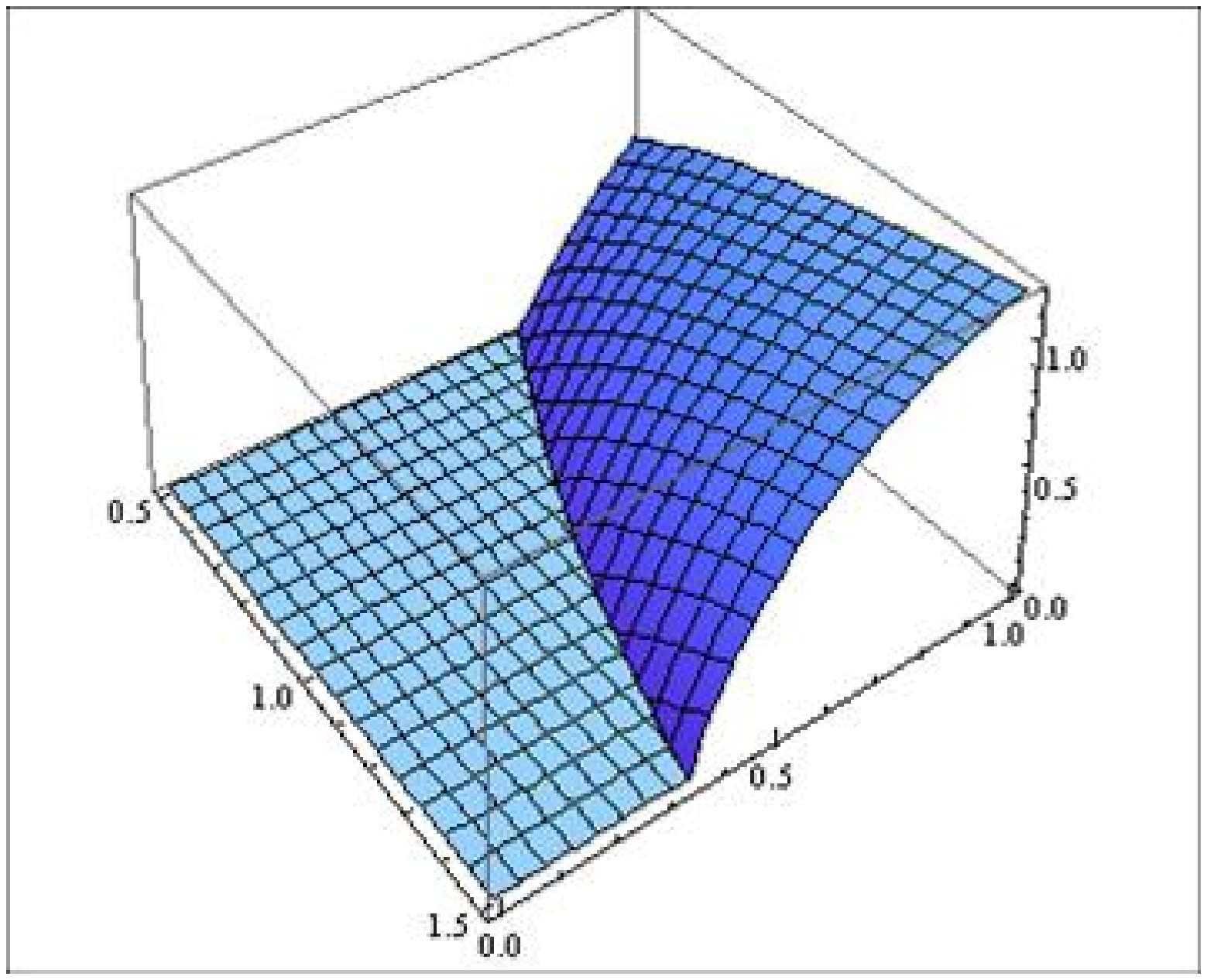,width=0.8\linewidth}}
\end{minipage}\vfill
\begin{minipage}{\linewidth}
\centerline{\epsfig{file=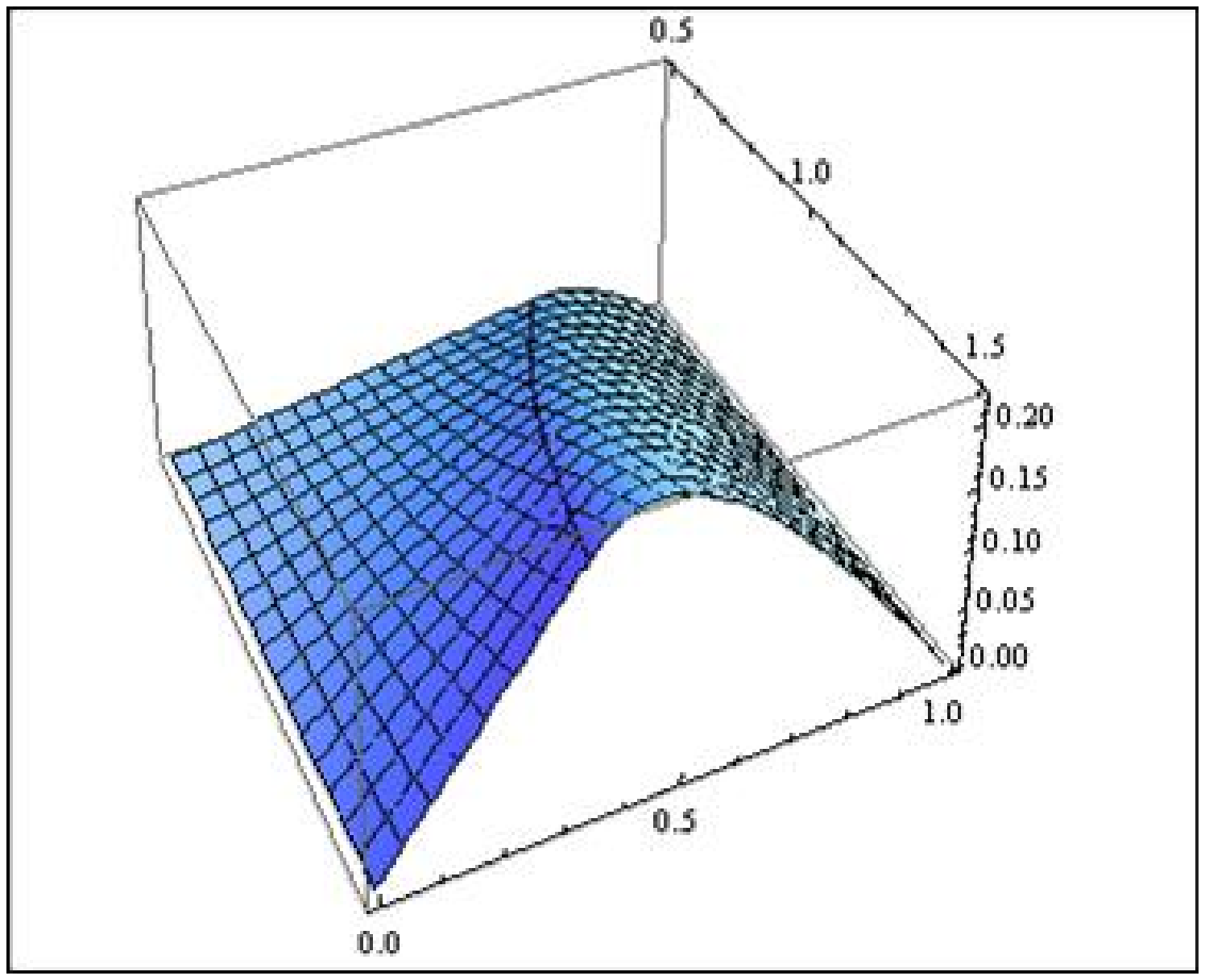,width=0.8\linewidth}}
\end{minipage}
\caption[]{\small The value of the deformation parameter $\rho_{0}$ 
and the ground state energy $E_0$ as a function of $0 \leq \chi \leq 1$ 
and $0.5 \leq R^2 \leq 1.5$.}
\label{phasetrans}
\end{figure}

\section{Three-body clusters}
 
The ACM for three identical clusters was introduced in hadron physics to describe 
the relative motion of three constituent quarks in a baryon \cite{BIL}, but more 
recently it has also found an interesting application in nuclear physics in which 
the properties of $^{12}$C are treated in terms of a cluster of three $\alpha$ 
particles \cite{C12,ACM}. 

In this case, the building blocks of the model consist of two vector boson operators 
(one for each relative Jacobi coordinate) and a scalar boson  
\ba
b^{\dagger}_{1,m} ~, \; b^{\dagger}_{2,m} ~, \;
s^{\dagger} ~. 
\label{bb3}
\ea
The set of 49 bilinear products of creation and annihilation operators spans the Lie algebra 
of $U(7)$. The model space contains the oscillator shells with $n=n_1+n_2=0,1,2,\ldots, N$.  
The ACM describes three-body cluster states in terms of a system of $N=n_s+n_1+n_2$ interacting 
bosons with angular momentum and parity $L^P=1^-$ (two vector bosons, $b_1^{\dagger}$ and 
$b_2^{\dagger}$) and $L^P=0^+$ (scalar boson, $s^{\dagger}$).  

\subsection{Geometrical symmetry}

For three identical particles, such as $X_3$ molecules, $3\alpha$ clusters or $qqq$ baryons, 
the Hamiltonian has to be invariant under the permutation group $S_3$. All permutations of 
three objects can be expressed in terms of two elementary ones, the transposition 
$P(12)$ and the cyclic permutation $P(123)$. The transformation properties under $S_3$ of 
all operators in the model follow from those of the building block,   
\ba
P(12) \left( \begin{array}{l} s^{\dagger} \\ b^{\dagger}_1 \\ b^{\dagger}_2 
\end{array} \right) &=& U_{\rm tr} 
\left( \begin{array}{l} s^{\dagger} \\ b^{\dagger}_1 \\ b^{\dagger}_2 \end{array} 
\right) U^{-1}_{\rm tr} 
\nonumber\\
&=& \left( \begin{array}{rrr} 1 & 0 & 0 \\ 0 & -1 & 0 \\ 0 & 0 & 1 \end{array} \right) 
\left( \begin{array}{l} s^{\dagger} \\ b^{\dagger}_1 \\ b^{\dagger}_2 \end{array} \right) ~,  
\ea
with 
\ba
U_{\rm tr} = \mbox{e}^{i \pi b^{\dagger}_1 b_1} ~, 
\label{p12}
\ea 
for the transposition $P(12)$, and 
\ba
P(123) \left( \begin{array}{l} s^{\dagger} \\ b^{\dagger}_1 \\ b^{\dagger}_2 \end{array} \right) 
&=& U_{\rm cycl} \left( \begin{array}{l} s^{\dagger} \\ b^{\dagger}_1 \\ b^{\dagger}_2 \end{array} 
\right) U^{-1}_{\rm cycl} \hspace{2cm}
\nonumber\\ 
&=& \left( \begin{array}{ccc} 1 & 0 & 0 \\ 0 & -\frac{1}{2} & \frac{\sqrt{3}}{2} \\ 
0 & -\frac{\sqrt{3}}{2} & -\frac{1}{2} \end{array} \right)
\left( \begin{array}{l} s^{\dagger} \\ b^{\dagger}_1 \\ b^{\dagger}_2 \end{array} \right) ~, 
\ea
with
\ba
U_{\rm cycl} = \mbox{e}^{i \pi (b^{\dagger}_1 b_1 + b^{\dagger}_2 b_2)} \, 
\mbox{e}^{\theta (b^{\dagger}_1 b_2 - b^{\dagger}_2 b_1)} ~, 
\label{p123}
\ea
and $\theta=\arctan \sqrt{3}$ for the cyclic permutation $P(123)$.   
The scalar boson, $s^{\dagger}$, transforms as the symmetric representation $[3]$ of $S_3$, 
whereas the vector Jacobi bosons, $b^{\dagger}_1$ and $b^{\dagger}_2$, 
transform as the two components of the mixed symmetry representation.

There are three different symmetry classes for the permutation of three objects 
characterized by the Young tableaux
\ba
\begin{array}{ccccc}
\setlength{\unitlength}{1pt}
\begin{picture}(30,20)(0,0)
\put( 0,10) {\line(1,0){30}}
\put( 0, 0) {\line(1,0){30}}
\put( 0, 0) {\line(0,1){10}}
\put(10, 0) {\line(0,1){10}}
\put(20, 0) {\line(0,1){10}}
\put(30, 0) {\line(0,1){10}}
\end{picture} & : & [3] & \sim & A_1 \\
\setlength{\unitlength}{1pt}
\begin{picture}(30,20)(0,0)
\put( 0, 10) {\line(1,0){20}}
\put( 0,  0) {\line(1,0){20}}
\put( 0,-10) {\line(1,0){10}}
\put( 0,-10) {\line(0,1){20}}
\put(10,-10) {\line(0,1){20}}
\put(20,  0) {\line(0,1){10}}
\end{picture} & : & [21] & \sim & E \\ 
\setlength{\unitlength}{1pt}
\begin{picture}(30,30)(0,0)
\put( 0, 10) {\line(1,0){10}}
\put( 0,  0) {\line(1,0){10}}
\put( 0,-10) {\line(1,0){10}}
\put( 0,-20) {\line(1,0){10}}
\put( 0,-20) {\line(0,1){30}}
\put(10,-20) {\line(0,1){30}}
\end{picture} & : & [111] & \sim & A_2 \\
& & & & \\ & & & &
\end{array}
\ea
The triangular configuration with three identical objects has point group symmetry ${\cal D}_{3h}$ 
\cite{ACM}. Since ${\cal D}_{3h} \sim {\cal D}_{3} \times P$, the transformation properties under 
${\cal D}_{3h}$ are labeled by parity $P$ and the representations of ${\cal D}_{3} \sim S_{3}$. 
Since $S_3$ is isomorphic to the point group ${\cal D}_3$, the irreducible representations 
can also be labeled by $[3] \sim A_1$ for the symmetric, $[111] \sim A_2$ for the antisymmetric, 
and $[21] \sim E$ for the mixed symmetry representations. 
Next, one can use the multiplication rules for $S_3$ (or ${\cal D}_3$) to construct physical operators 
with the appropriate symmetry properties. For example, for the bilinear products of the three vector 
Jacobi bosons, one finds  
\ba
[21] \otimes [21] = [3] \oplus [21] \oplus [111] ~,
\ea
or, equivalently,
\ba
E \otimes E = A_1 \oplus E \oplus A_2 ~. 
\label{ee}
\ea
$A_1$ and $A_2$ are one-dimensional representations, whereas $E$ is a two-fold degenerate representation. 

\subsection{Hamiltonian}

For the case of three identical clusters, the Hamiltonian has to be invariant under the point group 
${\cal D}_3 \sim S_{3}$, and can be written as \cite{BIL} 
\ba
H &=& \epsilon _{0} \, s^{\dagger} \tilde{s} 
-\epsilon_{1} \, (b_1^{\dagger} \cdot \tilde{b}_1 + b_2^{\dagger} \cdot \tilde{b}_2) 
\nonumber\\
&& + u_{0} \,(s^{\dagger} s^{\dagger} \tilde{s} \tilde{s}) 
- u_{1} \, s^{\dagger} (b_1^{\dagger} \cdot \tilde{b}_1 
+ b_2^{\dagger} \cdot \tilde{b}_2) \tilde{s}  
\nonumber \\
&& + v_{0} \left[ (b_1^{\dagger} \cdot b_1^{\dagger} 
+ b_2^{\dagger} \cdot b_2^{\dagger}) \tilde{s} \tilde{s} \right.
\nonumber\\
&& \hspace{1cm} \left.
+ s^{\dagger} s^{\dagger} (\tilde{b}_1 \cdot \tilde{b}_1 
+ \tilde{b}_2 \cdot \tilde{b}_2) \right]    
\nonumber\\
&& + \sum_{L=0,2} a_{L} \,(b_1^{\dagger} b_1^{\dagger} 
+ b_2^{\dagger} b_2^{\dagger})^{(L)} \cdot 
(\tilde{b}_1 \tilde{b}_1 + \tilde{b}_2 \tilde{b}_2)^{(L)}  
\nonumber\\
&& + \sum_{L=0,2} b_{L} \left[ (b_1^{\dagger} b_1^{\dagger} 
- b_2^{\dagger} b_2^{\dagger})^{(L)} \cdot \right.
(\tilde{b}_1 \tilde{b}_1 - \tilde{b}_2 \tilde{b}_2)^{(L)} 
\nonumber\\
&& \hspace{2cm} \left. + 4 \, (b_1^{\dagger} b_2^{\dagger})^{(L)} 
\cdot (\tilde{b}_2 \tilde{b}_1)^{(L)} \right] 
\nonumber\\
&& + c_{1} \, (b_1^{\dagger} b_2^{\dagger})^{(1)} 
\cdot (\tilde{b}_2 \tilde{b}_1)^{(1)} ~. 
\label{HS3}
\ea
The first five terms are equivalent to those in Eq.~(\ref{HSk}), whereas the last three terms 
proportional to $a_L$, $b_L$ and $c_1$, correspond to interactions involving pairs of vector 
bosons with $A_1$, $E$ and $A_2$ symmetry, respectively (see Eq.~(\ref{ee})). 
Since the cluster states arise entirely from the relative motion of the three clusters, and not 
from the internal excitations of the clusters, the allowed states have to be symmetric $t=A_1$. 
The discrete symmetry of the eigenfunctions can be determined from the matrix elements of $P(12)$ 
and $P(123)$ \cite{BIL}. The invariance of $H$ under $P(12)$ allows one to distinguish between 
wave functions which are even and odd under $P(12)$
\ba
\left< \psi_t \right| P(12) \left| \psi_t \right> = 
\left< \psi_t \right| U_{\rm tr} \left| \psi_t \right> = \pm 1 ~.
\ea
Algebraically, the matrix elements of the cyclic permutation can be written as 
\ba
\left< \psi_t \right| P(123) \left| \psi_t \right> =  
\left< \psi_t \right| U_{\rm cycl} \left| \psi_t \right> ~. 
\ea
The first term in $U_{\rm cycl}$ of Eq.~(\ref{p123}) gives rise to a phase factor $(-1)^{n_1+n_2}$ 
which is $+1$ ($-1$) for states with even (odd) parity. The second term corresponds to a transformation 
among oscillator coordinates, and therefore its matrix elements are given in terms of Talmi-Moshinksy 
brackets \cite{Talmi,Moshinsky} which were calculated with the program TMB \cite{Dobes}. 
In practice, the wave functions $\psi_t$ are obtained numerically by diagonalization, 
and hence are determined up to a sign. The relative phases of the degenerate representation  
$E$ with $E_{\rho}$ and $E_{\lambda}$, can be fixed by calculating the off-diagonal matrix elements 
of $P(123)$ and requiring that they transform like the components of the degenerate representations 
(see \ref{app3b}). Summarizing, the symmetry character under ${\cal D}_3 \sim S_3$ of any given wave 
function can be determined by comparing the matrix elements of $P(12)$ and $P(123)$ with 
the transformation properties listed in \ref{app3b}.  

\subsection{Oblate top}

A study of the classical limit of the $U(6)$ and $SO(7)$ dynamical symmetries of the Hamiltonian of 
Eq.~(\ref{HS3}) shows that the corresponding potential energy surfaces only depend on the hyperspherical 
radius $\rho$ with equilibrium values $\rho_0=0$ and $\rho_0=1$, respectively (see Section~\ref{limits}). 
However, these are not the only possible equilibrium shapes.  
For example, consider the Hamiltonian \cite{ACM,BIL} 
\ba
H_{\rm 3b} &=& \xi_{1} \, P^{\dagger} \tilde{P}
+ \xi_{2} \left[ (b_1^{\dagger} \cdot b_1^{\dagger} - b_2^{\dagger} \cdot b_2^{\dagger}) \, 
(\tilde{b}_1 \cdot \tilde{b}_1 - \tilde{b}_2 \cdot \tilde{b}_2) \right.
\nonumber\\
&& \hspace{1cm} \left. + 4 \, (b_1^{\dagger} \cdot b_2^{\dagger}) \, (\tilde{b}_2 \cdot \tilde{b}_1) \right] 
+ \kappa \, \vec{L} \cdot \vec{L} ~,   
\label{H3b}
\ea
where $P^{\dagger}$ is the generalized pairing operator 
\ba
P^{\dagger} = R^{2} \, s^{\dagger} s^{\dagger}
- b_1^{\dagger} \cdot b_1^{\dagger} - b_2^{\dagger} \cdot b_2^{\dagger} ~,
\label{pair3}
\ea
and $\vec{L}$ the angular momentum 
\ba
\hat{L}_m = \sqrt{2} \, ( b^{\dagger}_1 \tilde{b}_1 + b^{\dagger}_2 \tilde{b}_2 )^{(1)}_m ~. 
\ea
For $R^{2}=1$ and $\xi_{2}=0$ the Hamiltonian has $U(7) \supset SO(7)$ symmetry and 
corresponds to a six-dimensional deformed oscillator. For $R^{2}=0$ there is no mixing 
between oscillator shells and the Hamiltonian is that of an anharmonic vibrator. 
In this section I will show that the general case with $R^{2} \neq 0$ and $\xi_{1}$, $\xi_{2}>0$  
corresponds to an oblate symmetric top. Even though in this case the Hamiltonian does not have
a dynamical symmetry, it is still possible to derive an approximate energy formula. 
The potential energy surface associated with the Hamiltonian of Eq.~(\ref{H3b}) is obtained by 
taking the static limit of its classical limit, {\it i.e.} by setting all momenta equal to zero 
\ba
V_{\rm cl}(\rho,\eta,\zeta) = \xi_1(N-1) \left[ R^2-\frac{1}{2}\rho^2(1+R^2) \right]^2 
\nonumber\\
\hspace{1cm} + \xi_2(N-1) \frac{\rho^4}{4} \left( \cos^2 2\eta + \sin^2 2\eta \cos^2 2\zeta \right) ~. 
\ea
Its equilibrium configuration is characterized by coordinates that have equal length 
($\rho_{1,0}=\rho_{2,0}=\rho_0/\sqrt{2}$) and are mutually perpendicular 
\numparts
\ba
\rho_{0} &=& \sqrt{2R^{2}/(1+R^{2})} ~, \\
\eta_{0} &=& \pi/4 ~, \\
\zeta_{0} &=& \pi/4 ~,
\label{angles3}
\ea
\endnumparts
where $2\zeta$ denotes the relative angle between $\vec{\alpha}_1$ and 
$\vec{\alpha}_2$. Geometrically, this equilibrium shape corresponds to three clusters 
located at the vertices of an equilateral triangle. In the limit of small oscillations
around the minimum $\rho=\rho_{0}+\Delta \rho$, $\eta=\eta_{0}+\Delta \eta$ and 
$\zeta=\zeta_{0}+\Delta \zeta$, the intrinsic degrees of freedom 
$\rho$, $\eta$ and $\zeta$ decouple and become harmonic. To leading order in $N$ one finds 
\ba
H_{\rm 3b,cl} &=& \xi_{1} N \left[ \frac{2R^{2}}{1+R^{2}} p^{2} + 2R^{2}(1+R^{2}) (\Delta \rho)^{2} \right] 
\nonumber\\
&& + \xi_{2} N \left[ p_{\eta}^{2} + \frac{4R^{4}}{(1+R^{2})^{2}} (\Delta \eta)^{2} 
\right. 
\nonumber\\
&& \hspace{1cm} \left. 
+ p_{\zeta}^{2} + \frac{4R^{4}}{(1+R^{2})^{2}} (\Delta \zeta)^{2} \right] ~.
\ea
Standard quantization of the harmonic oscillator yields the vibrational
energy spectrum of an oblate top 
\ba
E_{\rm 3b,vib} &=& \omega_{1}(v_{1}+\tfrac{1}{2}) 
+ \omega_{2}(v_{2}+1) ~,
\label{e3vib}
\ea
with frequencies 
\numparts
\ba
\omega_{1} &=& 4NR^{2} \xi_{1} ~, \\ 
\omega_{2} &=& \frac{4NR^{2}}{1+R^{2}} \xi_{2} ~,
\ea
\endnumparts
in agreement with the results obtained in a normal mode analysis \cite{BIL}. 
The vibrational part of the Hamiltonian of Eq.~({\ref{HS3}) has a very simple 
physical interpretation in terms of the three fundamental vibrations of a 
triangular configuration (see Figure~\ref{fundvib3}): $v_{1}$ represents 
the vibrational quantum number for a symmetric stretching $A$ vibration, and 
$v_2=v_{2a}+v_{2b}$ for a two-fold degenerate $E$ vibration. 
The rotational spectrum is given by 
\ba
E_{\rm 3b,rot} = \kappa \, L(L+1) ~. 
\label{e3rot}
\ea

\begin{figure}
\centering
\setlength{\unitlength}{0.6pt}
\begin{picture}(420,200)(20,30)
\thinlines
\put ( 85, 50) {$v_1$}
\put ( 50,100) {\circle*{10}}
\put (130,100) {\circle*{10}}
\put ( 90,180) {\circle*{10}}
\put ( 50,100) {\line ( 4, 3){40}} 
\put (130,100) {\line (-4, 3){40}} 
\put ( 90,180) {\line ( 0,-1){50}} 
\thicklines
\put ( 50,100) {\vector(-4,-3){16}}
\put (130,100) {\vector( 4,-3){16}}
\put ( 90,180) {\vector( 0, 1){20}}
\thinlines
\put (215, 50) {$v_{2a}$}
\put (180,100) {\circle*{10}}
\put (260,100) {\circle*{10}}
\put (220,180) {\circle*{10}}
\put (180,100) {\line ( 4, 3){40}} 
\put (260,100) {\line (-4, 3){40}} 
\put (220,180) {\line ( 0,-1){50}} 
\thicklines
\put (180,100) {\vector( 4,-3){16}}
\put (260,100) {\vector(-4,-3){16}}
\put (220,180) {\vector( 0, 1){20}}
\thinlines
\put (345, 50) {$v_{2b}$}
\put (310,100) {\circle*{10}}
\put (390,100) {\circle*{10}}
\put (350,180) {\circle*{10}}
\put (310,100) {\line ( 4, 3){40}} 
\put (390,100) {\line (-4, 3){40}} 
\put (350,180) {\line ( 0,-1){50}} 
\thicklines
\put (310,100) {\vector(-1,-2){ 8}}
\put (390,100) {\vector(-1, 2){ 8}}
\put (350,180) {\vector( 1, 0){20}}
\end{picture}
\caption[]{ 
Fundamental vibrations of a triangular configuration (point group ${\cal D}_{3h}$).}
\label{fundvib3}
\end{figure}
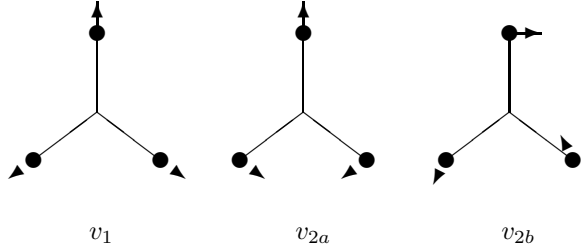

Fig.~\ref{obltop} shows the structure of the rotational excitations of the ground-state 
vibrational band $(v_1v_2)=(00)$. For identical bosons, as is the case for a cluster of 
three $\alpha$-particles, the allowed states are symmetric with $t=A_1$, and therefore the 
rotational structure of the ground-state band is a sequence of states with angular momentum 
and parity $L^P=0^+$, $2^+$, $3^-$, $4^\pm$, $5^-$, $\ldots$, as shown in the right panel  
of Fig.~\ref{obltop}. 

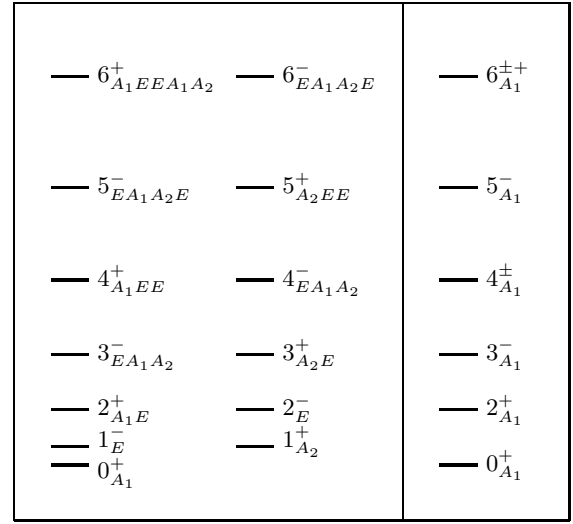
\begin{figure}
\centering
\setlength{\unitlength}{0.7pt} 
\begin{picture}(300,280)(0,0)
\thinlines
\put (  0,  0) {\line(1,0){300}}
\put (  0,280) {\line(1,0){300}}
\put (  0,  0) {\line(0,1){280}}
\put (210,  0) {\line(0,1){280}}
\put (300,  0) {\line(0,1){280}}
\thicklines
\put ( 20, 30) {\line(1,0){20}}
\put ( 20, 40) {\line(1,0){20}}
\put ( 20, 60) {\line(1,0){20}}
\put ( 20, 90) {\line(1,0){20}}
\put ( 20,130) {\line(1,0){20}}
\put ( 20,180) {\line(1,0){20}}
\put ( 20,240) {\line(1,0){20}}

\put (120, 40) {\line(1,0){20}}
\put (120, 60) {\line(1,0){20}}
\put (120, 90) {\line(1,0){20}}
\put (120,130) {\line(1,0){20}}
\put (120,180) {\line(1,0){20}}
\put (120,240) {\line(1,0){20}}

\put (230, 30) {\line(1,0){20}}
\put (230, 60) {\line(1,0){20}}
\put (230, 90) {\line(1,0){20}}
\put (230,130) {\line(1,0){20}}
\put (230,180) {\line(1,0){20}}
\put (230,240) {\line(1,0){20}}
\small
\put ( 45, 23) {$0^+_{A_1}$}
\put ( 45, 40) {$1^-_{E}$}
\put ( 45, 57) {$2^+_{A_1 E}$}
\put ( 45, 87) {$3^-_{E A_1 A_2}$}
\put ( 45,127) {$4^+_{A_1 E E}$}
\put ( 45,177) {$5^-_{E A_1 A_2 E}$}
\put ( 45,237) {$6^+_{A_1 E E A_1 A_2}$}

\put (145, 40) {$1^+_{A_2}$}
\put (145, 57) {$2^-_{E}$}
\put (145, 87) {$3^+_{A_2 E}$}
\put (145,127) {$4^-_{E A_1 A_2}$}
\put (145,177) {$5^+_{A_2 E E}$}
\put (145,237) {$6^-_{E A_1 A_2 E}$}

\put (255, 27) {$0^+_{A_1}$}
\put (255, 57) {$2^+_{A_1}$}
\put (255, 87) {$3^-_{A_1}$}
\put (255,127) {$4^{\pm}_{A_1}$}
\put (255,177) {$5^-_{A_1}$}
\put (255,237) {$6^{\pm+}_{A_1}$}

\end{picture}
\caption{Schematic spectrum of the rotational states of the ground stand vibrational 
band with $(v_1 v_2)=(00)$. The states are labeled by $L^P_t$. 
The right-hand panel shows the symmetric states with $t=A_1$.} 
\label{obltop}
\end{figure}

Fig.~\ref{oblatetop} shows the rotation-vibration spectrum of the oblate top according to 
Eqs.~(\ref{e3vib}) and (\ref{e3rot}). The energy spectrum consists of a series of rotational bands 
labeled by $(v_1,v_2)$. The bands with $(v_1,0)$ have angular momenta and parity 
$L^{P}=0^{+}$, $2^{+}$, $3^{-}$, $4^{\pm}$, $5^{-}$, $\ldots$, whereas the doubly degenerate 
vibrations with $(v_1,1)$ have $L^{P}=1^{-}$, $2^{\mp}$, $3^{\mp}$, $\ldots$, in agreement with 
Ref.~\cite{Herzberg}. 

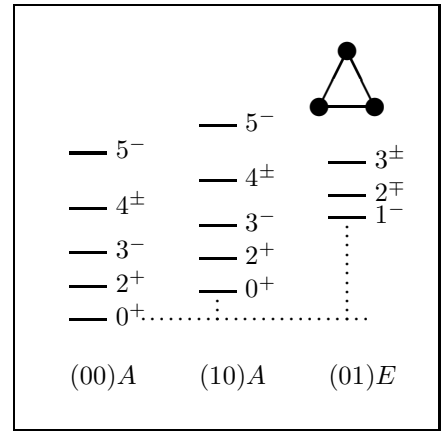
\begin{figure}
\centering
\setlength{\unitlength}{0.7pt} 
\begin{picture}(230,230)(0,0)
\thinlines
\put (  0,  0) {\line(1,0){230}}
\put (  0,230) {\line(1,0){230}}
\put (  0,  0) {\line(0,1){230}}
\put (230,  0) {\line(0,1){230}}
\thicklines
\put ( 30, 60) {\line(1,0){20}}
\put ( 30, 78) {\line(1,0){20}}
\put ( 30, 96) {\line(1,0){20}}
\put ( 30,120) {\line(1,0){20}}
\put ( 30,150) {\line(1,0){20}}
\multiput ( 70, 60)(5,0){25}{\circle*{0.1}}
\thinlines
\put ( 30, 25) {$(00)A$}
\put ( 55, 57) {$0^+$}
\put ( 55, 75) {$2^+$}
\put ( 55, 93) {$3^-$}
\put ( 55,117) {$4^{\pm}$}
\put ( 55,147) {$5^-$}
\thicklines
\put (100, 75) {\line(1,0){20}}
\put (100, 93) {\line(1,0){20}}
\put (100,111) {\line(1,0){20}}
\put (100,135) {\line(1,0){20}}
\put (100,165) {\line(1,0){20}}
\multiput (110, 60)(0,5){3}{\circle*{0.1}}
\thinlines
\put (100, 25) {$(10)A$}
\put (125, 72) {$0^+$}
\put (125, 90) {$2^+$}
\put (125,108) {$3^-$}
\put (125,132) {$4^{\pm}$}
\put (125,162) {$5^-$}
\thicklines
\put (170,115) {\line(1,0){20}}
\put (170,127) {\line(1,0){20}}
\put (170,145) {\line(1,0){20}}
\multiput (180, 60)(0,5){11}{\circle*{0.1}}
\thinlines
\put (170, 25) {$(01)E$}
\put (195,112) {$1^-$}
\put (195,124) {$2^{\mp}$}
\put (195,142) {$3^{\pm}$}
\thicklines
\put(165,175) {\circle*{10}} 
\put(195,175) {\circle*{10}}
\put(180,205) {\circle*{10}}
\put(165,175) {\line( 1,0){30}}
\put(165,175) {\line( 1,2){15}}
\put(195,175) {\line(-1,2){15}}
\end{picture}
\caption[]{Schematic rotation-vibration spectrum of an oblate symmetric top. 
The rotational bands are labeled by $(v_1v_2)$ (bottom), and the rotational 
excitations by $L^P$ (right). All states are symmetric under ${\cal D}_3 \sim S_3$.} 
\label{oblatetop}
\end{figure}

\subsection{Electromagnetic couplings}

For the case of three-body clusters, transition form factors and 
$B(EL)$ values are obtained from the matrix elements of the transition operator 
\numparts
\ba
\hat T(\epsilon) &=& e^{-iq\beta \hat{D}_{2,z}/X_{D}} ~, \\
\hat D_{2,m} &=& (b^{\dagger}_2 \tilde{s} - s^{\dagger} \tilde{b}_2)^{(1)}_m ~. 
\ea
\endnumparts
For the oblate top, the form factors can be obtained in closed form in the large $N$ limit. 
In this case, the normalization factor $X_D$ is given by 
\ba
X_D = \frac{2NR}{(1+R^2)\sqrt{2}} ~.
\ea 
Just as for the case of two-body clusters, the transition form factors for transitions along the 
ground state band $(v_1 v_2)=(00)$ are given in terms of a spherical Bessel function 
$c_L j_L(q \beta)$, but now with
\ba
c_L^2 = \frac{2L+1}{3} \left[ 1+2P_{L}(-\tfrac{1}{2}) \right] ~. 
\label{cl3}
\ea
The coefficient $c_1^2$ vanishes as a consequence of the triangular symmetry.  
Some values which are relevant to the lowest states are $c_0^2=1$, $c_2^2=5/4$, 
$c_3^2=35/8$, $c_4^2=81/64$ and $c_5^2=385/128$. 
For the extended charge distribution of Eq.~(\ref{rhor1}) the form factors 
are multiplied by an exponential factor $\exp(-q^{2}/4\alpha)$. 

The $B(EL;0 \rightarrow L)$ values along the ground state band are given by 
the same formula as for the case of the axial rotor $(Ze c_L \beta^L)^2/4\pi$, the only difference 
being in the value of the coefficients $c_L^2$. In this case they are given by Eq.~(\ref{cl3}). 
The resulting $B(EL)$ values are in agreement with the results of Eq.~(\ref{BEL3}). 

Form factors and $B(EL)$ values depend on $\alpha$, $\beta$ and $c_L$.  
The coefficients $\alpha$ and $\beta$ can be determined from the charge radius and the first 
minimum in the elastic form factor, whereas the $c_L$'s arise from the discrete 
symmetry of the triangular configuration of the three $\alpha$ particles. 
The analytic results given in this section provide a set of closed expressions which can be 
used to analyze and interpret the experimental data. 

\subsection{Shape-phase transition}

The ACM describes the relative dynamics of a three-body system and contains both  
the harmonic oscillator, the deformed oscillator and the oblate top as special limits, 
as well as the region in between these limiting cases. 
The transitional region can be described by the schematic Hamiltonian
\ba
H &=& (1-\chi) \sum_{m} (b_{1,m}^{\dagger} b_{1,m} + b_{2,m}^{\dagger} b_{2,m})  
\nonumber\\
&& + \frac{\chi}{4(N-1)} \, P^{\dagger} \tilde{P} 
\nonumber\\
&& + \frac{\xi}{N-1} \left[ (b_1^{\dagger} \cdot b_1^{\dagger} - b_2^{\dagger} \cdot b_2^{\dagger}) \, 
(\tilde{b}_1 \cdot \tilde{b}_1 - \tilde{b}_2 \cdot \tilde{b}_2) \right.
\nonumber\\
&& \hspace{2cm} \left. + 4 \, (b_1^{\dagger} \cdot b_2^{\dagger}) \, (\tilde{b}_2 \cdot \tilde{b}_1) \right] ~, 
\label{trans3b}
\ea
with $0 \leq \chi \leq 1$ and $\xi > 0$. For $\chi=0$ it reduces to the harmonic oscillator with anharmonic 
terms proportional to $\xi$, and for $\chi=1$ to the oblate top of the previous section. 
For $\chi=1$, $R^2=1$ and $\xi=0$ it reduces to the deformed oscillator discussed in Section~\ref{dosc}. 
The transitional region between any of these special cases can be studied in the classical limit 
of the Hamiltonian following the methods already discussed in Section~\ref{phases}. 
In this case, the potential energy surface is given by
\ba
V_{\rm cl}(\rho,\eta,\zeta) &=& \frac{1-\chi}{2} \, \rho^2 
+ \frac{\chi}{4} \left[ R^2-\frac{1}{2}\rho^2(1+R^2) \right]^2 
\nonumber\\
&& + \frac{\xi \rho^4}{4} \left( \cos^2 2\eta + \sin^2 2\eta \cos^2 2\zeta \right) ~. 
\ea
An analysis of the classical limit of the Hamiltonian of Eq.~(\ref{trans3b}) shows that $H$ 
exhibits a second-order phase transition between the spherical and deformed shapes. As before, 
the spherical shape is characterized by $\rho_{0}=0$, but in this case the deformed shape 
corresponds to three clusters located at the vertices of an equilateral triangle characterized by 
$\rho^2_{0} > 0$, $\eta_{0} = \pi/4$ and $\zeta_{0} = \pi/4$. The value of the deformation parameter 
$\rho_0^2$ and the dependence of the critical point $\chi_c$ on $R^2$ are given by 
Eqs.~(\ref{deformation}) and (\ref{critical}), respectively.  

\section{Four-body clusters}

The extension of the algebraic cluster model to four-body systems was introduced 
recently in an application to $^{16}$O as a cluster of four $\alpha$ particles 
\cite{O16,RB}. The ACM introduces a vector boson with $L^P=1^-$ for each independent 
relative Jacobi coordinate, together with a scalar boson with $L^P=0^+$
\ba 
b^{\dagger}_{1,m} ~, \; b^{\dagger}_{2,m} ~, \; b^{\dagger}_{3,m} ~, \; 
s^{\dagger} ~. \label{bb4}
\ea
This procedure leads to a compact spectrum generating algebra of $U(10)$ whose model 
space is spanned by the symmetric irreducible representation $[N]$ which contains 
the oscillator shells with $n=n_1+n_2+n_3=0,1,2,\ldots, N$. 

\subsection{Discrete symmetry}

For four identical objects ({\it e.g.} for X$_4$ molecules and $4\alpha$ clusters) 
the Hamiltonian has to be invariant under the permuation group $S_4$. All permutations 
can be expressed in terms of the transposition $P(12)$ and the cyclic permutation $P(1234)$ 
\cite{KM}. The transformation properties under $S_4$ of all operators in the model follow 
from those of the building blocks,  
\ba
P(12) \left( \begin{array}{l} s^{\dagger} \\ b^{\dagger}_1 \\ b^{\dagger}_2 \\ 
b^{\dagger}_3 \end{array} \right) = U_{\rm tr} 
\left( \begin{array}{l} s^{\dagger} \\ b^{\dagger}_1 \\ b^{\dagger}_2 \\ 
b^{\dagger}_3 \end{array} \right) U^{-1}_{\rm tr} 
\nonumber\\
\hspace{1cm} = \left( \begin{array}{rrrr} 1 & 0 & 0 & 0 \\ 0 & -1 & 0 & 0 \\ 0 & 0 & 1 & 0 \\ 
0 & 0 & 0 & 1 \end{array} \right) \left( \begin{array}{l} s^{\dagger} \\ b^{\dagger}_1 \\ 
b^{\dagger}_2 \\ b^{\dagger}_3 \end{array} \right) ~, 
\ea
with $U_{\rm tr}$ defined in Eq.~(\ref{p12}) for the transposition $P(12)$, and 
\ba
P(1234) \left( \begin{array}{l} s^{\dagger} \\ b^{\dagger}_1 \\ b^{\dagger}_2 \\ 
b^{\dagger}_3 \end{array} \right) = U_{\rm cycl} 
\left( \begin{array}{l} s^{\dagger} \\ b^{\dagger}_1 \\ b^{\dagger}_2 \\ 
b^{\dagger}_3 \end{array} \right) U^{-1}_{\rm cycl}
\nonumber\\ 
\hspace{1cm} = \left( \begin{array}{cccc} 1 & 0 & 0 & 0 \\ 0 & -\frac{1}{2} & \frac{\sqrt{3}}{2} & 0 \\ 
0 & -\frac{1}{2\sqrt{3}} & -\frac{1}{6} & \frac{\sqrt{8}}{3} \\ 
0 & -\frac{\sqrt{2}}{\sqrt{3}} & -\frac{\sqrt{2}}{3} & -\frac{1}{3} \end{array} \right)  
\left( \begin{array}{l} s^{\dagger} \\ b^{\dagger}_1 \\ b^{\dagger}_2 \\ 
b^{\dagger}_3 \end{array} \right) ~,
\ea
with
\ba
U_{\rm cycl} &=& \mbox{e}^{i \pi (b^{\dagger}_1 b_1 + b^{\dagger}_2 b_2 + b^{\dagger}_3 b_3)} \,  
\mbox{e}^{\theta_1 (b^{\dagger}_1 b_2 - b^{\dagger}_2 b_1)} \,
\mbox{e}^{\theta_2 (b^{\dagger}_2 b_3 - b^{\dagger}_3 b_2)} ~, 
\nonumber\\
\mbox{}
\label{p1234}
\ea
and $\theta_1=\arctan \sqrt{3}$ and $\theta_2=\arctan \sqrt{8}$, for the cyclic permutation 
$P(1234)$. 
The scalar boson, $s^{\dagger}$, transforms as the symmetric representation $[4]$ of $S_4$, whereas the 
three vector Jacobi bosons, $b^{\dagger}_1$, $b^{\dagger}_2$ and $b^{\dagger}_3$, 
transform as the three components of the mixed symmetry representation $[31]$. 

There are five different symmetry classes for the permutation of four objects 
characterized by the Young tableaux
\ba
\begin{array}{ccccc}
\setlength{\unitlength}{1pt}
\begin{picture}(40,20)(0,0)
\put( 0,10) {\line(1,0){40}}
\put( 0, 0) {\line(1,0){40}}
\put( 0, 0) {\line(0,1){10}}
\put(10, 0) {\line(0,1){10}}
\put(20, 0) {\line(0,1){10}}
\put(30, 0) {\line(0,1){10}}
\put(40, 0) {\line(0,1){10}}
\end{picture} & : & [4] & \sim & A_1 \\
\setlength{\unitlength}{1pt}
\begin{picture}(40,20)(0,0)
\put( 0, 10) {\line(1,0){30}}
\put( 0,  0) {\line(1,0){30}}
\put( 0,-10) {\line(1,0){10}}
\put( 0,-10) {\line(0,1){20}}
\put(10,-10) {\line(0,1){20}}
\put(20,  0) {\line(0,1){10}}
\put(30,  0) {\line(0,1){10}}
\end{picture} & : & [31] & \sim & F_2 \\ & & & & \\ 
\setlength{\unitlength}{1pt}
\begin{picture}(40,20)(0,0)
\put( 0, 10) {\line(1,0){20}}
\put( 0,  0) {\line(1,0){20}}
\put( 0,-10) {\line(1,0){20}}
\put( 0,-10) {\line(0,1){20}}
\put(10,-10) {\line(0,1){20}}
\put(20,-10) {\line(0,1){20}}
\end{picture} & : & [22] & \sim & E \\ 
\setlength{\unitlength}{1pt}
\begin{picture}(40,30)(0,0)
\put( 0, 10) {\line(1,0){20}}
\put( 0,  0) {\line(1,0){20}}
\put( 0,-10) {\line(1,0){10}}
\put( 0,-20) {\line(1,0){10}}
\put( 0,-20) {\line(0,1){30}}
\put(10,-20) {\line(0,1){30}}
\put(20,  0) {\line(0,1){10}}
\end{picture} & : & [211] & \sim & F_1 \\
\setlength{\unitlength}{1pt}
\begin{picture}(40,40)(0,0)
\put( 0, 10) {\line(1,0){10}}
\put( 0,  0) {\line(1,0){10}}
\put( 0,-10) {\line(1,0){10}}
\put( 0,-20) {\line(1,0){10}}
\put( 0,-30) {\line(1,0){10}}
\put( 0,-30) {\line(0,1){40}}
\put(10,-30) {\line(0,1){40}}
\end{picture} & : & [1111] & \sim & A_2 \\
& & & & \\ & & & & \\ & & & & 
\end{array}
\ea
Since $S_4$ is isomorphic to the tetrahedral group ${\cal T}_d$, the irreducible representations 
can also be labeled by $[4] \sim A_1$ for the symmetric, $[1111] \sim A_2$ for the antisymmetric, 
and $[31] \sim F_2$, $[22] \sim E$ and $[211] \sim F_1$ for the mixed symmetry representations. 
Next, one can use the multiplication rules for $S_4$ (or ${\cal T}_d$) to construct physical operators 
with the appropriate symmetry properties. For example, for the bilinear products of the three vector 
Jacobi bosons, one finds  
\ba
[31] \otimes [31] = [4] \oplus [31] \oplus [211] \oplus [22] ~,
\ea
or, equivalently,
\ba
F_2 \otimes F_2 = A_1 \oplus F_2 \oplus F_1 \oplus E ~. 
\label{f2f2}
\ea

\subsection{Hamiltonian}

The ACM Hamiltonian that describes the relative motion of a system of four identical clusters 
has to be invariant under the tetrahedral group ${\cal T}_d \sim S_4$, and can be written as 
\ba
H &=& \epsilon_{0} \, s^{\dagger} \tilde{s}
- \epsilon_{1} \, (b_1^{\dagger} \cdot \tilde{b}_1 
+ b_2^{\dagger} \cdot \tilde{b}_2 + b_3^{\dagger} \cdot \tilde{b}_3)
\nonumber\\ 
&& + u_0 \, s^{\dagger} s^{\dagger} \tilde{s} \tilde{s} 
\nonumber\\ 
&& - u_1 \, s^{\dagger} ( b_1^{\dagger} \cdot \tilde{b}_1 
+ b_2^{\dagger} \cdot \tilde{b}_2 + b_3^{\dagger} \cdot \tilde{b}_3 ) \tilde{s} 
\nonumber\\
&& + v_0 \left[ ( b_1^{\dagger} \cdot b_1^{\dagger} + b_2^{\dagger} \cdot b_2^{\dagger}  
+ b_3^{\dagger} \cdot b_3^{\dagger} ) \tilde{s} \tilde{s} + {\rm h.c.} \right]
\nonumber\\
&& + \sum_{L=0,2} a_{L} \, ( b_1^{\dagger} b_1^{\dagger} 
+ b_2^{\dagger} b_2^{\dagger} + b_3^{\dagger} b_3^{\dagger} )^{(L)} \cdot ( {\rm h.c.} )
\nonumber\\
&& + \sum_{L=0,2} b_{L} \left[ 
( -2\sqrt{2} \, b_1^{\dagger} b_3^{\dagger}
+ 2 b_1^{\dagger} b_2^{\dagger} )^{(L)} \cdot ( {\rm h.c.} ) \right.
\nonumber\\
&& \hspace{0.25cm} + \left. ( -2\sqrt{2} \, b_2^{\dagger} b_3^{\dagger} 
+ b_1^{\dagger} b_1^{\dagger} - b_2^{\dagger} b_2^{\dagger} )^{(L)} \cdot ( {\rm h.c.} ) \right]
\nonumber\\
&& + \sum_{L=0,2} c_{L} \left[ ( 2 b_1^{\dagger} b_3^{\dagger}
+ 2\sqrt{2} \, b_1^{\dagger} b_2^{\dagger} )^{(L)} 
\cdot ( {\rm h.c.} ) \right.
\nonumber\\
&& \hspace{0.25cm} + ( 2 b_2^{\dagger} b_3^{\dagger} 
+ \sqrt{2} ( b_1^{\dagger} b_1^{\dagger} - b_2^{\dagger} b_2^{\dagger} ) )^{(L)} \cdot ( {\rm h.c.} )
\nonumber\\
&& \hspace{0.25cm} \left. + ( b_1^{\dagger} b_1^{\dagger} 
     + b_2^{\dagger} b_2^{\dagger} 
   - 2 b_3^{\dagger} b_3^{\dagger} )^{(L)} \cdot ( {\rm h.c.} ) \right]
\nonumber\\
&& + c_1 \left[ 
  ( b_1^{\dagger} b_2^{\dagger} )^{(1)} \cdot ( \tilde{b}_2 \tilde{b}_1 )^{(1)} \right. 
+ ( b_2^{\dagger} b_3^{\dagger} )^{(1)} \cdot ( \tilde{b}_3 \tilde{b}_2 )^{(1)} 
\nonumber\\
&& \hspace{1cm} \left.  
+ ( b_3^{\dagger} b_1^{\dagger} )^{(1)} \cdot ( \tilde{b}_1 \tilde{b}_3 )^{(1)} \right] ~.
\label{HS4}
\ea
The first five terms are equivalent to those in Eq.~(\ref{HSk}), whereas the last four terms 
proportional to $a_L$, $b_L$, $c_L$ and $c_1$, according to the product of Eq.~(\ref{f2f2}), 
correspond to interactions involving pairs of vector bosons with $A_1$, $E$, $F_2$ and $F_1$ 
symmetry, respectively. 

In general, the eigenvalues and corresponding eigenvectors are obtained numerically. 
The wave functions are characterized by the quantum numbers $N$, $L^P_t$, where $N=n_s+n_1+n_2+n_3$ 
is the total number of bosons, $L$ the angular momentum, $P$ the parity, and $t$ the transformation 
property $t$ under the tetrahedral group ${\cal T}_d$. Since internal excitations of the clusters are not 
considered, the four-body wave functions arise solely from the relative motion and 
have to be symmetric ($t=A_1$). 

The discrete symmetry $t$ of a given wave function can be determined as follows. 
Since the invariance of the Hamiltonian of Eq.~(\ref{HS4}) under the transposition $P(12)$ 
implies that basis states with $n_1$ even and $n_1$ odd do not mix, wave functions  
which are even and odd under $P(12)$ can be distinguished by 
\ba
\left< \psi_t \right| P(12) \left| \psi_t \right> = 
\left< \psi_t \right| U_{\rm tr} \left| \psi_t \right> = \pm 1 ~.
\label{utr}
\ea
The expectation value of the cyclic permutation $P(1234)$ can be obtained from 
\ba
\left< \psi_t \right| P(1234) \left| \psi_t \right> = 
\left< \psi_t \right| U_{\rm cycl} \left| \psi_t \right> ~. 
\label{cyc}
\ea
The first term in $U_{\rm cycl}$ of Eq.~(\ref{p1234}) gives rise to a phase factor 
$(-1)^{n_1+n_2+n_3}$ which is $+1$ ($-1$) for states with even (odd) parity. 
The second and third terms in $U_{\rm cycl}$ can be expressed in terms of Talmi-Moshinksy 
brackets \cite{Talmi,Moshinsky}. The permutation symmetry of any given wave function is then 
determined from the matrix elements of $P(12)$ and $P1234)$ (see \ref{app4b}). 
In practice, the wave functions $\psi_t$ are obtained numerically by diagonalization, 
and hence are determined up to a sign. The relative phases of the degenerate representations 
(the two-dimensional $E$ with $E_{\rho}$ and $E_{\lambda}$, and the three-dimensional $F_2$ 
with $F_{2\rho}$, $F_{2\lambda}$ and $F_{2\eta}$, and $F_1$ with $F_{1\rho}$, $F_{1\lambda}$ 
and $F_{1\eta}$) can be determined from the off-diagonal matrix elements of $P(1234)$ requiring 
that they transform as the components of the degenerate representations (see \ref{app4b}).  

\subsection{Spherical top}

In a study of the classical limit of the $U(9)$ and $SO(10)$ dynamical symmetries of Eq.~(\ref{HS4}) 
presented in Section~\ref{limits} it was shown that the corresponding potential energy surfaces only 
depend on the hyperspherical radius $\rho$. In general, the potential energy surface of the 
Hamiltonian of Eq.~(\ref{HS4}) depends on the three Jacobi coordinates, $\rho$, $\eta_1$ and $\eta_2$ 
in the hypersperical notation of Eq.~(\ref{hyper1}-\ref{hyper}), and the three relative angles between the Jacobi 
vectors, $\zeta_{12}$, $\zeta_{23}$ and $\zeta_{31}$, In this section, I analyze the properties of the 
Hamiltonian introduced in \cite{O16} in a study of the $4\alpha$-cluster states of the nucleus $^{16}$O  
\ba
H_{4b} &=& \xi_{1} \, P^{\dagger} \tilde{P}   
\nonumber\\
&& + \xi_2 \left[ ( -2\sqrt{2} \, b_1^{\dagger} \cdot b_3^{\dagger} 
+ 2 b_1^{\dagger} \cdot b_2^{\dagger} ) \, ( {\rm h.c.} ) \right.
\nonumber\\
&& \hspace{0.25cm} + \left. ( -2\sqrt{2} \, b_2^{\dagger} \cdot b_3^{\dagger} 
+ b_1^{\dagger} \cdot b_1^{\dagger} 
- b_2^{\dagger} \cdot b_2^{\dagger} ) \, ( {\rm h.c.} ) \right]
\nonumber\\
&& + \xi_3 \left[ ( 2 b_1^{\dagger} \cdot b_3^{\dagger}
+ 2\sqrt{2} \, b_1^{\dagger} \cdot b_2^{\dagger} ) \, ( {\rm h.c.} ) \right.
\nonumber\\
&& \hspace{0.25cm} + ( 2 b_2^{\dagger} \cdot b_3^{\dagger} 
+ \sqrt{2} \, ( b_1^{\dagger} \cdot b^{\dagger}_1  
- b_2^{\dagger} \cdot b_2^{\dagger} )) \, ( {\rm h.c.} )
\nonumber\\
&& \hspace{0.25cm} \left. + ( b_1^{\dagger} \cdot b_1^{\dagger} 
     + b_2^{\dagger} \cdot b_2^{\dagger} 
   - 2 b_3^{\dagger} \cdot b_3^{\dagger} ) \, ( {\rm h.c.} ) \right]
\nonumber\\
&& + \kappa_1 \, \vec{L} \cdot \vec{L} 
+ \kappa_2 \, (\vec{L} \cdot \vec{L} - \vec{I} \cdot \vec{I})^2 ~,  
\label{H4b}
\ea
where $P^{\dagger}$ is the generalized pairing operator 
\ba
P^{\dagger} = R^{2} \, s^{\dagger} s^{\dagger}
- b_1^{\dagger} \cdot b_1^{\dagger} - b_2^{\dagger} \cdot b_2^{\dagger} - b_3^{\dagger} \cdot b_3^{\dagger} ~,
\label{pair4}
\ea
$\vec{L}$ denotes the angular momentum in coordinate space 
\ba
\hat{L}_m = \sqrt{2} \, ( b^{\dagger}_1 \tilde{b}_1 + b^{\dagger}_2 \tilde{b}_2  
+ b^{\dagger}_3 \tilde{b}_3 )^{(1)}_m ~,
\ea
and $\vec{I}$ the angular momentum in index space
\numparts
\ba
\hat{I}_1 &=& -i \sum_m \left( b^{\dagger}_{2,m} b_{3,m} - b^{\dagger}_{3,m} b_{2,m} \right) ~, \\
\hat{I}_2 &=& -i \sum_m \left( b^{\dagger}_{3,m} b_{1,m} - b^{\dagger}_{1,m} b_{3,m} \right) ~, \\
\hat{I}_3 &=& -i \sum_m \left( b^{\dagger}_{1,m} b_{2,m} - b^{\dagger}_{2,m} b_{1,m} \right) ~. 
\ea
\endnumparts
For $R^{2}=1$ and $\xi_2=\xi_3=0$ the Hamiltonian has $U(10)\supset SO(10)$ symmetry and 
corresponds to a nine-dimensional deformed oscillator. For $R^{2}=0$ there is no mixing 
between oscillator shells and the Hamiltonian is that of an anharmonic vibrator. 
In this section I discuss the general case with $R^{2} \neq 0$ and $\xi_{1}$, $\xi_{2}$, 
$\xi_{3}>0$, The potential energy surface associated with the Hamiltonian of Eq.~(\ref{H4b}) 
has an equilibrium configuration corresponding to coordinates that have equal length 
($\rho_{1,0}=\rho_{2,0}=\rho_{3,0}=\rho_0/\sqrt{3}$)  
\numparts
\ba
\rho_{0} &=& \sqrt{2R^{2}/(1+R^{2})} ~, \label{radii4} \\
\eta_{1,0} &=& \arctan \sqrt{2} ~, \\  
\eta_{2,0} &=& \pi/4 ~,
\ea
and are mutually perpendicular
\ba  
\zeta_{12,0} = \zeta_{23,0} = \zeta_{31,0} = \pi/4 ~,
\label{angles4}
\ea
\endnumparts
where $2\zeta_{ij}$ denotes the relative angle between $\vec{\alpha}_i$ and $\vec{\alpha}_j$. 
Just as in the case of the oblate top, the intrinsic degrees of freedom decouple in the limit 
of small oscillations around the equilibrium shape $\rho=\rho_0+\Delta \rho$, 
$\eta_i = \eta_{i,0}+\Delta \eta_i$ and $\zeta_{ij} = \zeta_{ij,0} + \Delta \zeta_{ij}$, 
and become harmonic. To leading order in $N$ one obtains
\ba
H_{\rm 4b,cl} &=& \xi_{1}N \left[ \frac{2R^{2}}{1+R^{2}} p^{2} 
+ 2R^{2}(1+R^{2}) (\Delta \rho)^{2} \right]  
\nonumber\\
&& + \xi_{2}N \left[ 
9p_{\nu_{2a}}^{2} + \frac{16R^{4}}{9(1+R^{2})^{2}} (\Delta \nu_{2a})^{2} \right. 
\nonumber\\
&& \hspace{1cm} \left. 
+ 9p_{\nu_{2b}}^{2} + \frac{16R^{4}}{9(1+R^{2})^{2}} (\Delta \nu_{2b})^{2} \right]  
\nonumber\\
&& +  \xi_{3}N \left[ 
9p_{\nu_{3a}}^{2} + \frac{16R^{4}}{9(1+R^{2})^{2}} (\Delta \nu_{3a})^{2} \right. 
\nonumber\\
&& \hspace{1cm} 
+ 9p_{\nu_{3b}}^{2} + \frac{16R^{4}}{9(1+R^{2})^{2}} (\Delta \nu_{3b})^{2} 
\nonumber\\
&& \hspace{1cm} \left. 2p_{\eta_1}^{2}+\frac{8R^{4}}{(1+R^{2})^{2}} (\Delta \eta_1)^{2} \right] ~, 
\ea
where the $\Delta \nu$'s denote the oscillations 
\numparts
\ba
\Delta \nu_{2a} &=& \Delta \zeta_{12} - \sqrt{2} \, \Delta \zeta_{31} ~, \\
\Delta \nu_{2b} &=& \Delta \eta_2 - \sqrt{2} \, \Delta \zeta_{23} ~, \\
\Delta \nu_{3a} &=& \Delta \zeta_{31} + \sqrt{2} \, \Delta \zeta_{12} ~, \\
\Delta \nu_{3b} &=& \Delta \zeta_{23} + \sqrt{2} \, \Delta \eta_2 ~,  
\ea
\endnumparts
and the $p_{\nu}$'s the canonically conjugate momenta. Standard quantization gives the 
vibrational energy spectrum of a spherical top with tetrahedral symmetry
\ba
E_{\rm 4b,vib} = \omega_{1}(v_{1}+\tfrac{1}{2}) 
+ \omega_{2}(v_{2}+1) 
+ \omega_{3}(v_{3}+\tfrac{3}{2}) 
\label{e4vib}
\ea
with frequencies 
\numparts
\ba
\omega_{1} &=&  4NR^{2} \xi_{1} ~, \\
\omega_{2} &=& \frac{8NR^{2}}{1+R^{2}} \xi_{2} ~, \\ 
\omega_{3} &=& \frac{8NR^{2}}{1+R^{2}} \xi_{3} ~, 
\ea
\endnumparts
in agreement with the results obtained in a normal mode analysis of a spherical top 
with tetrahedral symmetry (see Fig.~\ref{fundvib4}). 
Here $v_{1}$ represents the vibrational quantum number for a symmetric stretching 
vibration with $A_1$ symmetry, $v_2=v_{2a}+v_{2b}$ denotes a doubly degenerate 
vibration with $E$ symmetry, and $v_3=v_{3a}+v_{3b}+v_{3c}$ a three-fold degenerate 
vibration with $F_2$ symmetry. For rigid configurations, $R^2=1$ and 
$\omega_i=4N \xi_i$ with $i=1,2,3$. 

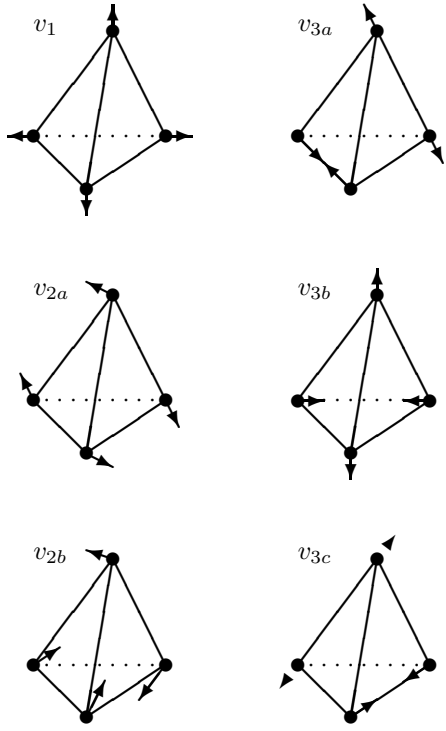
\begin{figure}
\centering
\vspace{15pt}
\setlength{\unitlength}{1pt}
\begin{picture}(210,300)(0,20)
\thicklines
\put( 50,230) {\circle*{5}} 
\put( 30,250) {\circle*{5}}
\put( 80,250) {\circle*{5}}
\put( 60,290) {\circle*{5}}
\put( 50,230) {\vector( 0,-1){10}} 
\put( 30,250) {\vector(-1, 0){10}}
\put( 80,250) {\vector( 1, 0){10}}
\put( 60,290) {\vector( 0, 1){10}}
\put( 50,230) {\line(-1, 1){20}}
\put( 50,230) {\line( 3, 2){30}}
\put( 60,290) {\line(-3,-4){30}}
\put( 60,290) {\line( 1,-2){20}}
\put( 60,290) {\line(-1,-6){10}}
\put( 30,290) {$v_1$}
\multiput( 30,250)(5,0){10}{\circle*{1}}

\put( 50,130) {\circle*{5}} 
\put( 30,150) {\circle*{5}}
\put( 80,150) {\circle*{5}}
\put( 60,190) {\circle*{5}}
\put( 50,130) {\vector( 2,-1){10}} 
\put( 30,150) {\vector(-1, 2){ 5}}
\put( 80,150) {\vector( 1,-2){ 5}}
\put( 60,190) {\vector(-2, 1){10}}
\put( 50,130) {\line(-1, 1){20}}
\put( 50,130) {\line( 3, 2){30}}
\put( 60,190) {\line(-3,-4){30}}
\put( 60,190) {\line( 1,-2){20}}
\put( 60,190) {\line(-1,-6){10}}
\put( 30,190) {$v_{2a}$}
\multiput( 30,150)(5,0){10}{\circle*{1}}

\put( 50, 30) {\circle*{5}} 
\put( 30, 50) {\circle*{5}}
\put( 80, 50) {\circle*{5}}
\put( 60, 90) {\circle*{5}}
\put( 50, 30) {\vector( 1, 2){ 7}} 
\put( 30, 50) {\vector( 3, 2){10}}
\put( 80, 50) {\vector(-3,-4){10}}
\put( 60, 90) {\vector(-3, 1){10}}
\put( 50, 30) {\line(-1, 1){20}}
\put( 50, 30) {\line( 3, 2){30}}
\put( 60, 90) {\line(-3,-4){30}}
\put( 60, 90) {\line( 1,-2){20}}
\put( 60, 90) {\line(-1,-6){10}}
\put( 30, 90) {$v_{2b}$}
\multiput( 30, 50)(5,0){10}{\circle*{1}}

\put(150,230) {\circle*{5}} 
\put(130,250) {\circle*{5}}
\put(180,250) {\circle*{5}}
\put(160,290) {\circle*{5}}
\put(150,230) {\vector(-1, 1){10}} 
\put(130,250) {\vector( 1,-1){10}}
\put(180,250) {\vector( 1,-2){ 5}}
\put(160,290) {\vector(-1, 2){ 5}}
\put(150,230) {\line(-1, 1){20}}
\put(150,230) {\line( 3, 2){30}}
\put(160,290) {\line(-3,-4){30}}
\put(160,290) {\line( 1,-2){20}}
\put(160,290) {\line(-1,-6){10}}
\put(130,290) {$v_{3a}$}
\multiput(130,250)(5,0){10}{\circle*{1}}

\put(150,130) {\circle*{5}} 
\put(130,150) {\circle*{5}}
\put(180,150) {\circle*{5}}
\put(160,190) {\circle*{5}}
\put(150,130) {\vector( 0,-1){10}} 
\put(130,150) {\vector( 1, 0){10}}
\put(180,150) {\vector(-1, 0){10}}
\put(160,190) {\vector( 0, 1){10}}
\put(150,130) {\line(-1, 1){20}}
\put(150,130) {\line( 3, 2){30}}
\put(160,190) {\line(-3,-4){30}}
\put(160,190) {\line( 1,-2){20}}
\put(160,190) {\line(-1,-6){10}}
\put(130,190) {$v_{3b}$}
\multiput(130,150)(5,0){10}{\circle*{1}}

\put(150, 30) {\circle*{5}} 
\put(130, 50) {\circle*{5}}
\put(180, 50) {\circle*{5}}
\put(160, 90) {\circle*{5}}
\put(150, 30) {\vector( 3, 2){10}} 
\put(130, 50) {\vector(-3,-4){ 7}}
\put(180, 50) {\vector(-3,-2){10}}
\put(160, 90) {\vector( 3, 4){ 7}}
\put(150, 30) {\line(-1, 1){20}}
\put(150, 30) {\line( 3, 2){30}}
\put(160, 90) {\line(-3,-4){30}}
\put(160, 90) {\line( 1,-2){20}}
\put(160, 90) {\line(-1,-6){10}}
\put(130, 90) {$v_{3c}$}
\multiput(130, 50)(5,0){10}{\circle*{1}}
\end{picture}
\caption[]{Fundamental vibrations of a tetrahedral configuration (point group ${\cal T}_d$).}
\label{fundvib4}
\end{figure}

Note, that in general the angular momentum in index space $I$ is not conserved by  
${\cal T}_d \sim S_4$ invariant interactions. Only if $a_L=c_L$ in Eq.~(\ref{HS4}) 
or $\xi_2=\xi_3$ in Eq.~(\ref{H4b}), does $I$ become a good quantum number for all 
eigenstates. Nevertheless, the general Hamiltonian of Eq.~(\ref{H4b}) still has some 
eigenstates with good $I$: the rotational excitations of the ground state band 
with $(v_1 v_2 v_3)=(000)$ are characterized by $I=L$. This can be understood from 
the fact that the operator $\vec{L} \cdot \vec{L} - \vec{I} \cdot \vec{I}$ annihilates 
the coherent (or intrinisic) state corresponding to the equilibrium shape of a 
regular tetrahedron, see Eqs.~(\ref{radii4}-\ref{angles4}). As a consequence, the 
rotational energies of the ground state band are given by 
\ba
E_{\rm 4b,rot} = \kappa_{1} L(L+1) ~.
\ea

\begin{figure}
\centering
\setlength{\unitlength}{0.7pt} 
\begin{picture}(320,280)(0,0)
\thinlines
\put (  0,  0) {\line(1,0){320}}
\put (  0,280) {\line(1,0){320}}
\put (  0,  0) {\line(0,1){280}}
\put (240,  0) {\line(0,1){280}}
\put (320,  0) {\line(0,1){280}}
\thicklines
\put ( 20, 30) {\line(1,0){20}}
\put ( 20, 40) {\line(1,0){20}}
\put ( 20, 60) {\line(1,0){20}}
\put ( 20, 90) {\line(1,0){20}}
\put ( 20,130) {\line(1,0){20}}
\put ( 20,180) {\line(1,0){20}}
\put ( 20,240) {\line(1,0){20}}

\put (120, 30) {\line(1,0){20}}
\put (120, 40) {\line(1,0){20}}
\put (120, 60) {\line(1,0){20}}
\put (120, 90) {\line(1,0){20}}
\put (120,130) {\line(1,0){20}}
\put (120,180) {\line(1,0){20}}
\put (120,240) {\line(1,0){20}}

\put (260, 30) {\line(1,0){20}}
\put (260, 90) {\line(1,0){20}}
\put (260,130) {\line(1,0){20}}
\put (260,240) {\line(1,0){20}}
\small
\put ( 45, 23) {$0^+_{A_1}$}
\put ( 45, 40) {$1^-_{F_2}$}
\put ( 45, 57) {$2^+_{E F_2}$}
\put ( 45, 87) {$3^-_{A_1 F_2 F_1}$}
\put ( 45,127) {$4^+_{A_1 E F_2 F_1}$}
\put ( 45,177) {$5^-_{E F_2 F_2 F_1}$}
\put ( 45,237) {$6^+_{A_1 A_2 E F_2 F_2 F_1}$}

\put (145, 23) {$0^-_{A_2}$}
\put (145, 40) {$1^+_{F_1}$}
\put (145, 57) {$2^-_{E F_1}$}
\put (145, 87) {$3^+_{A_2 F_1 F_2}$}
\put (145,127) {$4^-_{A_2 E F_1 F_2}$}
\put (145,177) {$5^+_{E F_1 F_1 F_2}$}
\put (145,237) {$6^-_{A_2 A_1 E F_1 F_1 F_2}$}

\put (285, 27) {$0^+_{A_1}$}
\put (285, 87) {$3^-_{A_1}$}
\put (285,127) {$4^+_{A_1}$}
\put (285,237) {$6^{\pm}_{A_1}$}

\end{picture}
\caption{Schematic spectrum of the rotational states of the ground stand vibrational 
band with $(v_1 v_2 v_3)=(000)$. All states have $L=I$ and are labeled by $L^P_t$. 
The right-hand panel shows the symmetric states with $t=A_1$.} 
\label{top}
\end{figure}
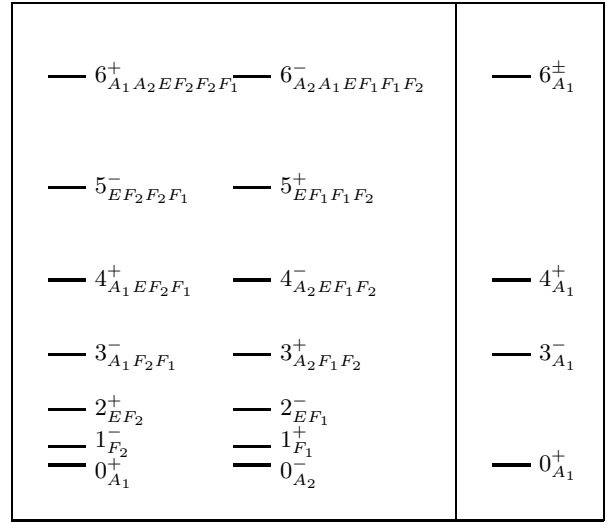

Fig.~\ref{top} shows the structure of the rotational states of the ground 
state band $(v_1 v_2 v_3)=(000)$. The rotational levels are doubled because of 
inversion doubling: for each value of the angular momentum $L$, one has doublets of 
states with ($A_1$, $A_2$), ($E$, $E$) and ($F_2$, $F_1$), in agreement with the 
classification of rotational levels of a spherical top with tetrahedral symmetry 
\cite{Herzberg}. 

For identical bosons, as is the case for a cluster of four $\alpha$-particles, 
the allowed rotational-vibrational states are symmetric with $t=A_1$, and therefore the 
states of the ground state band $(v_1 v_2 v_3)=(000)$ have angular momentum and 
parity $L^P=0^+$, $3^-$, $4^+$, $6^{\pm}$, $\ldots$, as shown in the right panel  
of Fig.~\ref{top}. A similar analysis can be done for the rotational bands built on the $(100)A_1$, 
$(010)E$ and $(001)F_2$ vibrations. For the $A_1$ vibration the values of angular momentum 
and parity are the same as for the ground state band  $L^P=0^+$, $3^-$, $4^+$, 
$6^{\pm}$, $\ldots$. For the double degenerate $E$ vibration they are $L^P=2^{\pm}$, 
$4^{\pm}$, $5^{\pm}$, $6^{\pm}$, $\ldots$, while for the triple degenerate $F_2$ vibration 
they are  $L^P=1^-$, $2^+$, $3^{\pm}$, $4^{\pm}$, $5^{-,\pm}$, $6^{+,\pm}$ $\ldots$. 
The situation is summarized in Fig.~\ref{sphtop} which shows the expected spectrum 
of a spherical top with tetrahedral symmetry and $\omega_1=\omega_2=\omega_3$. 

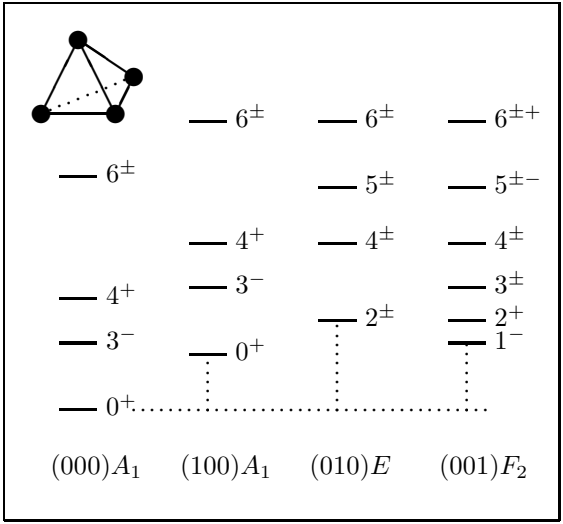
\begin{figure}
\centering
\setlength{\unitlength}{0.7pt} 
\begin{picture}(300,280)(0,0)
\thinlines
\put (  0,  0) {\line(1,0){300}}
\put (  0,280) {\line(1,0){300}}
\put (  0,  0) {\line(0,1){280}}
\put (300,  0) {\line(0,1){280}}
\thicklines
\put ( 30, 60) {\line(1,0){20}}
\put ( 30, 96) {\line(1,0){20}}
\put ( 30,120) {\line(1,0){20}}
\put ( 30,186) {\line(1,0){20}}
\multiput ( 70, 60)(5,0){39}{\circle*{0.1}}
\thinlines
\put ( 25, 25) {$(000)A_1$}
\put ( 55, 57) {$0^+$}
\put ( 55, 93) {$3^-$}
\put ( 55,117) {$4^+$}
\put ( 55,183) {$6^{\pm}$}
\thicklines
\put (100, 90) {\line(1,0){20}}
\put (100,126) {\line(1,0){20}}
\put (100,150) {\line(1,0){20}}
\put (100,216) {\line(1,0){20}}
\multiput (110, 60)(0,5){6}{\circle*{0.1}}
\thinlines
\put ( 95, 25) {$(100)A_1$}
\put (125, 87) {$0^+$}
\put (125,123) {$3^-$}
\put (125,147) {$4^+$}
\put (125,213) {$6^{\pm}$}
\thicklines
\put (170,108) {\line(1,0){20}}
\put (170,150) {\line(1,0){20}}
\put (170,180) {\line(1,0){20}}
\put (170,216) {\line(1,0){20}}
\multiput (180, 60)(0,5){10}{\circle*{0.1}}
\thinlines
\put (165, 25) {$(010)E$}
\put (195,105) {$2^{\pm}$}
\put (195,147) {$4^{\pm}$}
\put (195,177) {$5^{\pm}$}
\put (195,213) {$6^{\pm}$}
\thicklines
\put (240, 96) {\line(1,0){20}}
\put (240,108) {\line(1,0){20}}
\put (240,126) {\line(1,0){20}}
\put (240,150) {\line(1,0){20}}
\put (240,180) {\line(1,0){20}}
\put (240,216) {\line(1,0){20}}
\multiput (250, 60)(0,5){8}{\circle*{0.1}}
\thinlines
\put (235, 25) {$(001)F_2$}
\put (265, 93) {$1^-$}
\put (265,105) {$2^+$}
\put (265,123) {$3^{\pm}$}
\put (265,147) {$4^{\pm}$}
\put (265,177) {$5^{\pm-}$}
\put (265,213) {$6^{\pm+}$}
\thicklines
\put(20,220) {\circle*{10}} 
\put(60,220) {\circle*{10}}
\put(70,240) {\circle*{10}}
\put(40,260) {\circle*{10}}
\put(20,220) {\line( 1,0){40}}
\put(20,220) {\line( 1,2){20}}
\put(60,220) {\line(-1,2){20}}
\put(60,220) {\line( 1,2){10}}
\put(70,240) {\line(-3,2){30}}
\multiput(20,220)(5,2){11}{\circle*{2}}
\end{picture}
\caption{Schematic rotation-vibration spectrum of a spherical top with tetrahedral symmetry 
and $\omega_1=\omega_2=\omega_3$. The rotational bands are labeled by $(v_1 v_2 v_3)$ (bottom). 
All states are symmetric under $S_4 \sim {\cal T}_d$.} 
\label{sphtop}
\end{figure}

\subsection{Electromagnetic couplings}

For four-body clusters, electromagnetic couplings can be obtained from the matrix elements of 
the transition operator 
\numparts
\ba
\hat T(\epsilon) &=& e^{-iq\beta \hat{D}_{3,z}/X_{D}} ~, \\
\hat D_{3,m} &=& (b^{\dagger}_3 \tilde{s} - s^{\dagger} \tilde{b}_3)^{(1)}_m ~. 
\ea
\endnumparts
For the spherical top, the form factors can be obtained in closed form in the large $N$ limit. 
The normalization factor $X_D$ is given by 
\ba
X_D = \frac{2NR}{(1+R^2)\sqrt{3}} ~.
\ea 
Also in this case, the transition form factors for transitions along the ground state band 
$(v_1 v_2 v_3)=(000)$ are given in terms of a spherical Bessel function $c_L j_L(q \beta)$, but with
\ba
c_L^2 = \frac{2L+1}{4} \left[ 1+3P_{L}(-\tfrac{1}{3}) \right] ~. 
\label{cl4}
\ea
The coefficients $c_1^2$, $c_2^2$ and $c_5^2$ vanish as a consequence of the tetrahedral 
symmetry, Some values which are relevant to the lowest states are $c_0^2=1$, $c_3^2=35/9$, 
$c_4^2=7/3$ and $c_6^2=416/81$.
For the extended charge distribution of Eq.~(\ref{rhor1}) the form factors 
are multiplied by an exponential factor $\exp(-q^{2}/4\alpha)$. 

Just as for the oblate top, the transition probabilities $B(EL;0 \rightarrow L)$ along the ground 
state band are given by the same formula as for the case of the axial rotor $(Ze c_L \beta^L)^2/4\pi$, 
but with the coefficients $c_L^2$ from Eq.~(\ref{cl4}). 
The $B(EL)$ values are in agreement with the results of Eq.~(\ref{BEL4}). 

Form factors and $B(EL)$ values depend on the coefficients $\alpha$, $\beta$, and $c_L$.  
Just as for two- and three-body clusters, the coefficients $\alpha$ and $\beta$ can be 
determined from the charge radius and the first minimum in the elastic form factor. 
The $c_L$'s are the consequence of the tetrahedral symmetry of the geometrical configuration 
of the four $\alpha$ particles. 

\subsection{Shape-phase transition}

The ACM for four-body systems contains the harmonic oscillator, the deformed oscillator and the 
spherical top as special limits, as well as the region in between these limiting cases. 
The transitional region can be described by the schematic Hamiltonian
\ba
H &=& (1-\chi) \sum_{m} (b_{1,m}^{\dagger} b_{1,m} 
+ b_{2,m}^{\dagger} b_{2,m} + b_{3,m}^{\dagger} b_{3,m})  
\nonumber\\
&& + \frac{\chi}{4(N-1)} \, P^{\dagger} \tilde{P} 
\nonumber\\
&& + \frac{\xi'_2}{N-1} \left[ ( -2\sqrt{2} \, b_1^{\dagger} \cdot b_3^{\dagger} 
+ 2 b_1^{\dagger} \cdot b_2^{\dagger} ) \, ( {\rm h.c.} ) \right.
\nonumber\\
&& \hspace{0.25cm} + \left. ( -2\sqrt{2} \, b_2^{\dagger} \cdot b_3^{\dagger} 
+ b_1^{\dagger} \cdot b_1^{\dagger} 
- b_2^{\dagger} \cdot b_2^{\dagger} ) \, ( {\rm h.c.} ) \right]
\nonumber\\
&& + \frac{\xi'_3}{N-1} \left[ ( 2 b_1^{\dagger} \cdot b_3^{\dagger}
+ 2\sqrt{2} \, b_1^{\dagger} \cdot b_2^{\dagger} ) \, ( {\rm h.c.} ) \right.
\nonumber\\
&& \hspace{0.25cm} + ( 2 b_2^{\dagger} \cdot b_3^{\dagger} 
+ \sqrt{2} \, ( b_1^{\dagger} \cdot b^{\dagger}_1  
- b_2^{\dagger} \cdot b_2^{\dagger} )) \, ( {\rm h.c.} )
\nonumber\\
&& \hspace{0.25cm} \left. + ( b_1^{\dagger} \cdot b_1^{\dagger} 
     + b_2^{\dagger} \cdot b_2^{\dagger} 
   - 2 b_3^{\dagger} \cdot b_3^{\dagger} ) \, ( {\rm h.c.} ) \right] ~,
\label{trans4b}
\ea
with $0 \leq \chi \leq 1$, $\xi'_2 > 0$ and $\xi'_3 > 0$. For $\chi=0$ it 
reduces to the harmonic oscillator with anharmonic terms proportional to 
$\xi'_2$ and $\xi'_3$, and for $\chi=1$ to the spherical top of the previous section. 
For $\chi=1$, $R^2=1$ and $\xi'_2=\xi'_3=0$ it reduces to the deformed oscillator 
discussed in Section~\ref{dosc}. The transitional region can be studied by analyzing 
the properties of the corresponding potential energy surface. The terms proportional 
to $\xi'_2$ and $\xi'_3$ (both $>0$) favor an equilibrium shape in which the 
coordinates have equal length and are mutually perpendicular, 
see Eqs.(\ref{radii4}-\ref{angles4}). 
In addition, the equilibrium configuration is characterized by $\rho_{0}$ which changes 
from spherical $\rho_0=0$ to deformed $\rho_0^2 > 0$ as a function of $\chi$ according to 
Eq.~(\ref{deformation}).  

An analysis of the classical limit of the Hamiltonian of 
Eq.~({\ref{trans4b}) shows that $H$ exhibits a second-order phase transition between the 
spherical and deformed shapes. As in the previous examples for two- and three-body clusters, 
the critical point depends on $R^2$ and is given by Eq.~(\ref{critical}).  
The dependence of the equilibrium shape and the ground state energy on $\chi$ and $R^2$ is 
the same as for the two-body ACM shown in Fig.~\ref{phasetrans}. In this case, the equilibrium 
shape in the deformed region corresponds to four clusters located at the vertices of a regular 
tetrahedron.

\section{Applications to $\alpha$-cluster nuclei}

Evidence for $\alpha$ clustering in light nuclei can be found in a plot of the binding energy 
per nucleon which shows maxima for nuclei with $A=4k$, {\it i.e.} for the nuclei $^{4}$He, $^{8}$Be, 
$^{12}$C and $^{16}$O for $k=1$, 2, 3 and 4, respectively. In this secion, it is shown that further 
evidence can be found in the spectroscopy of $\alpha$-cluster states in $^{12}$C \cite{C12,ACM} and 
$^{16}$O \cite{O16}. 
It is discussed how the structure of rotational bands can be used to obtain information about the 
underlying geometric configuration of the $\alpha$-clusters, and hence to distinguish between 
different theoretical approaches of $\alpha$-cluster nuclei. 

\subsection{The nucleus $^{12}$C}

\begin{figure*}
\centering
\includegraphics[width=6in]{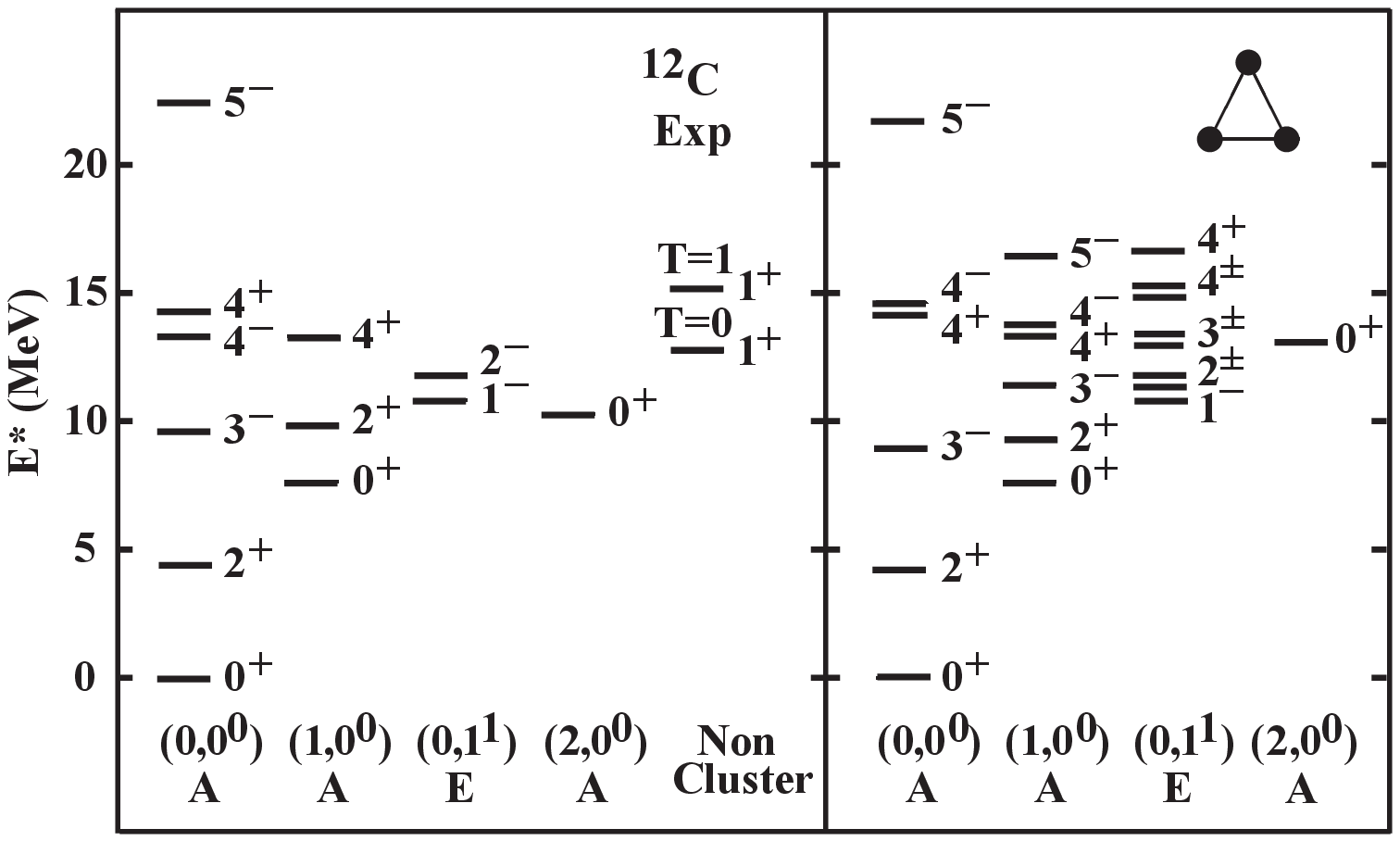}
\caption{\label{carbon12} Comparison between the low-lying experimental spectrum of $^{12}$C and the energies
of the oblate symmetric top calculated using Eq.~(\ref{ost}). The last column on
the left-hand side, shows the lowest observed non-cluster levels with $J^P=1^+$ and $T=0$, $1$.}.
\end{figure*}

Fig.~\ref{carbon12} shows a comparison of the cluster states of $^{12}$C with the spectrum of the oblate 
top according to the approximate energy formula \cite{C12}
\ba
E &=& E_0 + \omega_{1}(v_{1}+\tfrac{1}{2}) \left( 1-\frac{v_1+\tfrac{1}{2}}{N} \right)
\nonumber\\
&& + \omega_{2}(v_{2}+1) \left( 1-\frac{v_2+1}{N+\tfrac{1}{2}} \right)
\nonumber\\
&& + \kappa_{1} L(L+1) + \kappa_{2} (K \mp 2\ell_{2})^{2}
\nonumber\\
&& + \left[\lambda_{1} (v_{1}+\tfrac{1}{2}) + \lambda_{2} (v_{2}+1) \right] L(L+1) ~. 
\label{ost}
\ea
The coefficients $\kappa_1$, $\lambda_1$ and $\lambda_2$ determine the moments of inertia of the ground 
state band, the symmetric stretching or breathing vibration and the bending vibration. The value of the 
$\kappa_2$ term is determined from the relative energies of the positive and negative parity states in 
the ground state band. The vibrational energies $\omega_1$ and $\omega_2$ are obtained from the
excitation energies of the first excited $0^+$ and $1^-$ states, respectively. In $^{12}$C  
the vibrational and rotational energies are of the same order. Therefore, one expects sizeable 
rotation-vibration couplings. Eq.~(\ref{ost}) includes both the anharmonicities which depend on $N$ 
and the vibrational dependence of the moments of inertia. The rotation-vibration couplings and 
anharmonicities are large and therefore $N$ is small. Here it is taken to be $N=10$ \cite{ACM}.
The large anharmonicities lead to an increase of the rms radius of the vibrational
excitations relative to that of the ground state.

For the ground state band of $^{12}$C both the positive and negative parity states have been 
observed, including a nearly degenerate doublet of states with $L^P=4^{\pm}$, and the recently 
measured $5^-$ state \cite{C12}.

The $0^+$ Hoyle state in $^{12}$C at 7.654 MeV is interpreted as the bandhead of the $A$ symmetric
stretching vibration or breathing mode of the triangular configuration with the same geometrical
arrangement and rotational structure as the ground state rotational band, as shown in Fig.~\ref{carbon12}. 
Recent measurements identified the $2^+$ \cite{Itoh,Freer,Gai} and $4^+$ \cite{Fre11} members of the Hoyle 
band \cite{Physics} which raises the question of the location of the predicted negative parity states 
as shown in Fig.~\ref{carbon12}. It is interesting to note that a (broad) negative parity state was suggested 
to lie between 11 and 14 MeV \cite{Fre07} which is close to the predicted energy for the $3^-$ state of the 
Hoyle band in Fig.~\ref{carbon12}. In order to distinguish
between different geometric configurations of the Hoyle band, {\it e.g.} equilateral triangular \cite{C12,ACM} 
or bent-arm \cite{EFT}, the identification of the negative parity states $3^-$ and $4^-$ is crucial. 

The $1^-$ state at 10.84 MeV is assigned as the bandhead of the vibrational bending mode
whose lowest-lying rotational excitations consist of nearly degenerate parity doublets of
$2^\pm$ and $3^\pm$ states. So far, only the $2^-$ has been identified.

\begin{figure}
\includegraphics[width=3in]{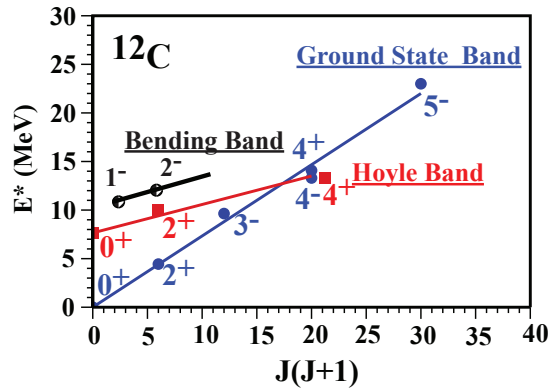}
\caption{\label{RotBands} (Color Online) Rotational band structure of the ground-state band, the Hoyle band and
the bending vibration in $^{12}$C.}
\end{figure}

Fig.~\ref{RotBands} shows that both the ground state rotational band and the Hoyle band follow a $J(J+1)$ 
trajectory albeit with a different moment of inertia. The moment of inertia of the bending vibration is 
almost the same as that of the Hoyle band.

Fig.~\ref{ffc12} shows a comparison between experimental and theoretical form factors. The coefficient $\beta$ is 
determined from the first minimum in the elastic form factor to be $\beta=1.74$ fm, and subsequently the coefficient 
$\alpha$ is obtained from the charge radius of $^{12}$C to be $\alpha=0.52$ fm$^{-2}$. 
The $q$ dependence of the form factors is consistent with experiment, indicating that the $2^{+}_1$ and  
$0^{+}_2$ states can be interpreted as rotational and vibrational excitations of a triangular configuration 
of three $\alpha$ particles with ${\cal D}_{3h}$ symmetry. 
 
\begin{figure*}
\begin{minipage}{.33\linewidth}
\centerline{\epsfig{file=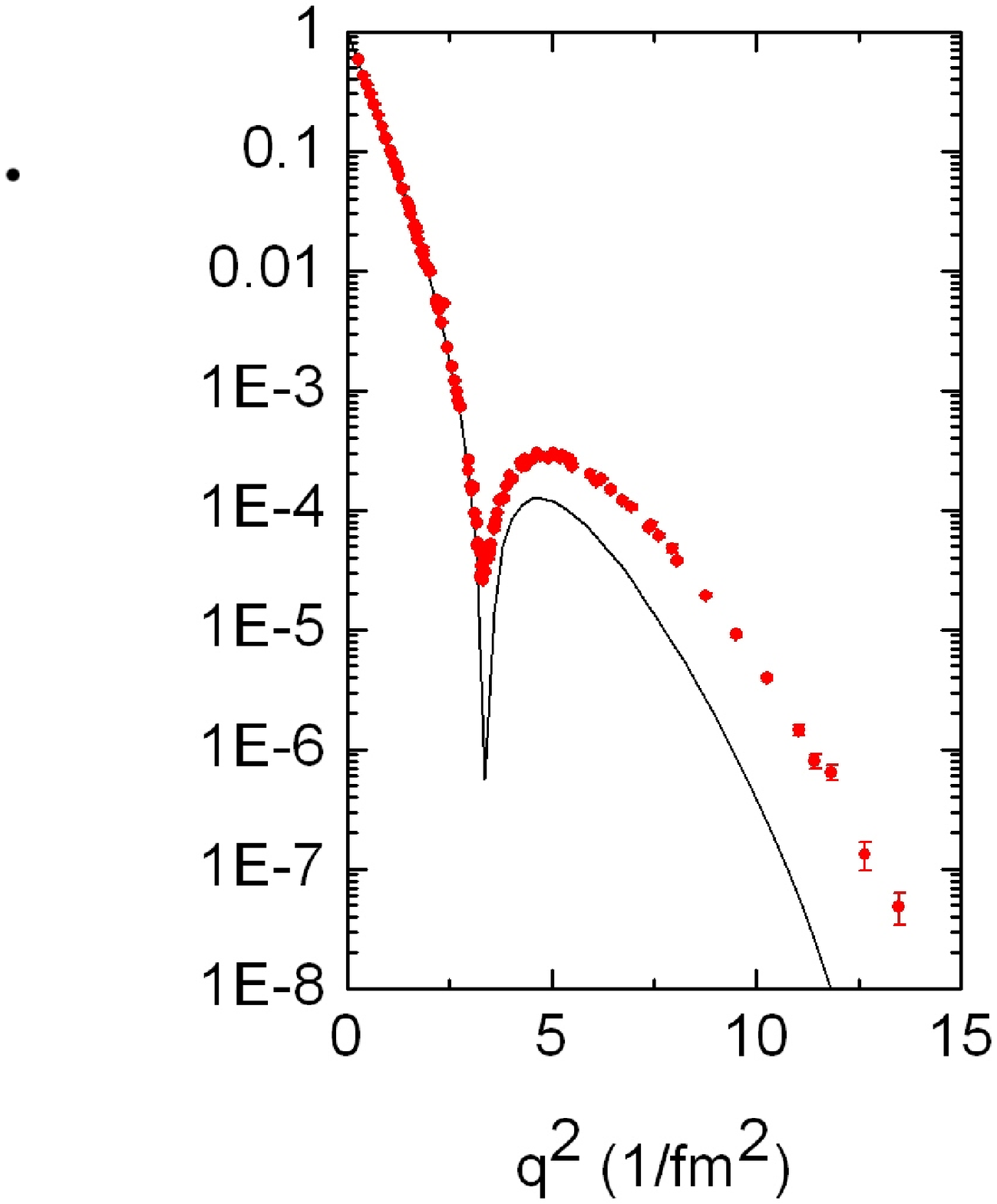,width=\linewidth}}
\end{minipage}\hfill
\begin{minipage}{.33\linewidth}
\centerline{\epsfig{file=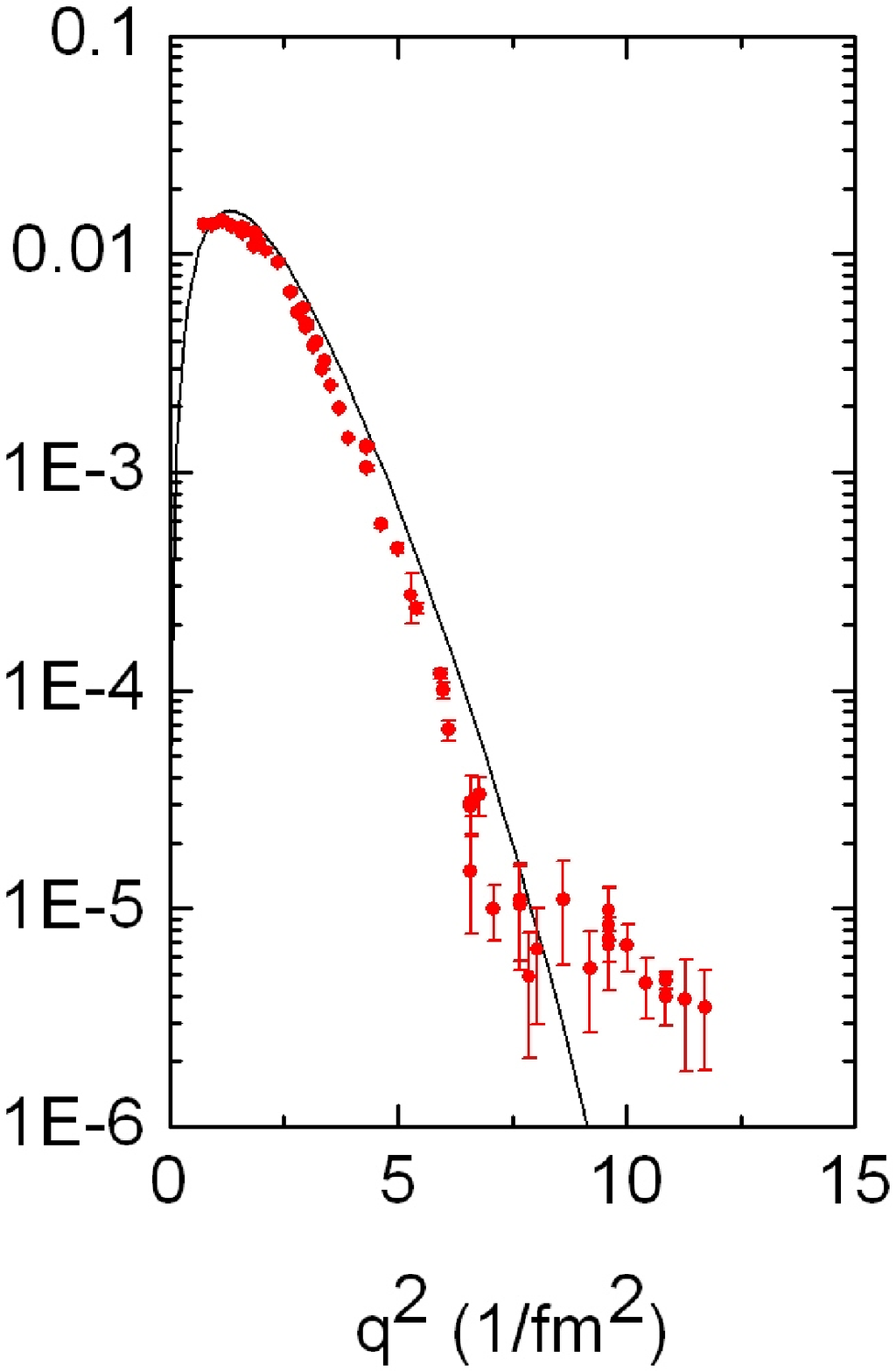,width=\linewidth}}
\end{minipage}
\begin{minipage}{.33\linewidth}
\centerline{\epsfig{file=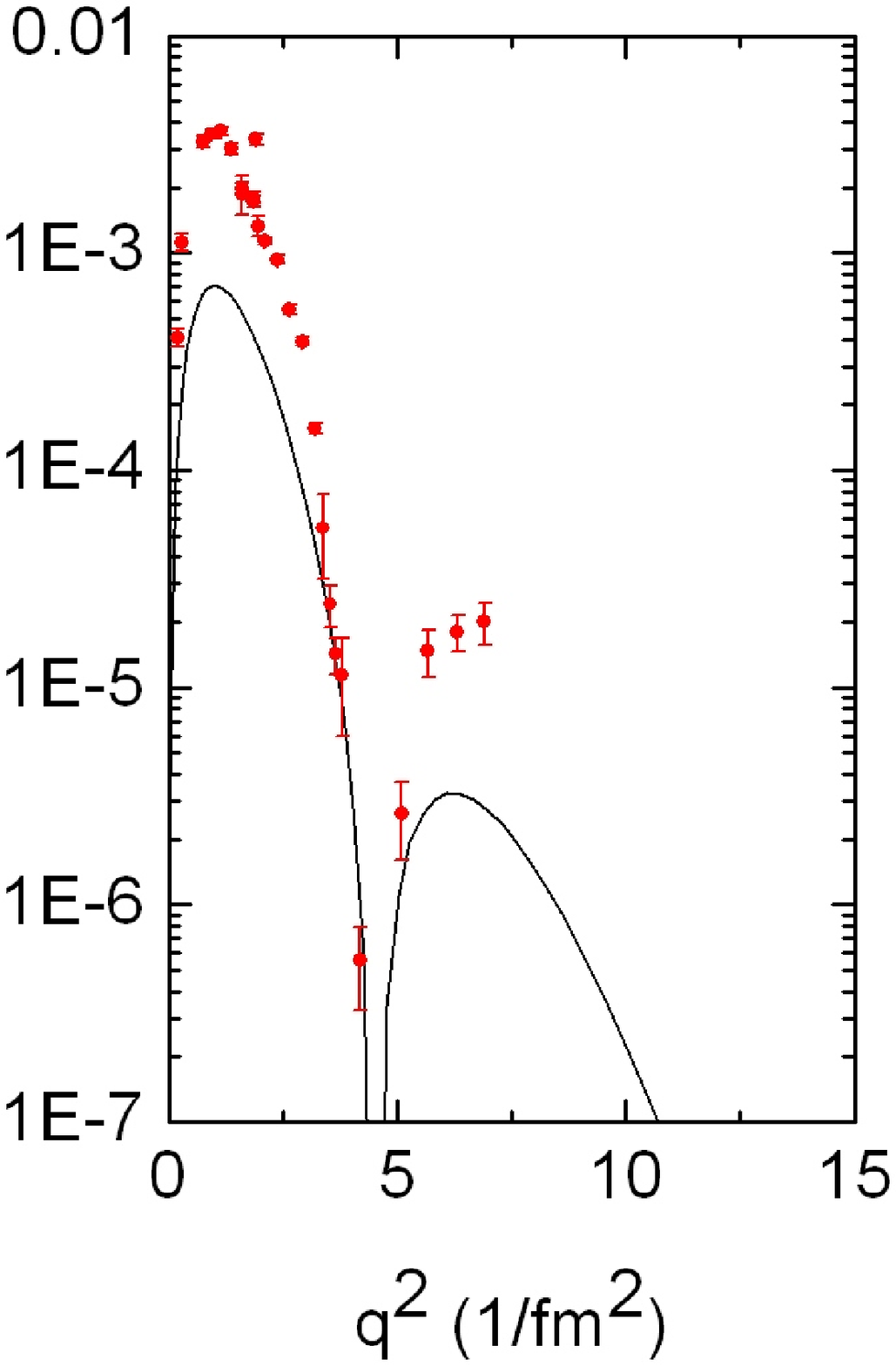,width=\linewidth}}
\end{minipage}
\caption[Form factors for $^{12}$C]
{Comparison between the experimental form factors $|{\cal F}(0^+_1 \rightarrow L^P_i;q)|^2$ of $^{12}$C 
for the final states with $L^P_i=0^+_1$, $2^+_1$ and $0^+_2$, and those obtained for the oblate top with $N=10$ 
and $R^2=1.40$. The experimental data are taken from \protect\cite{reuter}-\protect\cite{strehl}.}
\label{ffc12}
\end{figure*}

The $B(EL)$ values can be obtained by taking the long wavelength limit of the form factors according to 
Eq.~(\ref{belif}). The values extracted from the fit to the form factors are shown in Table~\ref{bem}. 
The large values of the electromagnetic transitions $B(E2;2_{1}^{+} \rightarrow 0_{1}^{+})$ and 
$B(E3;3_{1}^{-}\rightarrow 0_{1}^{+})$ indicate a collectivity which is not predicted for simple shell 
model states. The good agreement for the $B(EL)$ values and the transition form factors for the ground band 
shows that the positive and negative parity states merge to form a single rotational band.  
While the $B(E2;2_{1}^{+} \rightarrow 0_{1}^{+})$ and $B(E3;3_{1}^{-}\rightarrow 0_{1}^{+})$ values 
are in good agreement with experiment, the $B(E2;0_{2}^{+}\rightarrow 2_{1}^{+})$ value deviates by an 
order of magnitude. This indicates that the $0^{+}_2$ Hoyle state cannot be interpreted as a 
simple vibrational excitation of a rigid triangular configuration of three $\alpha$ particles, but rather 
corresponds to a more floppy configuration with large rotation-vibration couplings. A more detailed study 
of the electromagnetic properties of $\alpha$-cluster nuclei in the ACM for non-rigid configurations 
is in progress \cite{Bijker}. 

\begin{table}
\centering
\caption[]{Comparison between calculated and measured $B(EL)$ values in $^{12}$C. 
Experimental data are taken from \cite{reuter,strehl,ajz}.}
\label{bem}
\vspace{5pt}
\begin{tabular}{cccl}
\hline
\noalign{\smallskip}
& Th & Exp & \\
\noalign{\smallskip}
\hline
\noalign{\smallskip}
$\langle r^2 \rangle^{1/2}$ & 2.468 & $2.468 \pm 0.12$ & fm \\ 
$M(E0;0_{2}^{+} \rightarrow 0_{1}^{+})$ &  0.4 & $5.5 \pm 0.2$ & $\mbox{fm}^{2}$ \\
$B(E2;2_{1}^{+} \rightarrow 0_{1}^{+})$ &  8.4 & $7.6 \pm 0.4$ & $e^{2}\mbox{fm}^{4}$ \\
$B(E3;3_{1}^{-} \rightarrow 0_{1}^{+})$ & 44   & $103 \pm 17$ & $e^{2}\mbox{fm}^{6}$ \\
$B(E4;4_{1}^{+} \rightarrow 0_{1}^{+})$ & 73   & & $e^{2}\mbox{fm}^{8}$ \\  
\noalign{\smallskip}
\hline
\end{tabular}
\end{table}

\subsection{The nucleus $^{16}$O}

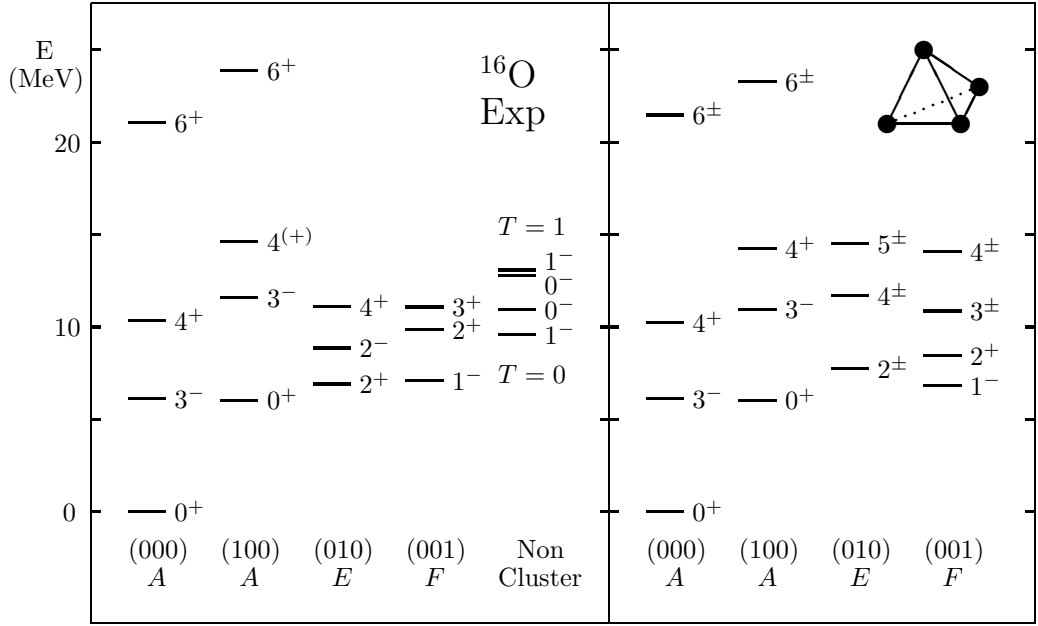
\begin{figure*}
\centering
\vspace{15pt}
\setlength{\unitlength}{0.7pt}
\begin{picture}(560,335)(-50,0)
\normalsize
\thinlines
\put (  0,  0) {\line(0,1){335}}
\put (  0,  0) {\line(1,0){510}}
\put (  0,335) {\line(1,0){510}}
\put (280,  0) {\line(0,1){335}}
\put (510,  0) {\line(0,1){335}}
\thicklines
\put (  0, 60) {\line(1,0){5}}
\put (  0,110) {\line(1,0){5}}
\put (  0,160) {\line(1,0){5}}
\put (  0,210) {\line(1,0){5}}
\put (  0,260) {\line(1,0){5}}
\put (  0,310) {\line(1,0){5}}
\put (275, 60) {\line(1,0){10}}
\put (275,110) {\line(1,0){10}}
\put (275,160) {\line(1,0){10}}
\put (275,210) {\line(1,0){10}}
\put (275,260) {\line(1,0){10}}
\put (275,310) {\line(1,0){10}}
\put (505, 60) {\line(1,0){5}}
\put (505,110) {\line(1,0){5}}
\put (505,160) {\line(1,0){5}}
\put (505,210) {\line(1,0){5}}
\put (505,260) {\line(1,0){5}}
\put (505,310) {\line(1,0){5}}
\put (-20, 55) { 0}
\put (-20,155) {10}
\put (-20,255) {20}
\put (-30,305) {E}
\put (-45,290) {(MeV)}
\put ( 20, 60.00) {\line(1,0){20}}
\put ( 20,121.30) {\line(1,0){20}}
\put ( 20,163.56) {\line(1,0){20}}
\put ( 20,270.52) {\line(1,0){20}}
\put ( 20, 35) {$(000)$}
\put ( 30, 20) {$A$}
\put ( 45, 55.00) {$0^+$}
\put ( 45,116.30) {$3^-$}
\put ( 45,158.56) {$4^+$}
\put ( 45,265.52) {$6^+$}
\put ( 70,120.49) {\line(1,0){20}}
\put ( 70,176.00) {\line(1,0){20}}
\put ( 70,206.20) {\line(1,0){20}}
\put ( 70,298.79) {\line(1,0){20}}
\put ( 70, 35) {$(100)$}
\put ( 80, 20) {$A$}
\put ( 95,115.49) {$0^+$}
\put ( 95,171.00) {$3^-$}
\put ( 95,201.20) {$4^{(+)}$}
\put ( 95,293.79) {$6^+$}
\put (120,129.17) {\line(1,0){20}}
\put (120,148.72) {\line(1,0){20}}
\put (120,170.97) {\line(1,0){20}}
\put (120, 35) {$(010)$}
\put (130, 20) {$E$}
\put (145,124.17) {$2^+$}
\put (145,143.72) {$2^-$}
\put (145,165.97) {$4^+$}
\put (170,131.16) {\line(1,0){20}}
\put (170,158.44) {\line(1,0){20}}
\put (170,170.80) {\line(1,0){20}}
\put (170, 35) {$(001)$}
\put (180, 20) {$F$}
\put (195,126.16) {$1^-$}
\put (195,153.44) {$2^+$}
\put (195,165.80) {$3^+$}
\put (220,155.85) {\line(1,0){20}}
\put (220,169.57) {\line(1,0){20}}
\put (220,187.96) {\line(1,0){20}}
\put (220,190.90) {\line(1,0){20}}
\put (230, 35) {Non}
\put (220, 20) {Cluster}
\put (245,150.85) {$1^-$}
\put (245,164.57) {$0^-$}
\put (220,130) {$T=0$}
\put (245,177.96) {$0^-$}
\put (245,190.90) {$1^-$}
\put (220,210) {$T=1$}
\Large
\put (210,270) {Exp}
\put (210,290) {$^{16}$O}
\normalsize
\put (300, 60.0) {\line(1,0){20}}
\put (300,121.3) {\line(1,0){20}}
\put (300,162.2) {\line(1,0){20}}
\put (300,274.6) {\line(1,0){20}}
\put (300, 35) {$(000)$}
\put (310, 20) {$A$}
\put (325, 55.00) {$0^+$}
\put (325,116.30) {$3^-$}
\put (325,157.2) {$4^+$}
\put (325,269.6) {$6^{\pm}$}
\put (350,120.5) {\line(1,0){20}}
\put (350,169.7) {\line(1,0){20}}
\put (350,202.5) {\line(1,0){20}}
\put (350,292.7) {\line(1,0){20}}
\put (350, 35) {$(100)$}
\put (360, 20) {$A$}
\put (375,115.5) {$0^+$}
\put (375,164.7) {$3^-$}
\put (375,197.5) {$4^+$}
\put (375,287.7) {$6^{\pm}$}
\put (400,137.4) {\line(1,0){20}}
\put (400,176.9) {\line(1,0){20}}
\put (400,205.1) {\line(1,0){20}}
\put (400, 35) {$(010)$}
\put (410, 20) {$E$}
\put (425,132.4) {$2^{\pm}$}
\put (425,171.9) {$4^{\pm}$}
\put (425,200.1) {$5^{\pm}$}
\put (450,128.5) {\line(1,0){20}}
\put (450,144.6) {\line(1,0){20}}
\put (450,168.7) {\line(1,0){20}}
\put (450,200.9) {\line(1,0){20}}
\put (450, 35) {$(001)$}
\put (460, 20) {$F$}
\put (475,123.5) {$1^-$}
\put (475,139.6) {$2^+$}
\put (475,163.7) {$3^{\pm}$}
\put (475,195.9) {$4^{\pm}$}
\put(430,270) {\circle*{10}} 
\put(470,270) {\circle*{10}}
\put(480,290) {\circle*{10}}
\put(450,310) {\circle*{10}}
\put(430,270) {\line( 1,0){40}}
\put(430,270) {\line( 1,2){20}}
\put(470,270) {\line(-1,2){20}}
\put(470,270) {\line( 1,2){10}}
\put(480,290) {\line(-3,2){30}}
\multiput(430,270)(5,2){11}{\circle*{2}}
\end{picture}
\caption{Comparison between the low-lying experimental spectrum of $^{16}$O and the energies
of the spherical top calculated using Eq.~(\ref{top}). The last column on
the left-hand side, shows the lowest observed non-cluster levels with $L^P=0^-$, $1^-$ and $T=0$, $1$.}.
\label{oxygen16}
\end{figure*}

As early as 1954, Dennison suggested that cluster states in $^{16}$O could be understood in terms of an 
$\alpha$-particle model with ${\cal T}_d$ symmetry \cite{Dennison}. This idea was adopted by Kameny \cite{Kameny}, 
Brink \cite{Brink1,Brink2}, and especially by Robson \cite{Robson}. In recent years, $^{16}$O has been again the 
subject of several investigations, both within the framework of the no-core shell model \cite{Navratil}, {\it ab initio} 
lattice calculations \cite{lattice} and the algebraic cluster model \cite{O16}.  

In this contribution, the spherical top limit of the $U(10)$ ACM with ${\cal T}_d$ symmetry is used to study cluster 
states in $^{16}$O. Fig.~\ref{oxygen16} shows a comparison of the cluster states of $^{16}$O with the spectrum of the 
spherical top with tetrahedral symmetry given by the energy formula 
\ba
E &=& E_0 + \omega_{1}(v_{1}+\tfrac{1}{2}) + \omega_{2}(v_{2}+1) 
\nonumber\\
&& + \omega_{3}(v_{1}+\tfrac{3}{2}) + \kappa_{(v)} \, L(L+1)  ~. 
\label{etop}
\ea
The coefficient $\kappa_{(v)}$ determines the the moments of inertia of the ground state band and 
the vibrational bands characterized by $(v)=(v_1 v_2 v_3)$. The vibrational energies $\omega_1$, 
$\omega_2$ and $\omega_3$ are obtained from the excitation energies of the first excited $0^+$, 
$2^+$ and $1^-$ states, respectively. 

\begin{figure}
\centering
\includegraphics[width=3in]{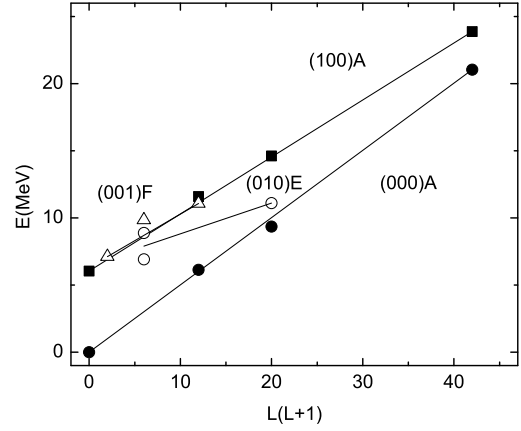}
\vspace{-0.5cm}
\caption[]{The excitation energies of cluster states in $^{16}$O 
plotted as a function of $L(L+1)$: closed circles for the ground state band, 
closed squares for the $A$ vibration, open circles for the $E$ vibration 
and open triangles for the $F$ vibration.} 
\label{bands}
\end{figure}

For the ground state rotational band the states with angular momentum and parity $L^P=0^+$, $3^-$, 
$4^+$, $6^+$ have been observed, only the $6^-$ state is missing. Another sequence of states $0^+$, 
$3^-$, $4^+$, $6^+$ has been identified as a candidate for the symmetric stretching or breathing 
vibration $(100)A$ with a somewhat larger moment of inertia than the ground state band which is 
to be expected for a breathing vibration. All three fundamental vibrations, $(100)A$, $(010)E$ 
and $(001)F$, have been observed with comparable energies, $\sim 6$ MeV. 
The band structure is summarized in Fig.~\ref{bands}. 

\begin{figure*}
\begin{minipage}{.33\linewidth}
\centerline{\epsfig{file=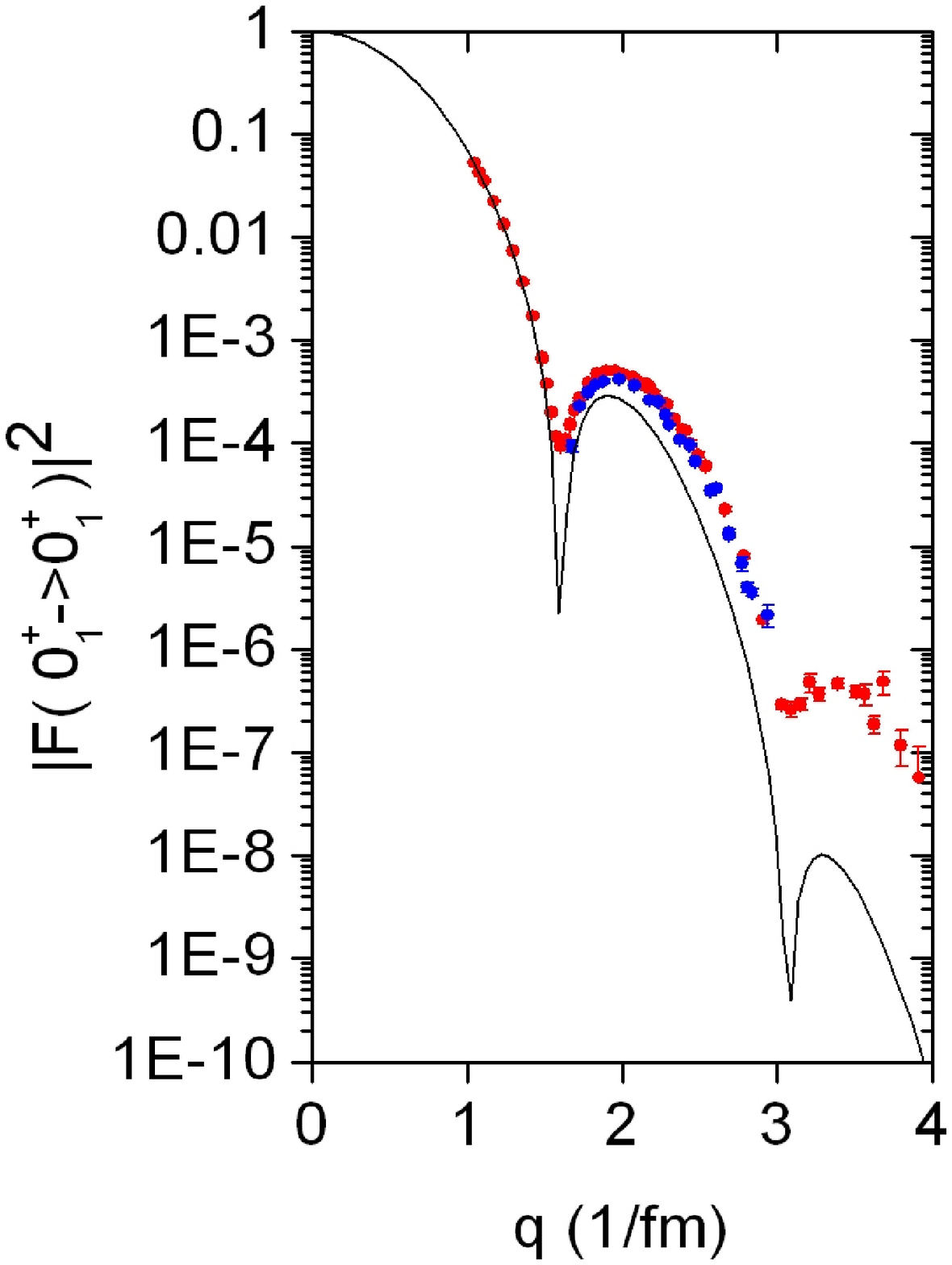,width=\linewidth}}
\end{minipage}\hfill
\begin{minipage}{.33\linewidth}
\centerline{\epsfig{file=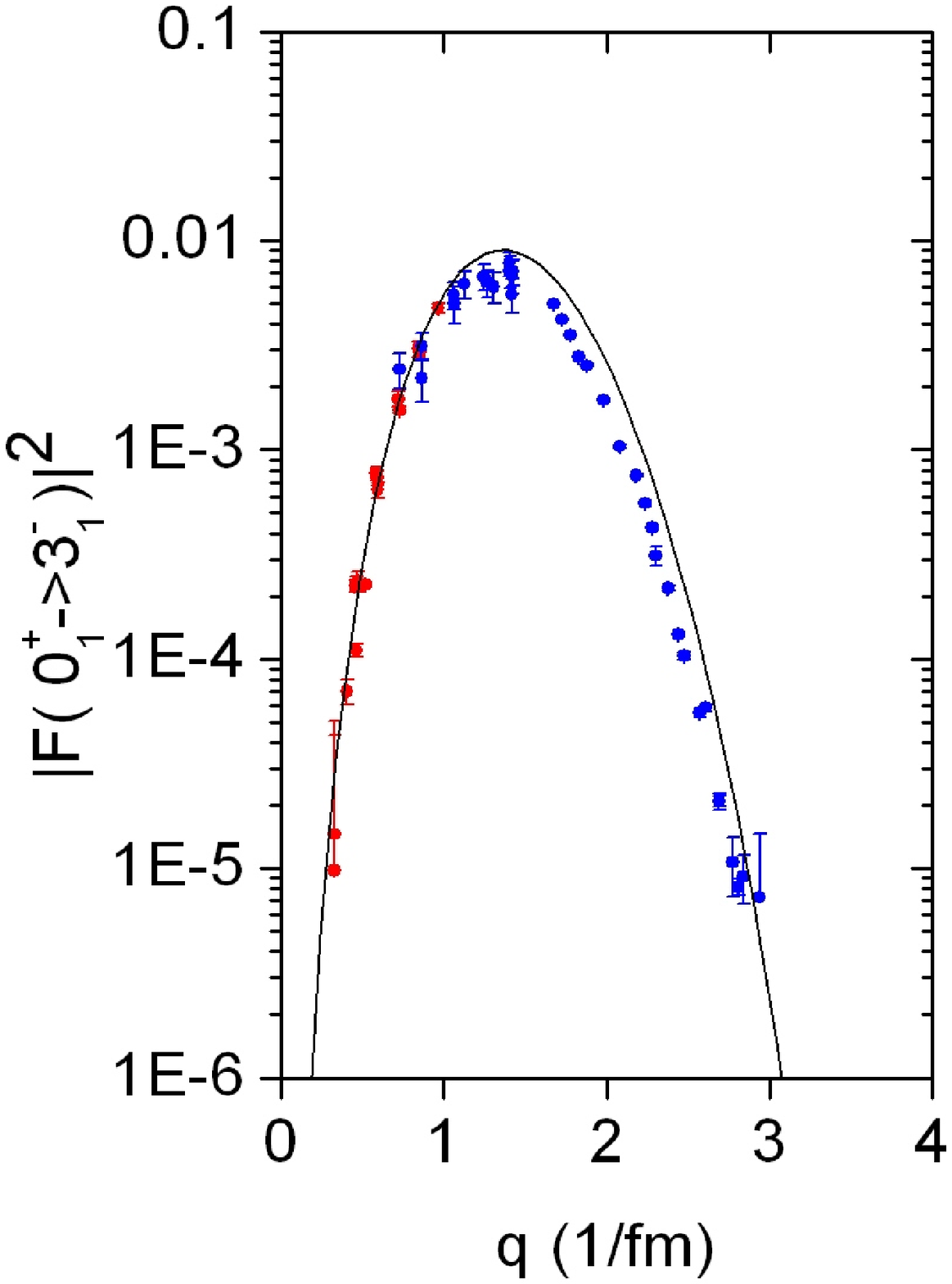,width=\linewidth}}
\end{minipage}
\begin{minipage}{.33\linewidth}
\centerline{\epsfig{file=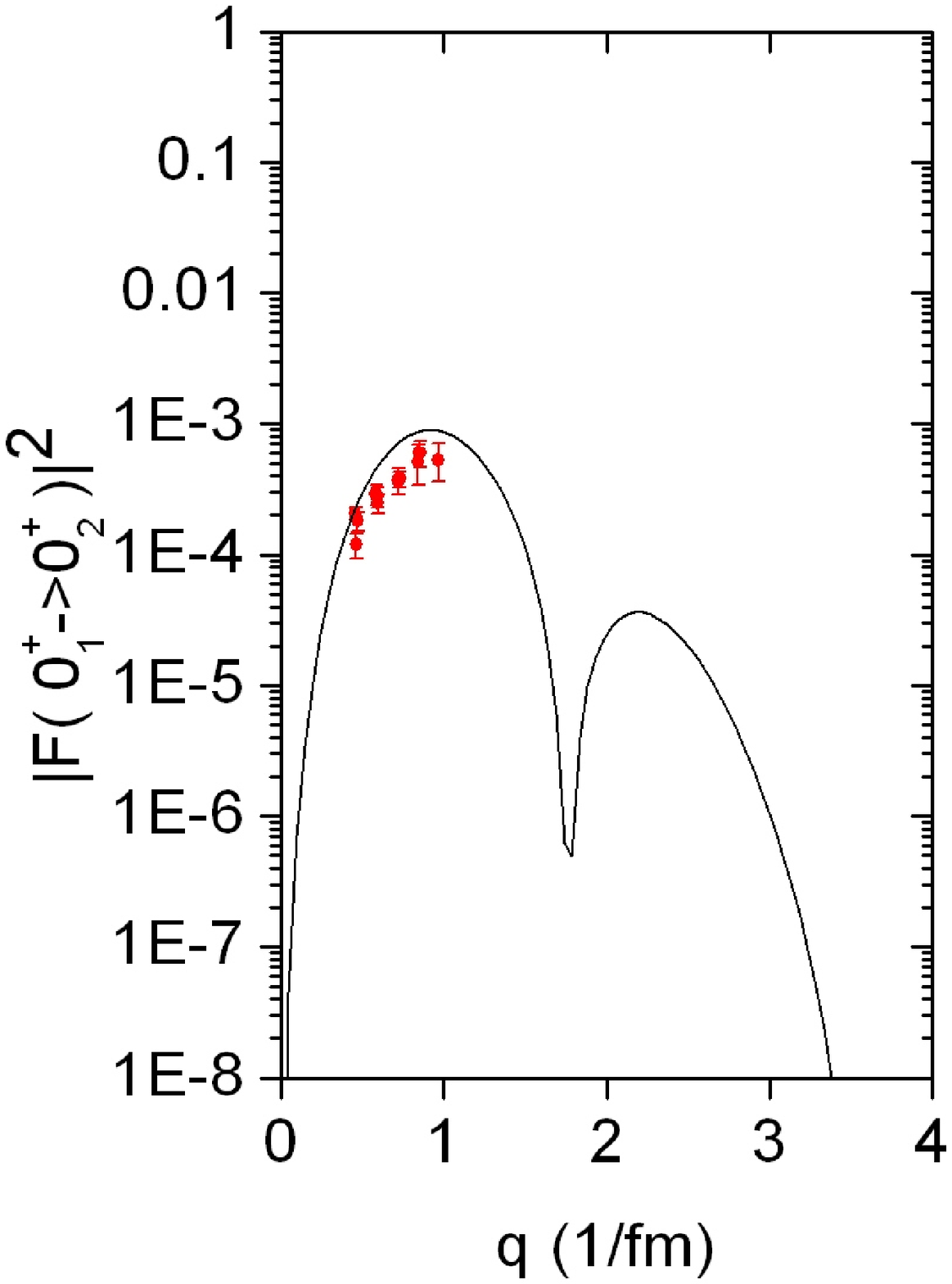,width=\linewidth}}
\end{minipage}\hfill
\caption[]{Comparison between the experimental form factors $|{\cal F}(0_1^+ \rightarrow L^P_i)|^2$ 
of $^{16}$O for the final states with $L^P_i=0^+_1$, $3^-_1$ and $0^+_2$ and those obtained 
for the spherical top with $N=10$ and $R^2=1.0$. The experimental data are taken from 
Refs.~\cite{sick,crannell2,Bergstrom1,Stroetzel,Bishop}.}
\label{ffgsb}
\end{figure*}

After having identified the cluster states, the ${\cal T}_d$ symmetry can be tested further by 
means of electromagnetic form factors and $B(EL)$ values. The value of $\beta$ was determined 
from the first minimum in the elastic form factor to be $\beta=2.071$ fm, and subsequently the 
value of $\alpha=0.605$ fm$^{-2}$ from the charge radius.  
Fig.~\ref{ffgsb} shows a comparison between experimental and calculated form factors.  
The experimental data for the excitation of the $3^-_1$ state (red) \cite{Bergstrom1,Stroetzel} are 
combined with those of the unresolved doublet of the $3^-_1$ and $0^+_2$ states at 6.1 MeV (blue) 
\cite{crannell2,Bishop}. Table~\ref{BELtable} shows the results for the $B(EL)$ values. 

\begin{table}
\caption{Comparison between calculated and measured $B(EL)$ values in $^{16}$O.
The experimental values are taken from \cite{NDS}.}
\label{BELtable}
\vspace{5pt}
\begin{tabular}{cccl}
\hline
\noalign{\smallskip}
& Th & Exp & \\
\noalign{\smallskip}
\hline
\noalign{\smallskip}
$\langle r^2 \rangle^{1/2}$ & 2.710 & $2.710 \pm 0.015$ & fm \\ 
$M(E0;0_{2}^{+} \rightarrow 0_{1}^{+})$ &  0.54 & $3.55 \pm 0.21$ & $\mbox{fm}^{2}$ \\
$B(E3;3_1^- \rightarrow 0_1^+)$ &  215 & $205 \pm  10$ & $e^{2}\mbox{fm}^{6}$ \\
$B(E4;4_1^+ \rightarrow 0_1^+)$ &  425 & $378 \pm 133$ & $e^{2}\mbox{fm}^{8}$ \\
$B(E6;6_1^+ \rightarrow 0_1^+)$ & 9626 &               & $e^{2}\mbox{fm}^{12}$ \\ 
\noalign{\smallskip}
\hline
\end{tabular}
\end{table}

In conclusion, the evidence for the occurrence of the tetrahedral symmetry in the low-lying spectrum 
of $^{16}$O presented long ago by Dennison \cite{Dennison}, Kameny \cite{Kameny} and Robson \cite{Robson}. 
was confirmed in a study of the energy spectrum in the ACM for four-body clusters albeit with some 
differences in the assignments of the states. A study of $B(EL)$ values along the ground state 
band provides additional evidence for ${\cal T}_d$ symmetry. 
Finally, the results presented here for $^{12}$C and $^{16}$O emphasize the occurrence 
of $\alpha$-cluster states in light nuclei with ${\cal D}_{3h}$ and ${\cal T}_d$ point 
group symmetries, respectively.  

\section{Summary and conclusions}

In this contribution, I presented a review of the algebraic cluster model for two-, three- 
and four-body clusters. The ACM is an interacting boson model that describes the relative motion 
of cluster configurations in which all vibrational and rotational degrees of freedom are present 
from the outset. Special attention was paid to the case of identical clusters, and the consequences 
of their geometrical configuration on the structure of rotational bands (for a summary see 
Table~\ref{ACMsummary}). For three identical clusters located at the vertices of an equilateral triangle, 
the ground state rotational band has $L^P=0^+$, $2^+$, $3^-$, $4^{\pm}$, $5^-$, $\ldots$, 
all of which have been observed in $^{12}$C. For four identical clusters located at the vertices 
of a tetrahedron, the sequence is given by $L^P=0^+$, $3^-$, $4^+$, $6^{\pm}$, $\ldots$. With the 
exception of the $6^-$ state, all have been observed in $^{16}$O. 

\begin{table}
\caption[]{Algebraic Cluster Model} 
\label{ACMsummary}
\vspace{5pt}
\begin{tabular}{cccc}
\hline
\hline
\noalign{\smallskip}
& $2\alpha$ & $3\alpha$ & $4\alpha$ \\
\noalign{\smallskip}
\hline
\noalign{\smallskip}
ACM & $U(4)$ & $U(7)$ & $U(10)$ \\
Point group   & ${\cal C}_2$  & ${\cal D}_{3h}$ & ${\cal T}_{d}$ \\
Geom. conf. & Linear  & Triangle & Tetrahedron \\
Model & Rotor & Obl. top & Sph. top \\
Vibrations & $1$  & $3$ & $6$ \\
Rotations & $2$  & $3$  & $3$ \\
\noalign{\smallskip}
\hline
\noalign{\smallskip}
G.s. band & $0^+$ & $0^+$ & $0^+$ \\
          & $2^+$ & $2^+$ &       \\
          &       & $3^-$ & $3^-$ \\
          & $4^+$ & $4^{\pm}$ & $4^+$ \\
          &       & $5^-$ &       \\
          & $6^+$ & $6^{\pm+}$ & $6^{\pm}$ \\  
\noalign{\smallskip}
\hline
\hline
\end{tabular}
\end{table}

The structure of the rotational bands can be considered as the fingerprint of the underlying 
geometrical cluster configuration. Whereas most theoretical models and {\it ab initio} calculations 
agree on a triangular configuration of $\alpha$ particles for the ground state band in $^{12}$C 
\cite{C12,EFT,ACM} and a tetrahedral configuration for the ground state band in $^{16}$O 
\cite{O16,lattice}, important differences are found in the predictions for the structure of the 
rotational band built on the first excited $0^+$ state. The so-called Hoyle band in $^{12}$C,  
{\it i.e.} the rotational excitations of the Hoyle state, is of particular interest \cite{Physics}. 
While the observed moment of inertia of the Hoyle band excludes the proposed linear chain structure 
of the Hoyle state \cite{Morinaga}, there are two alternatives for the geometrical arrangement of 
the three alpha particles in the Hoyle state of $^{12}$C, either an equilateral triangular arrangement 
\cite{C12,ACM} or a bent-arm configuration as suggested by EFT lattice calculations \cite{EFT}. 
In order to distinguish between these two different geometrical configurations of the Hoyle band, 
the identification of the negative parity states $3^-$ and $4^-$ is crucial. The selectivity of 
$\gamma$-ray beams as well as electron beams could help to populate the states of interest and 
resolve the broad interfering states. These new capabilities should initiate an extensive experimental 
program for the search of the predicted (``missing") states and promises to shed new light on the 
clustering phenomena in light nuclei.

A similar situation exists for the first excited $0^+$ state in $^{16}$O. Whereas the ACM is based 
on a tetrahedral configuration of the $\alpha$ particles, {\it ab initio} lattice calculations 
suggested a square configuration \cite{lattice}. The latter configuration would imply a large breaking 
of the ${\cal T}_d$ symmetry for the vibrations in Fig.~\ref{oxygen16}. 

As a final remark, the ACM is able to account for many different geometrical configurations other than the 
rigid structures for identical clusters as discussed in this contribution. It can be applied both  
for identical and non-identical clusters, and for rigid and floppy structures.  
One of the most challenging problems in clustering in nuclei is to understand what type of configurations 
are present, and to provide unambiguous experimental evidence for these configurations. The algebraic
method provides a general theoretical framework in which calculations can be performed easily in a clear 
and transparent manner. 

\section*{Acknowledgments}

It is a pleasure to thank Franco Iachello and Moshe Gai for many interesting and stimulating 
discussions on $\alpha$-clustering in light nuclei. 
This work was supported in part by research grant IN107314 from PAPIIT-DGAPA, UNAM.  

\appendix

\section{Triangular symmetry}
\label{app3b}

There are three different symmetry classes for the permutation of three objects. 
Due to the isomorphism with the dihedral group $S_3 \sim {\cal D}_3$, the three 
symmetry classes can also be labeled by the irreducible representations of the point 
group ${\cal D}_3$ as $[3] \sim A_{1}$, $[21] \sim E$, and $[111] \sim A_2$, 
with dimensions 1, 2 and 1, respectively. 

The permutation symmetry can be determined by considering the transposition $P(12)$ and the 
cyclic permutation $P(123)$. The transformation properties of the three different symmetry 
classes under $P(12)$ and $P(123)$ are given by
\ba
P(12) \left( \begin{array}{c} \psi_{A_1} \\ \psi_{A_2} \end{array} \right) 
= \left( \begin{array}{rr} 1 & 0 \\ 0 & -1 \end{array} \right) 
\left( \begin{array}{c} \psi_{A_1} \\ \psi_{A_2} \end{array} \right) , \\
P(12) \left( \begin{array}{c} 
\psi_{E_{\rho}} \\ \psi_{E_{\lambda}} \end{array} \right) 
= \left( \begin{array}{rr} -1 & 0 \\ 0 & 1 \end{array} \right) 
\left( \begin{array}{c} \psi_{E_{\rho}} \\ \psi_{E_{\lambda}} \end{array} \right) ,
\ea
and 
\ba
P(123) \left( \begin{array}{c} \psi_{A_1} \\ \psi_{A_2} \end{array} \right) 
= \left( \begin{array}{cc} 1 & 0 \\ 0 & 1 \end{array} \right) 
\left( \begin{array}{c} \psi_{A_1} \\ \psi_{A_2} \end{array} \right) , \\
P(123) \left( \begin{array}{c} 
\psi_{E_{\rho}} \\ \psi_{E_{\lambda}} \end{array} \right) 
= \left( \begin{array}{cc} -\frac{1}{2} & \frac{\sqrt{3}}{2} \\ 
-\frac{\sqrt{3}}{2} & -\frac{1}{2} \end{array} \right)  
\left( \begin{array}{c} \psi_{E_{\rho}} \\ \psi_{E_{\lambda}} \end{array} \right) .
\ea

\section{Tetrahedral symmetry}
\label{app4b}

There are five different symmetry classes for the permutation of four objects. 
Due to the isomorphism with the tetrahedral group $S_4 \sim {\cal T}_d$, the five 
symmetry classes can also be labeled by the irreducible representations of the point 
group ${\cal T}_d$ as $[4] \sim A_{1}$, $[31] \sim F_2$, $[22] \sim E$, $[211] \sim F_1$ 
and $[1111] \sim A_2$, with dimensions 1, 3, 2, 3 and 1, respectively. 

The permutation symmetry can be determined by considering the transposition $P(12)$ and the 
cyclic permutation $P(1234)$. The transformation properties of the five different symmetry 
classes under $P(12)$ and $P(1234)$ are given by
\ba
P(12) \left( \begin{array}{c} \psi_{A_1} \\ \psi_{A_2} \end{array} \right) 
&=& \left( \begin{array}{rr} 1 & 0 \\ 0 & -1 \end{array} \right) 
\left( \begin{array}{c} \psi_{A_1} \\ \psi_{A_2} \end{array} \right) , \\
P(12) \left( \begin{array}{c} 
\psi_{E_{\rho}} \\ \psi_{E_{\lambda}} \end{array} \right) 
&=& \left( \begin{array}{rr} -1 & 0 \\ 0 & 1 \end{array} \right) 
\left( \begin{array}{c} \psi_{E_{\rho}} \\ \psi_{E_{\lambda}} \end{array} \right) , \\
P(12) \left( \begin{array}{c} 
\psi_{F_{2\rho}} \\ \psi_{F_{2\lambda}} \\ \psi_{F_{2\eta}} \end{array} \right) 
&=& \left( \begin{array}{rrr} -1 & 0 & 0 \\ 0 & 1 & 0 \\ 0 & 0 & 1 \end{array} \right) 
\left( \begin{array}{c} \psi_{F_{2\rho}} \\ \psi_{F_{2\lambda}} \\ \psi_{F_{2\eta}} \end{array} \right) , \\
P(12) \left( \begin{array}{c} 
\psi_{F_{1\rho}} \\ \psi_{F_{1\lambda}} \\ \psi_{F_{1\eta}} \end{array} \right) 
&=& \left( \begin{array}{rrr} 1 & 0 & 0 \\ 0 & -1 & 0 \\ 0 & 0 & -1 \end{array} \right) 
\left( \begin{array}{c} \psi_{F_{1\rho}} \\ \psi_{F_{1\lambda}} \\ \psi_{F_{1\eta}} \end{array} \right) ,
\nonumber\\
\mbox{}
\ea
and 
\ba
P(1234) \left( \begin{array}{c} \psi_{A_1} \\ \psi_{A_2} \end{array} \right) 
&=& \left( \begin{array}{cc} 1 & 0 \\ 0 & 1 \end{array} \right) 
\left( \begin{array}{c} \psi_{A_1} \\ \psi_{A_2} \end{array} \right) , \\
P(1234) \left( \begin{array}{c} 
\psi_{E_{\rho}} \\ \psi_{E_{\lambda}} \end{array} \right) 
&=& \left( \begin{array}{cc} \frac{1}{2} & -\frac{\sqrt{3}}{2} \\ 
-\frac{\sqrt{3}}{2} & -\frac{1}{2} \end{array} \right)  
\left( \begin{array}{c} \psi_{E_{\rho}} \\ \psi_{E_{\lambda}} \end{array} \right) ,
\nonumber\\
\mbox{}
\ea
\ba
P(1234) \left( \begin{array}{c} 
\psi_{F_{2\rho}} \\ \psi_{F_{2\lambda}} \\ \psi_{F_{2\eta}} \end{array} \right) 
\nonumber\\
\hspace{1cm} = \left( \begin{array}{ccc} -\frac{1}{2} & \frac{\sqrt{3}}{2} & 0 \\ 
-\frac{1}{2\sqrt{3}} & -\frac{1}{6} & \frac{\sqrt{8}}{3} \\ 
-\frac{\sqrt{2}}{\sqrt{3}} & -\frac{\sqrt{2}}{3} & -\frac{1}{3} \end{array} \right)  
\left( \begin{array}{c} \psi_{F_{2\rho}} \\ \psi_{F_{2\lambda}} \\ \psi_{F_{2\eta}} 
\end{array} \right) , \\
P(1234) \left( \begin{array}{c} 
\psi_{F_{1\rho}} \\ \psi_{F_{1\lambda}} \\ \psi_{F_{1\eta}} \end{array} \right) 
\nonumber\\ 
\hspace{1cm} = \left( \begin{array}{ccc} \frac{1}{2} & -\frac{\sqrt{3}}{2} & 0 \\ 
\frac{1}{2\sqrt{3}} & \frac{1}{6} & -\frac{\sqrt{8}}{3} \\ 
\frac{\sqrt{2}}{\sqrt{3}} & \frac{\sqrt{2}}{3} & \frac{1}{3} \end{array} \right)  
\left( \begin{array}{c} \psi_{F_{1\rho}} \\ \psi_{F_{1\lambda}} \\ \psi_{F_{1\eta}} \end{array} \right) .
\ea

\end{document}